\documentclass{aa}
\usepackage[pdftex]{graphicx}
\usepackage{epsfig}
\usepackage{amsmath}
\usepackage{wasysym}

\usepackage{marvosym}
\usepackage{txfonts}
\pdfoutput=1
\usepackage[pdftex]{color,xcolor}
\usepackage{hyperref}
\hypersetup{pdfauthor=Christoph Mordasini}
\hypersetup{backref=true, pagebackref=true, hyperindex=true, breaklinks=true,colorlinks=true,urlcolor=blue, linkcolor=blue,  citecolor=blue,pagecolor=red, bookmarks=true, bookmarksopen=true}
\usepackage{epstopdf}

\def\mearth{M_\oplus}
\def\rearth{R_\oplus}

\def\msun{M_\odot}
\def\rhill{R_{\rm H}}
\def\mcore{M_{\rm core}}

\def\fpg{f_{\rm D/G}} 

\def\f1{f_{\rm I}}
\def\mj{M_{\textrm{\tiny \jupiter }}}
\newcommand{\lj}{L_{\textrm{\tiny \jupiter}}}
\newcommand{\rj}{R_{\textrm{\tiny \jupiter}}}

\def\mstar{M_*}
\def\mdisk{M_{\rm disk}}

\def\rcore{R_{\rm core}}

\def\beq{\begin{equation}}
\def\eeq{\end{equation}}

\def\fopa{f_{\rm opa}}
\def\t2{\tau_{\rm II}}

\def\sigmas0{\Sigma_{\rm s,0}}

\newcommand{\mz}{M_{\rm Z}}
\newcommand{\mxy}{M_{\rm XY}}

\newcommand{\mdotz}{\dot{M}_{\rm Z}}

\def\araa{ARA\&A}             
\def\apj{ApJ}                 
\def\apjl{ApJ}                
\def\apjs{ApJS}               
\def\aap{A\&A}                
\def\mnras{MNRAS}             
\def\pasp{PASP}               
\def\ssr{Space~Sci.Rev.}      

\def\({\left(}
\def\){\right)}
\def\<{\left<}
\def\>{\right>}

\begin{document}

\title{Characterization of exoplanets from their formation II:\\ The planetary mass-radius relationship}

\author{C. Mordasini\inst{1}  \and  Y. Alibert\inst{2}  \and C. Georgy\inst{3} \and K.-M. Dittkrist\inst{1} \and H. Klahr\inst{1}  \and T. Henning\inst{1}}

\institute{ Max-Planck-Institut f\"ur Astronomie, K\"onigstuhl 17, D-69117 Heidelberg, Germany \and
Center for space and habitability, Physikalisches Institut, University of Bern, Sidlerstrasse 5, CH-3012 Bern, Switzerland \and Centre de Recherche Astrophysique de Lyon (CRAL), Ecole Nationale Sup\'erieure, 46, All\'ee d'Italie, 69364 Lyon Cedex 07, France  }

\offprints{Christoph MORDASINI, \email{mordasini@mpia.de}}

\date{Received 15.11.2011 / Accepted 27.8.2012}

\abstract
{The research of extrasolar planets has entered an era in which we characterize extrasolar planets. This has become possible with measurements of the radii of transiting planets, and the luminosity of planets observed by direct imaging.  {Meanwhile, the} precision of  radial velocity surveys allows to discover not only giant planets, but also very low-mass planets.}
{Uniting all these different observational constraints  into one coherent picture to better understand planet formation is an important, but also difficult undertaking. One  {approach} is to develop a theoretical model which can make testable predictions for all  {these} observational techniques. Our goal is to have such a model, and to use it in population synthesis calculations.}
{In a companion paper, we described how we have extended our formation model into a self-consistently coupled formation \textit{and} evolution model. In this second paper, we  {first} continue with the model description. We describe how we calculate the internal structure of the solid core of the planet, and include radiogenic heating. We also introduce an upgrade of the  {protoplanetary} disk model.  {Finally, we use the upgraded model in population synthesis calculations.}}
{ {We present how the planetary mass-radius relationship of planets with primordial H$_{2}$/He envelopes forms and evolves in time. The basic shape of the mass-radius relationship can be understood from the core accretion model. Low-mass planets cannot bind massive envelopes, while super-critical cores necessarily trigger runway gas accretion, leading to ``forbidden'' zones in the $M-R$ plane.  For a given mass, there is a considerable diversity of radii, mainly due to different bulk compositions, reflecting different formation histories. We compare the synthetic $M-R$ plane with the observed one, finding good agreement for $a>0.1$ AU. The synthetic planetary radius distribution is characterized by a  strong increase towards small $R$, and a second, lower local maximum at about $1\rj$. The increase towards small radii comes from the increase of the mass function towards low $M$. The second local maximum is due to the fact that radii are nearly independent of mass for giant planets. A comparison of the synthetic radius distribution with \textit{Kepler} data shows good agreement for $R\gtrsim2\rearth$, but divergence for smaller radii. This indicates that for $R\gtrsim2\rearth$, the radius distribution can be described with planets with primordial H$_{2}$/He atmospheres, while at smaller radii, planets of a different nature dominate. We predict that in the next few years, \textit{Kepler} should find the second, local maximum at about 1 $\rj$.}}
{With the updated model, we can compute  the most important  quantities like mass, semimajor axis, radius and luminosity characterizing an extrasolar planet  {self-consistently from its formation}.  {The comparison of the radii of the synthetic planets with observations allows to better constrain this formation process, and to distinguish between fundamental types of planets. } }

\titlerunning{Characterization of exoplanets from their formation II}
\authorrunning{C. Mordasini et al.}

\maketitle

\section{Introduction}
In the  {first part of this} paper we  continue to present the improvement of our formation model in order to extend it into a combined planet formation \textit{and} evolution model.  {With these extensions, and those introduced and in the companion paper (Mordasini et al. \cite{mordasinialibert2011}, hereafter Paper I), we are  able to synthesize planetary populations for which we can calculate in a self-consistent way not only the mass and semimajor axis as in previous works (e.g. Mordasini et al. \cite{mordasinialibert2009a}, Alibert et al. \cite{alibertmordasini2011}) but also the radius and luminosity. This is possible during the entire formation and evolution, starting at a tiny sub-Earth mass seed embryo and ending at a Gigayear old planet. In the second part of the paper, we use population synthesis calculations to study different aspects of  the planetary mass-radius relationship and the distribution of the radii of the planets.} 

{The background to motivate this work is that the number of known transiting extrasolar planets has increased very rapidly in the past few years. Recently, the space missions \textit{CoRoT} and \textit{Kepler} have yielded very precise measurements of the radius of a very large number of planet candidates (L\'eger et al. \cite{legerrouan2009}; Borucki et al. \cite{boruckikoch2011}, Batalha et al. \cite{batalharowe2012}). The data from the \textit{Kepler} satellite has stimulated a multitude of studies like for example Howard et al. (\cite{howardmarcy2011}); Traub (\cite{traub2011});  Demory \& Seager (\cite{demoryseager2011}); Youdin  (\cite{youdin2011}); Wolfgang \& Laughlin (\cite{wolfganglaughlin2012}) or Gaidos et al. (\cite{gaidosfischer2012}) to name just a few. The results of these studies will be used in this work to compare the radii of the actual and the synthetic population.}

{For a number of transiting planets, masses were measured using the radial velocity method (e.g. Bouchy et al. \cite{bouchybonomo2011}), or by transit timing variations (e.g. Cochran et al. \cite{cochranfabrycky2011}). By the combination of the mass and the radius of the planet one gets the planetary mass-radius diagram, which is an observational result of similar importance as the planetary semimajor axis-mass diagram. The reason for this is that one can derive the mean density of the planet, which constrains, at least to some extent, the internal structure which is crucial to understand the nature (e.g. Chabrier et al. \cite{chabrierleconte2011}) and, as we shall see in this work, the formation of a planet.  We thus make an important step towards a physical characterization of planets based on their formation history.} 

{The radius adds a whole new class of observational constraints to which the output of the theoretical model in a population synthesis  can be compared to. Important examples are the radius distribution, the mass-radius relationship, or the impact of semimajor axis, metallicity and stellar mass on the radii. Due to the complexity of planet formation and evolution with a multitude of interacting effects (often only understood in a rudimentary way), it is of highest importance to have as many observational constraints which can concurrently and self-consistently be used for comparison with the theoretical results. This is the central motivation for the work here and in the companion paper. The combination of results of different observational techniques allows to obtain insights which could not be obtained from the data of one technique alone (e.g. Wolfgang \& Laughlin \cite{wolfganglaughlin2012}; Gaidos et al. \cite{gaidosfischer2012}).}

{The reason to improve the model of the protoplanetary disk is on one hand the observational progress that has been made in the last few years also in this field (e.g. Andrews et al. \cite{andrewswilner2009}, \cite{andrewswilner2010}; Fedele et al. \cite{fedeleancker2010}), but also the theoretical result that the disk structure is very important for the final properties of a synthetic planetary population (Mordasini et al. \cite{mordasinialibert2012})}

{Planetary population synthesis has become possible thanks to the large number of extrasolar planets that have been detected in the last decade. Apart from our series of works, several other studies have been presented (e.g. Ida \& Lin \cite{idalin2004}-\cite{idalin2010}; Thommes et al. \cite{thommesmatsumura2008}; Miguel et al. \cite{migueletal2011}; Hellary \& Nelson \cite{hellarynelson2012}). These studies can roughly be divided into two classes, one emphasizing more the coupling of several processes (solid and gas accretion, migration, disk evolution), the other treating in details the N-body interaction.   Here, we add the long-term evolution in order to allow a physical characterization as an additional aspect.  A natural extension of this approach will be in future a better description of the atmospheric structure and the chemical compositions of the planets, as well as the inclusion of  atmospheric evaporation and the formation of secondary atmospheres for low-mass planets.  This will finally lead to the study of planetary habitability.}

{It is the  goal of population synthesis  to eventually calculate from first physical principles a synthetic  population that is able to reproduce in a statistically significant way the observed properties of the actual population in terms of its major characteristics like mass, semimajor axis, eccentricity, composition, radius and luminosity.  The formation model used here for the syntheses combines several simple standard descriptions for important physical mechanism acting during planet formation ($\alpha$-disk model, core accretion mechanism, tidal migration, 1D planetary structure equations). It is important to understand to what extent this much simplified approach is sufficient to reach the mentioned goal. Comparisons with as many observational results as possible should allow to identify which aspects of current planet formation and evolution theory need most improvements. We also want to understand if physical mechanisms which are currently not at all included in the model are necessary in order to explain the observed properties.  Examples are - among may others - the interaction of planetary systems with third, external bodies causing Kozai migration  (Fabrycky \& Tremaine \cite{fabryckytremaine2007}) or dynamical instabilities (Malmberg et al. \cite{malmbergdavies2011}).}

In related papers we present how we have concurrently improved also other aspects of our original formation model (Alibert, Mordasini \& Benz \cite{alibertmordasini2004}, Alibert et al. \cite{alibertmordasini2005}). We worked on the inclusion of stellar irradiation for the temperature structure of the protoplanetary disk (Fouchet et al. \cite{fouchetalibert2011}), the description of disk migration  taking  into account  non-isothermal type I migration (Dittkrist et al. in prep.),  the concurrent formation of many planets in one disk (Alibert et al. in prep.) and a more realistic planetesimal disk and solid accretion rate (Fortier et al. submitted).

\subsection{Structure of the papers}
In Paper I we have presented the improvements and extensions regarding the calculation of the gaseous envelope of a planet. We have introduced a new, simple and stable method to calculate planetary cooling tracks. We have also described  the boundary conditions which are used to solve the structure equations.

In Paper I we have used the upgraded model to study  the in situ formation and evolution of Jupiter, the  {radius of giant planets as function of their mass and age}, the influence of the core mass on the radius of giant planets and the planetary luminosity both in the ``hot start'' and the ``cold start'' scenario. We put special emphasis on the comparison of our results with those of other models of planet formation and evolution like Burrows et al. (\cite{burrowsmarley1997}), Baraffe et al. (\cite{baraffechabrier2003}), Fortney et al. (\cite{fortneymarley2007}) or Lissauer et al. (\cite{lissauerhubickyj2009}). We found that our results agree very well with those of the more complex models, despite a number of simplifications we make. 

In this second paper, we introduce further improvements: In Section \ref{sect:planetmodel} we describe two upgrades of the computational module which describes the planetary interior structure. First, we now solve the internal structure equations of the solid core\footnote{We use the expression core in the astrophysical ``giant planet'' sense denoting the part of the planet not consisting of hydrogen and helium. This is different from the geophysical meaning where the core is the central part of a planet consisting of iron-nickel, but not including the silicates (the mantle) and possibly ices.} which yields a variable, realistic core density as a function of core mass, core composition and external pressure by the surrounding envelope (Sect. \ref{sect:variablecoredensity}).  We assume that the planet is differentiated, consisting of iron, silicates, and, if it accreted outside the ice line, ices.  {The ice line corresponds to the orbital distance in the protoplanetary disk where the temperature drops below the level necessary for ices to condense, see e.g. Hayashi (\cite{hayashi1981}). } Second, we include radiogenic heating due to the decay of long-lived radionuclides in the mantle of the planet (Sect. \ref{sect:upgradelradio}).  {These extensions should in particular allow to calculate the evolution also of planets of a relatively low mass, provided that they have a primordial H$_{2}$/He atmosphere.} In Section \ref{sect:disk model} we show how we have improved the disk model by using more realistic initial and boundary conditions, and a detailed model for the photoevaporation of the disk. 

{In the following sections, we present our results.} We first use the upgrades to study the effect of a variable core density on the in situ formation and evolution of Jupiter (Sect. \ref{result:coredensity}), and the formation and evolution of a close-in super-Earth\footnote{With super-Earth, we denote planets with masses between 1 and 10 $\mearth$, independent of their composition.} planet under the influence of radiogenic heating (Sect. \ref{result:lradio}). As in Paper I, we  want to validate the model by comparison with previous studies. Our goal is also to understand how these mechanisms potentially influence  population synthesis calculations.  {The degree of uncertainty which affects our radius calculation is estimated in Sect. \ref{sect:uncertainty}.}  {The most important result, which is the application of the updated model in population synthesis calculations to obtain the planetary mass-radius relationship is shown in  Sect. \ref{sect:formationevolutionmrr}. We study  how the planetary mass-radius relationship forms and evolves (Sect. \ref{sect:mrrformation}). We show the predicted planetary radius distribution and compare with the results of the \textit{Kepler} satellite (Sect. \ref{sect:planetaryrdist}). We  study the impact of the semimajor axis on the radius distribution and give, for comparison with observations, an analytical expression for the mean planetary radius as function of mass  (Sect. \ref{sect:meanrasfctofmass}). We quantify the spread in radii for a given mass interval in Sect. \ref{sect:impacta} and compare core and total radii in Sect. \ref{coreandtotalradius}. }  In Section \ref{sect:summaryconclusion} we summarize our results and finally present the conclusions in Section \ref{sect:conclusionsfin}.

{In the appendix we show our results concerning the disk module which are mainly comparisons with previous works.} The stability of $\alpha$ disks with the new initial profile against self-gravity is discussed in  {Appendix \ref{result:stabilitydisk}, while} the evolution of the disks with the new photoevaporation model  {is shown in Appendix \ref{sect:diskcharactevo}}.

\section{Planet model}\label{sect:planetmodel}

\subsection{Variable core density}\label{sect:variablecoredensity}
In previous versions of the formation model, we assumed as other formation models a constant core   density of 3.2 g/cm$^{3}$ independent of the size and composition of the core. In order to obtain realistic radii for the core we have replaced this assumption with an  internal structure model. Our model is  simpler than the one presented, e.g., by Valencia et al. (\cite{valenciaoconnell2006}), and includes for example no internal temperature profile, but, as noted by Seager et al. (\cite{seagerkuchner2007}), nevertheless yields reasonably accurate results for the radius  {(see Sect. \ref{sect:uncertainty} for a discussion of the uncertainties)}.

For low-mass, core-dominated planets (or planets without a significant atmosphere) this means  that we  have realistic total radii which is important for the comparison with, e.g., \textit{Kepler} results (Sect. \ref{sect:comprdistkepler}).  

The model calculating the core radii was originally developed for the study of GJ 436b in Figueira et al. (\cite{figueirapont2009}) where the methods are described. Our approach is very similar to the one described in Fortney et al. (\cite{fortneymarley2007}) for solid planets. We assume that the core is differentiated and consists of layers of iron/nickel, silicates, and, if the embryo accreted planetesimals beyond the iceline, (water) ice.  The silicate-iron/nickel ratio is fixed to  {2:1 in mass} as found for the Earth (we shall call such a ratio ``rocky''), and the ice fraction is given by the formation model as it is  known whether the planet accretes in- or outside the iceline. 

Ordinary  chondrites have  typical total iron weight fractions of roughly 20 to 30\% (Kallemeyn et al. \cite{kallemeynrubin1989}), and carbonaceous chondrites have Fe/Ni weight fractions around 25\% (Mason, \cite{mason1963}). Therefore, assuming 33\%  as a typical (not strongly altered by giant impacts, cf. Marcus et al. \cite{marcussteward2009}) iron fraction for rocky planets is probably not far from what is expected for the typical solar abundance mix, although at the upper end (cf. also Weidenschilling \cite{weidenschilling1977}).
  
 \begin{figure*}
\begin{center}
\includegraphics[width=0.965\columnwidth]{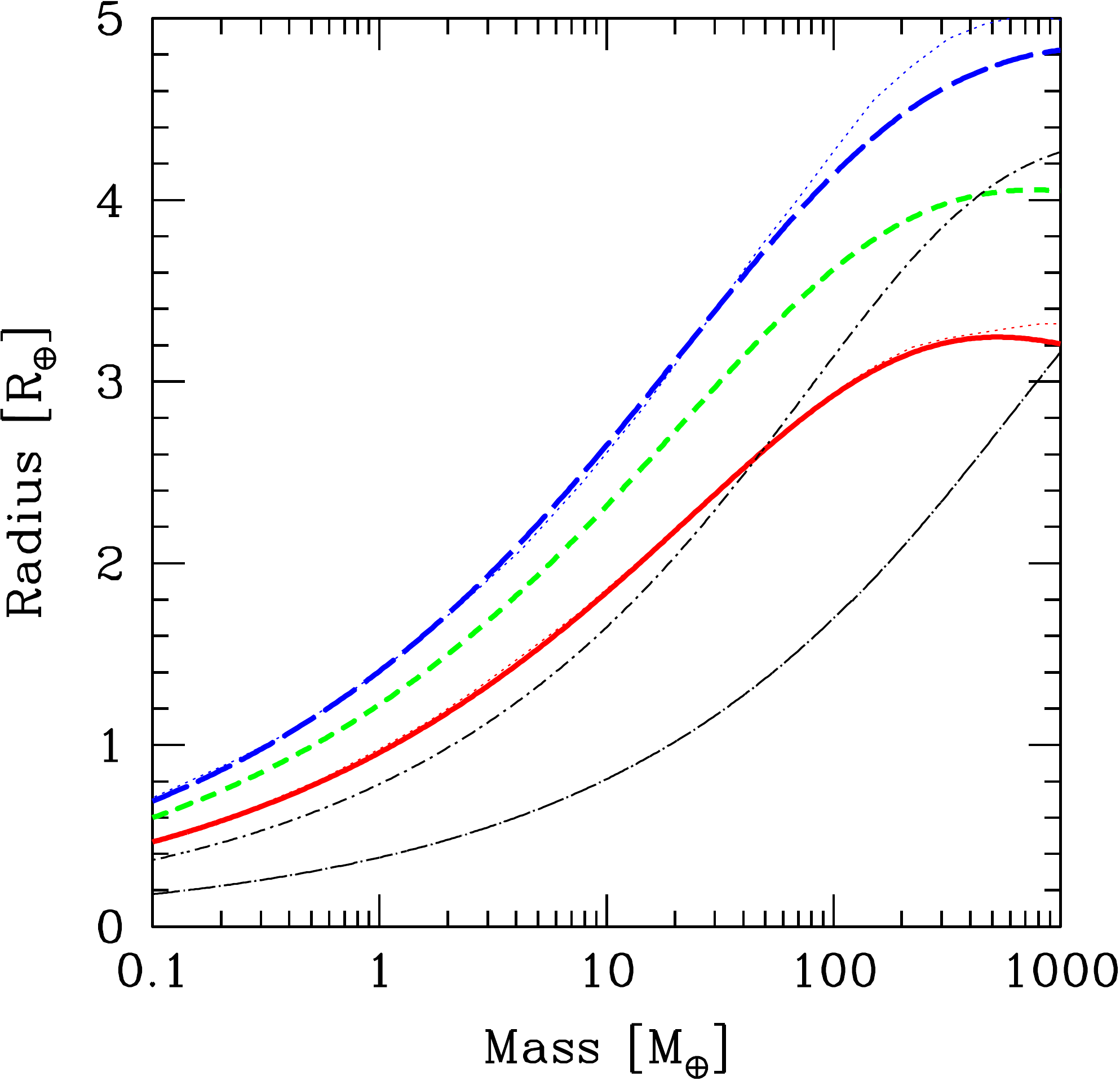}
\includegraphics[width=\columnwidth]{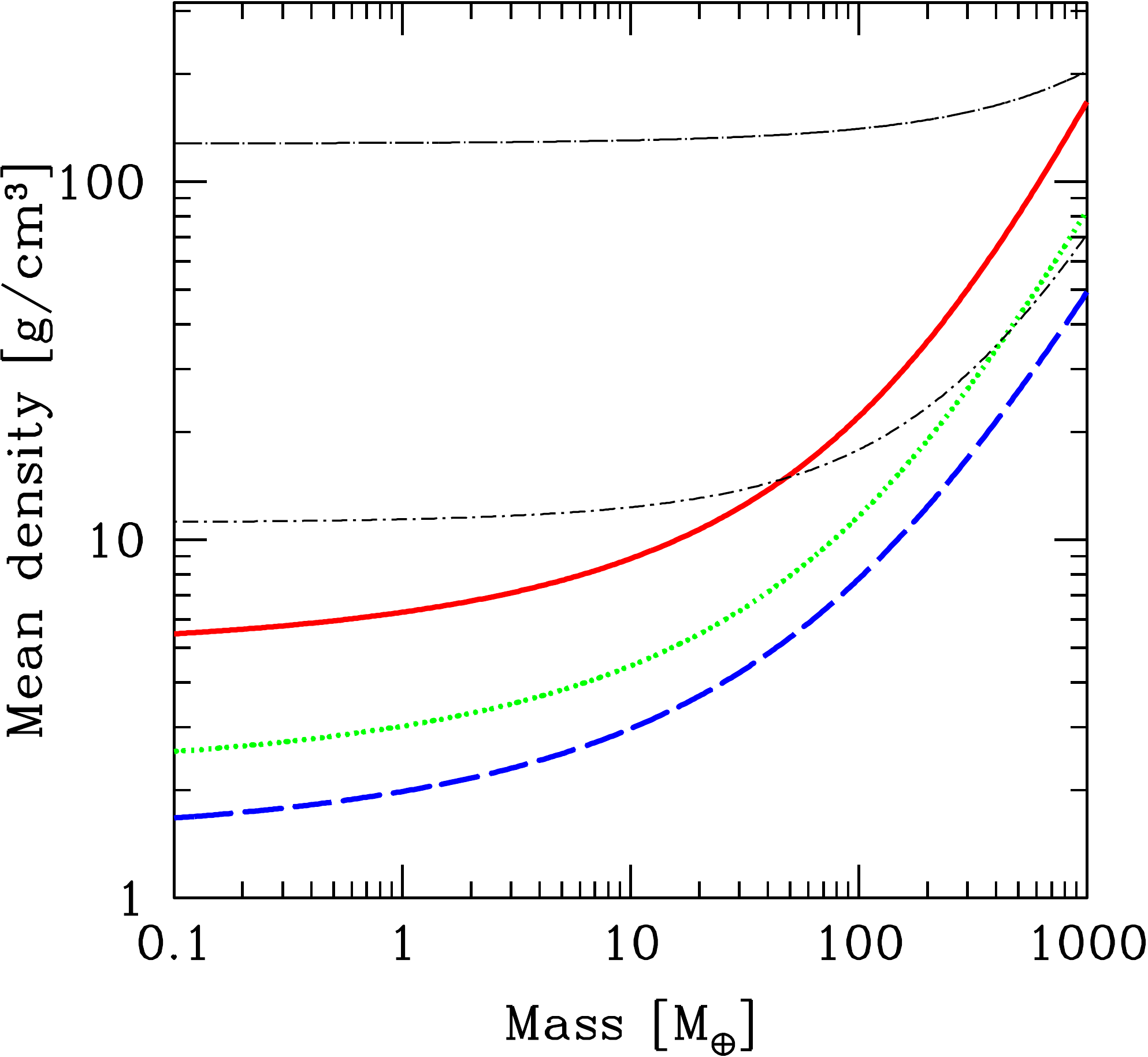}
\caption{Total planetary or core radii and mean densities as a function of mass for different compositions. The red solid curve corresponds to a ``rocky'' planet  {with a 2:1 silicate-to-iron ratio}, similar to Earth. The green dotted curve is for planets consisting of 50\% ice, 33\% silicates and 17\% iron. The blue dashed line shows radii for planets consisting entirely of ice. These three models are calculated without an external pressure on the surface and, therefore, apply to low-mass planets without significant envelopes.  {The thin red and blue  dotted lines show the result from Seager et al. (\cite{seagerkuchner2007}) for comparison}. The thin black lines correspond to cores of giant planets. The short-dashed-dotted line is for a 75\% ice, 25\% ``rocky'' core inside an old, Jovian mass planet ($P_{\rm ext}=4 \times 10^{13}$ dyn/cm$^{2}$). The long-dashed-dotted line is for a core of the same composition inside an old, 10 $\mj$ Super-Jupiter planet ($P_{\rm ext}=6\times 10^{15}$ dyn/cm$^{2}$).   }\label{fig:rhocore}
\end{center}
\end{figure*} 
  
An important effect for the size of cores of giant planets is the external pressure exerted by the envelope. This leads to significant compression of the core for massive planets (Baraffe, Chabrier, \& Barman \cite{baraffechabrier2008}) where the core is buried below hundreds, or even thousands of Earth masses of gas, and subject to huge pressures of thousands of GPa. This effect is included in our calculations.

While we used in Figueira et al. (\cite{figueirapont2009}) the EOS for solid materials from Fortney et al. (\cite{fortneymarley2007}), which are more accurate, but applicable in a limited pressure domain only, we   now use the slightly less accurate, but more widely applicable (also at very high pressures) modified polytropic EOS from Seager et al. (\cite{seagerkuchner2007}) which has a form ($\rho$ is the density, $P$ the pressure)
\beq
\rho(P)=\rho_{0}+c P^{n}
\eeq
where the parameters $\rho_{0}$, $c$ and $n$ are given in Seager et al. (\cite{seagerkuchner2007}).  {For silicates, we use the parameters appropriate for MgSiO$_{3}$ (perovskite).} At high pressures, such an EOS becomes equivalent to a completely degenerate, non-relativistic electron gas (Zapolsky \& Salpeter \cite{zapolskysalpeter1969}) which can be used to solve the Lane-Emden equation with an external pressure (Milne \cite{milne1936}).  Regarding the cores of giant planets, we assume for simplicity that they do not dissolve in the envelope even during the long-term evolution, though this seems possible (Guillot et al. \cite{guillotstevenson2004};  Wilson \& Militzer \cite{wilsonmilitzer2010}). 

\subsubsection{Core radius as a function of core mass}\label{sect:rcoremass}
In Figure \ref{fig:rhocore}, left panel we show the classical plot of (core) radius versus mass for various compositions (e.g. Zapolsky \& Salpeter \cite{zapolskysalpeter1969}; Stevenson \cite{stevenson1982}; Seager et al. \cite{seagerkuchner2007}; Fortney et al. \cite{fortneymarley2007}). It shows first the increase of the radius  with mass,  until a  maximum is reached and then a decrease as the electrons become increasingly degenerated, and the material more compressible.  {The right panel shows the corresponding mean densities.}

The red solid curve applies to planets with a purely ``rocky'' composition ({2:1 silicate-to-iron ratio}), close to the terrestrial composition. For $M= 1 \mearth$, the models yields a radius of 0.96 $\rearth$. The green dotted curve is for  50\% ice and 50\% ``rocky'', while the dashed blue line is for a planet made purely of water ice, which is a composition that is however not expected to exist in nature.  All  three cases were calculated without applying an external pressure $P_{\rm ext}$, and thus correspond to terrestrial (or ice) planets without significant envelopes.   {In the left panel, we also show the result of Seager et al. (\cite{seagerkuchner2007})  for planets with a ``rocky'' and pure water composition calculated with a more accurate EOS. One sees that for masses less than $\sim200\mearth$ and $\sim60\mearth$ for ``rocky'' and pure ice planets, respectively, we find  very similar radii and recover their results to typically within 1\% to 2\%. For a pure ice planet, at a mass of $100\mearth$, we find a radius that is 3\% smaller than obtained by Seager et al. (\cite{seagerkuchner2007}). At such a high mass, it is however expected that a non-negligible amount of gas is accreted during the presence of the protoplanetary disk, which has a much stronger effect on the (total) radius.  It is indeed found in the synthesis that at a core mass of $100 \mearth$, even the planets which accreted the lowest amount of gas have a total radius that is 65\% larger than the core radius (cf. Fig. \ref{fig:rcorertot}).}

The two black lines  correspond to cores of giant planets. Both lines show the radius for objects consisting of 75\% ice and 25\% ``rocky'' material as expected for the composition  {of solids in the Solar System} beyond the iceline  {(we shall call such a composition ``icy'')}.   {This 3:1 ice-to-rock ratio comes from the classical minimum mass solar nebula model of Hayashi (\cite{hayashi1981}). More recent works  (Lodders et al. \cite{lodders2003}; Min et al. \cite{mindullemond2011}) indicate a ratio that is closer to unity for water ice. At sufficiently low temperatures, the ices of CH$_{4}$ and NH$_{3}$, which are the most important ices after H$_{2}$O ice, would  approximately lead to a 2:1 ratio (Min et al. \cite{mindullemond2011}). }

Without an external pressure, the radii (respectively density) of planets with such a composition would lie approximately  in the middle between the green and the blue line.  With the external pressure, the core gets compressed. The short-dashed-dotted line is for an external pressure  $P_{\rm ext}$ of 4000 GPa ($4\times 10^{13}$ dyn/cm$^{2}$), which is the estimated central pressure of a Jupiter today (Guillot \& Gautier \cite{guillotgautier2009}). The long-dashed-dotted line shows the even more extreme conditions at the core-envelope boundary of an 5 Gyr old, 10 $\mj$ planet. For this case we find (Paper I) an external pressure as high as $\sim$600 TPa ($6\times 10^{15}$ dynes/cm$^{2}$). Such a central pressure is roughly expected from the simple estimate based on hydrostatic equilibrium which scales as $M^{2}/R^{4}$ (e.g. Kippenhahn \& Weigert \cite{kippenhahnweigert1990}), and the fact that the radius $R$ only  changes (decreases) weakly between 1 and 10 Jupiter masses (e.g. Chabrier et al. \cite{chabrierbaraffe2009}). 

Figure \ref{fig:rhocore} shows that a typical giant planet core with a mass of 10 $\mearth$ inside a Jupiter mass planet gets compressed to a radius of about 1.6 $\rearth$ corresponding to a mean density of about 12 to 14 g/cm$^{3}$. Inside the Super-Jupiter, the core gets squeezed to a radius of about 0.75 $\rearth$, corresponding to a very high mean density of about 130 g/cm$^{3}$. 

\begin{figure*}
\begin{center}
\includegraphics[width=\columnwidth]{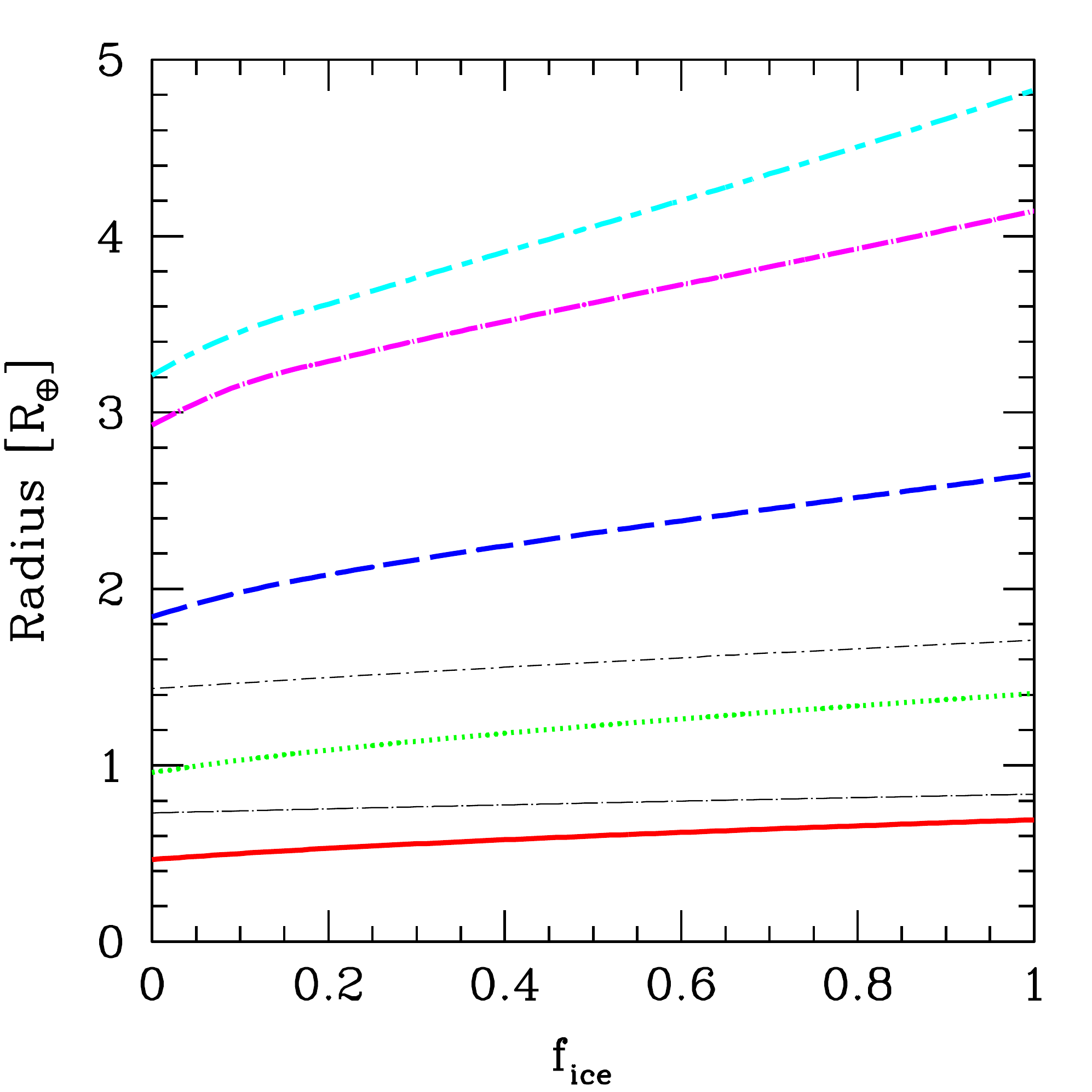}
\includegraphics[width=\columnwidth]{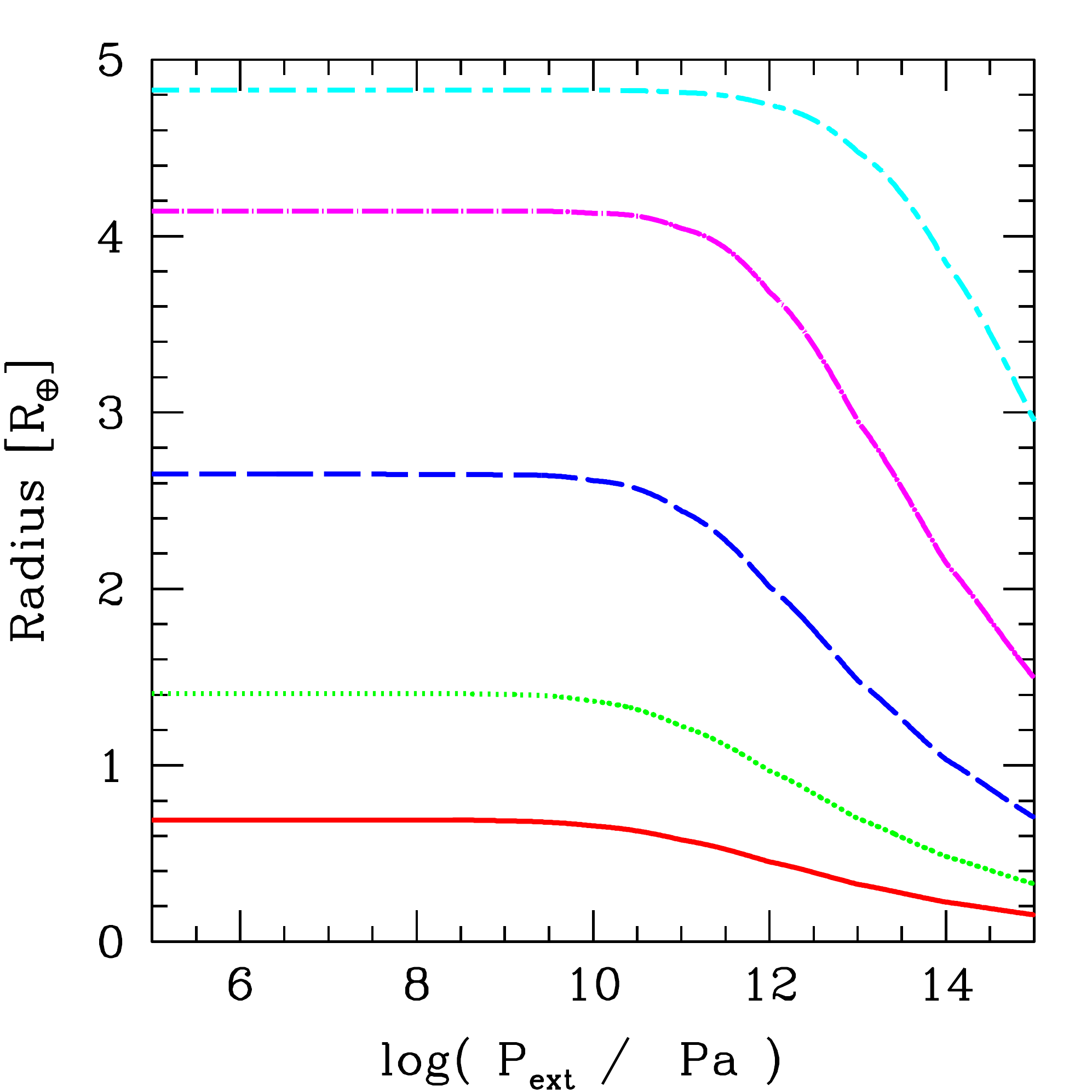}
\caption{Left panel: Influence of the ice mass fraction $f_{\rm ice}$ on the radius. The reminder of the mass has a ``rocky'' composition. Radii are plotted for planets of $M=0.1\mearth$ (red solid line), 1 $\mearth$ (green dashed line), 10 $\mearth$ (blue dashed),  100 $\mearth$ (magenta dashed-dotted) and 1000 $\mearth$ (cyan short-long dashed), all without an external pressure. The thin black short-dashed-dotted line is for a 10 $\mearth$ core of an old 1 $\mj$ giant planet, and the thin black long-dashed-dotted line is for the 10 $\mearth$ core of a 10 $\mj$ giant. Right panel: Radius as a function of external pressure $P_{\rm ext}$ for pure ice planets. The (thick) lines represent the same planet masses as in the left panel. }\label{fig:rhocoreficePext}
\end{center}
\end{figure*}

\subsubsection{Core radius as function of ice mass fraction}
In figure \ref{fig:rhocoreficePext}, left panel, we show the impact of the ice mass fraction in the planet $f_{\rm ice}$ on the radius, for planets of different masses, with and without  external pressure. The remainder of the mass is assumed to have a ``rocky'' composition i.e.  {a 2:1 silicate-to-iron ratio}.

The plot shows a planet of a mass $M=0.1\mearth$ (red solid line), 1 $\mearth$ (green dotted), 10 $\mearth$ (blue dashed), 100 $\mearth$ (magenta dashed-dotted) and  finally  for a hypothetical  1000 $\mearth$ solid planet (cyan short-long dashed line). All these cases were calculated without an external pressure. One can see that with increasing ice fraction, the radius increases, as expected. For most of the domain of $f_{\rm ice}$, $R$ increases about linearly with the ice fraction, which is in agreement with the results of Fortney et al. (\cite{fortneymarley2007}).  

The thin black short and long dashed-dotted lines are, as before, for a 10 $\mearth$ core inside a 1 and 10 $\mj$ planet, respectively. Compared to the blue line (also 10 $\mearth$, but without external pressure), these lines lie  at clearly smaller radii. One also notes that the composition becomes less important: For $ P_{\rm ext}=0$, $R$ increases by a factor 1.44 when going from $f_{\rm ice}=0$ to $f_{\rm ice}=1$, which is in good agreement with the fitting formula from Fortney et al. (\cite{fortneymarley2007}) and just corresponds to the result expected when the mean density of  ``rocky'' material is three times higher than ice.  With the high external pressures, we find in contrast a radius increase by only a factor of about 1.19 and 1.15 for the Jovian and Super-Jupiter case, respectively. This is due to the fact that for the (completely) degenerate electron gas, the density for a given pressure depends on the material only via the mean molecular weight per free electron  $\mu_{\rm e}$ (Cox \& Giuli \cite{coxgiuli1968}).  {The mean molecular weight per free electron (and thus the density in the limit of complete degeneracy) of  ``rocky'' material is however only about 1.14 times higher than the $\mu_{\rm e}$ of pure water ice. This explains why at very high pressures, the dependence of the radius on the specific core material becomes weaker.} 

\subsubsection{Core radius as function of external pressure}
In Figure \ref{fig:rhocoreficePext}, right panel we show how the radius decreases as the pressure acting on the surface of the planet increases. The external pressure is varied between the (low) pressure on the Earth's surface, and the extremely high pressure found at the core-envelope boundary of a $\sim20\mj$ planet. The lines represent the same planet masses as in the left panel. A composition of 100\% ice is assumed. The results for ``rocky'' material are qualitatively identical.  

We see that up to a $P_{\rm ext}$ of about $10^{10}$ Pa (for lower core masses) to $10^{12}$ Pa (at high core masses), the radius is independent of $P_{\rm ext}$.  Such a critical pressure is roughly expected (Seager et al. \cite{seagerkuchner2007}), because the bulk modulus of ice is of order $10^{10}$ (Benz \& Asphaug  \cite{benzasphaug1999}) to $10^{11}$ Pa (for the EOS used here, see Seager et al. \cite{seagerkuchner2007}). After a pressure of this order is reached, the chemical bonds in the material start to get crushed, and the radius decreases. With increasing density, the pressure of the degenerate electrons becomes increasingly important. 

In Sect. \ref{result:coredensity} we study the effect of a variable core density on the formation and evolution of Jupiter,  {while in Sect. \ref{coreandtotalradius} we look at core radii as found in a population synthesis calculation}.

\subsection{Uncertainties in the radius calculations}\label{sect:uncertainty}
{The different sources of uncertainty which affect theoretical models calculating planetary radii have been addressed in many works (e.g. Seager et al. \cite{seagerkuchner2007}; Baraffe et al. \cite{baraffechabrier2008}; Grasset et al. \cite{grassetschneider2009}). We here discuss simplifications which are specific to our model. In view of the population synthesis calculations presented below where comparison are made with the observed radii, it is important to estimate the uncertainties which are introduced by these simplifications.   For the solid core, we have neglected the internal temperature profile, used a simple modified polytropic equation of state (EOS) and a Si/Fe ratio fixed at the terrestrial value. For the gaseous envelope, we assume that all solid reside in the core (no heavy element mixing with the envelope), and use simple gray atmospheric boundary conditions  (Paper I). We next discuss the possible impact of these simplifications.}

\begin{figure*}
\begin{center}
\begin{minipage}{0.35\textwidth}
	      \centering
        \includegraphics[width=0.98\textwidth,height=5.75cm]{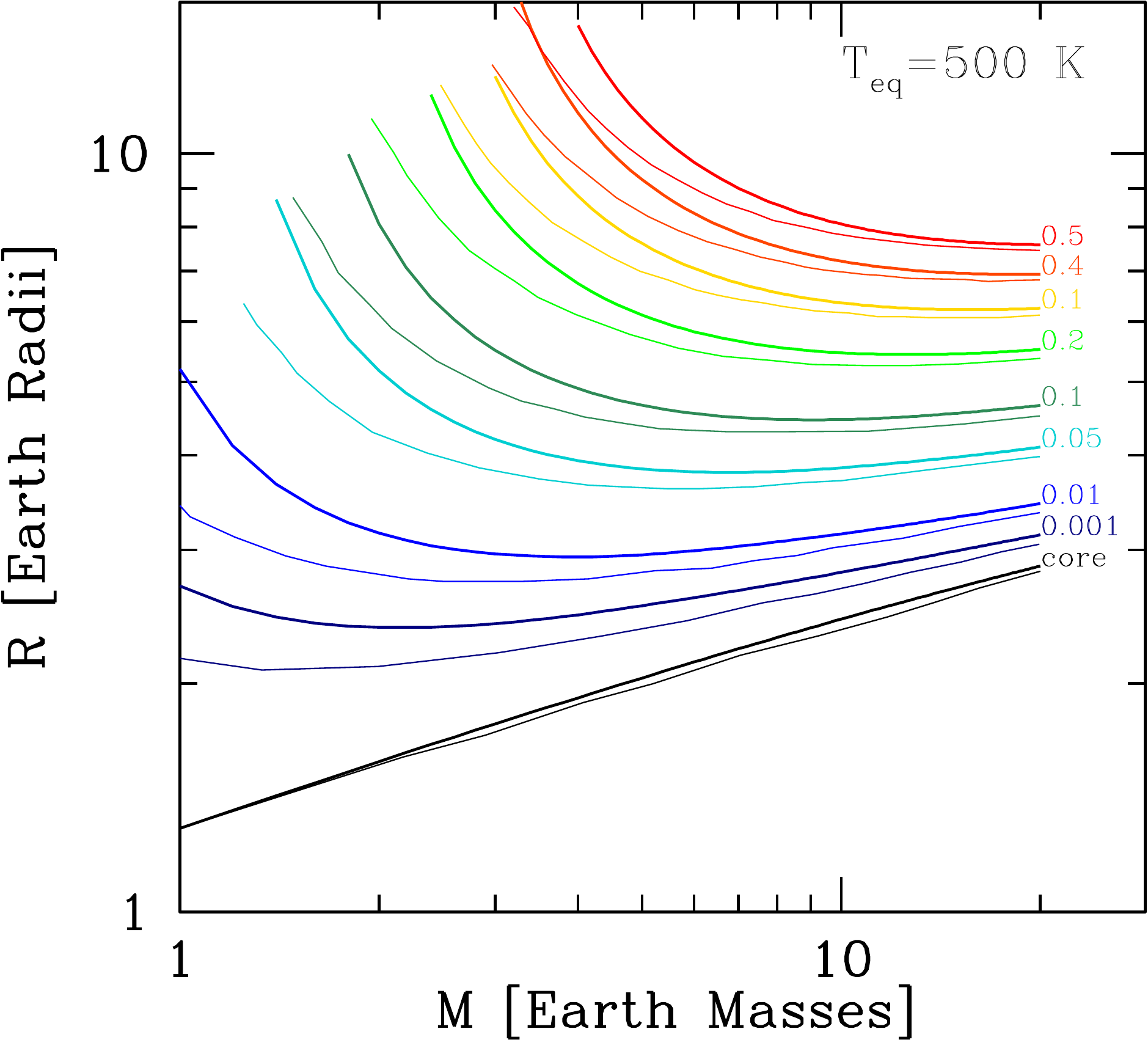}
     \end{minipage}\hfill
     \begin{minipage}{0.32\textwidth}
      \centering
       \includegraphics[width=0.96\textwidth]{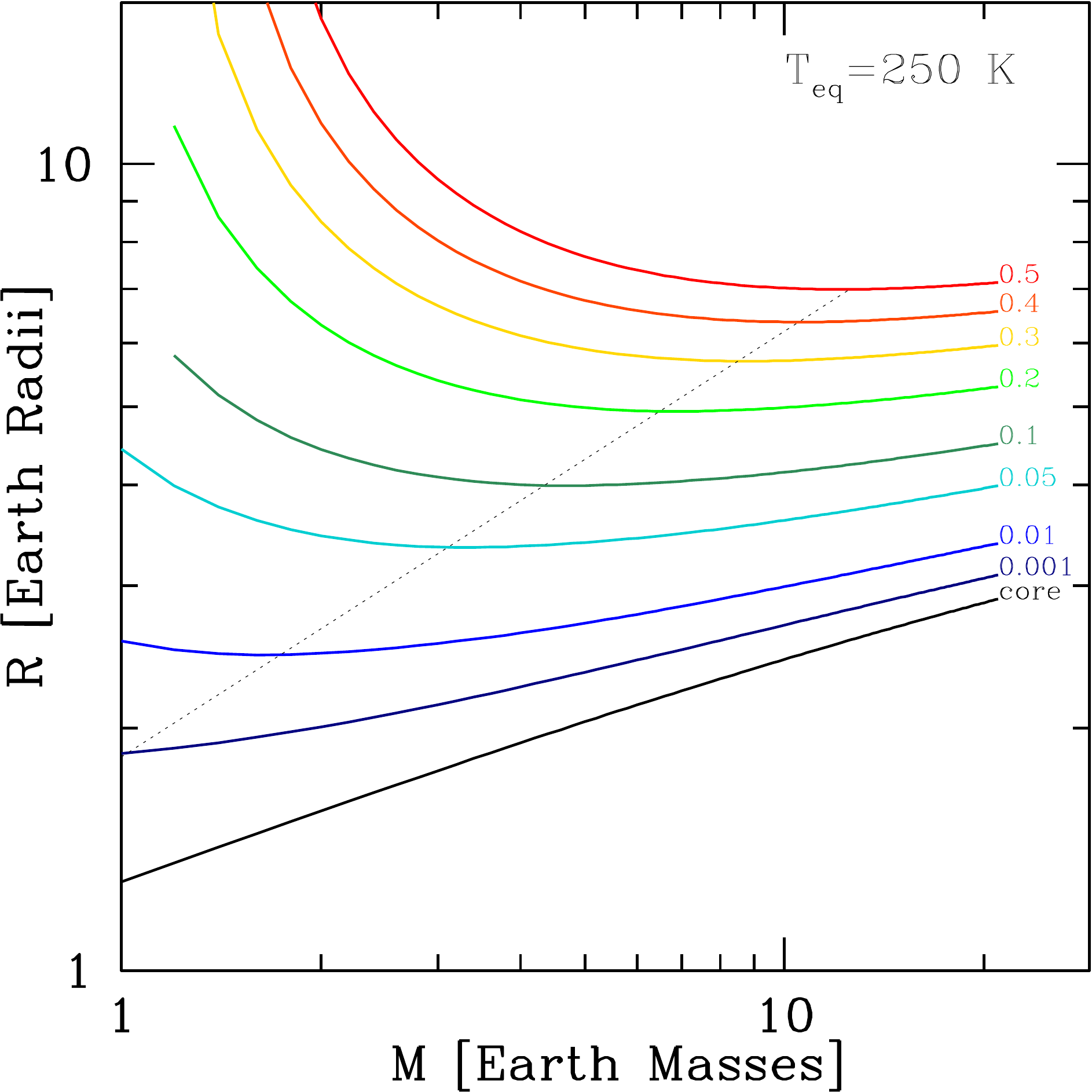}
     \end{minipage}
          \begin{minipage}{0.32\textwidth}
      \centering
       \includegraphics[width=0.96\textwidth]{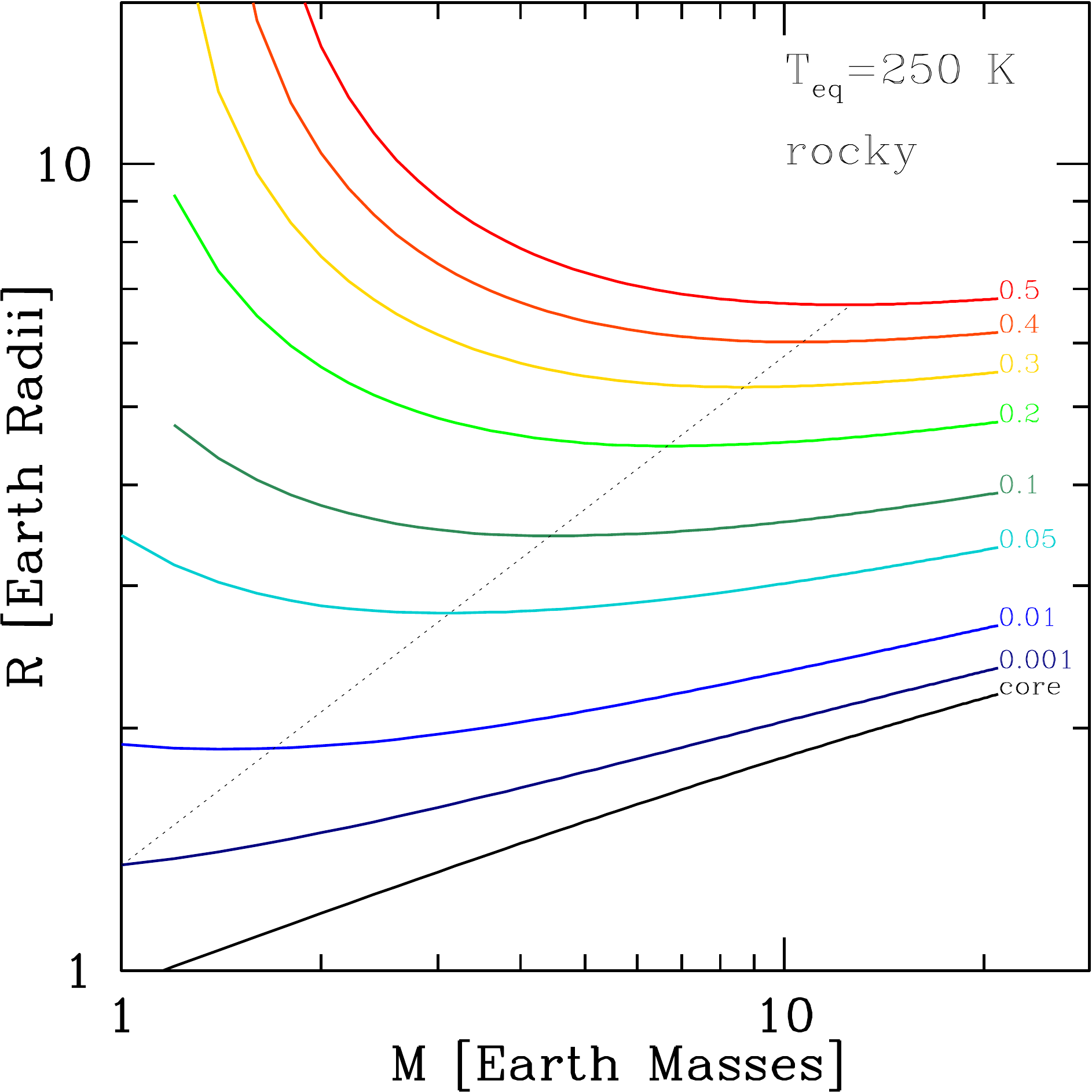}
     \end{minipage}
\caption{{Radii of low-mass planets with H$_{2}$/He atmospheres. In all panels, the core radius is shown, and the total radius for different envelope mass fractions $f$, as labeled in the figures. In the left panel, thick lines show the result from this work (simple gray atmospheres), while thin lines are the results  of Rogers et al. (\cite{rogersbodenheimer2011}) calculated with the more accurate ``two-stream'' approximation of Guillot (\cite{guillot2010}). Planets in the right panel have a purely ``rocky'' core, otherwise an ice mass fraction $f_{\rm ice}=0.67$ is used. The intrinsic luminosity per mass is  $10^{-6.5}$ erg g$^{-1}$ s$^{-1}$ in all cases. The equilibrium temperature is indicated in the panels. To the right of the black dotted line, the gray atmosphere should not introduce errors larger than $\sim15$\% in the calculations with $T_{\rm eq}=250$ K. }}\label{fig:mrrogerscomp}
\end{center}
\end{figure*}

\subsubsection{Solid core}
\textit{No temperature structure}  {Grasset et al. (\cite{grassetschneider2009}) found that very large thermal variations within iron-silicate planets affect the radius only on a low, $\sim$1.5\% level, in agreement with Valencia et al. (\cite{valenciaoconnell2006}). For water ice, the thermal pressure could be more important (Fortney et al. \cite{fortneymarley2007}), but as found by Seager et al. (\cite{seagerkuchner2007}) also for icy planets, the errors in the radius introduced by neglecting it are small, as the thermal pressure decreases the density by typically 3\% or less in the relevant domain. Even if the mean density $\bar{\rho}$ would decrease by 10\%, it would cause an increase of the radius by only about 3.6\% thanks to the weak dependency of $R\propto 1/\bar{\rho}^{1/3}$.} 

\textit{Simplified EOS}  {As shown in Fig. \ref{fig:rhocore}, the simple polytropic EOS leads to radii of solid planets which agree on a 1-3\% level for $M\lesssim 100 \mearth$ with the results obtained when combining the Birch-Murnagham EOS, density functional theory, and a modified Thomas-Fermi-Dirac EOS (Zapolsky \& Salpeter \cite{zapolskysalpeter1969}). This agrees with the findings of   Seager et al. (\cite{seagerkuchner2007})}

\textit{Fixed silicate-iron ratio}  {Grasset at al. (\cite{grassetschneider2009})  studied in details the effects of different [Mg/Si] and [Fe/Si] values. Using observed abundances of extrasolar host stars, and assuming that the planets have the same relative abundances in Fe, Mg, and Si as their hosts, they found that the radii of ``rocky'' planets only change by less than 1.5\% when the composition varies over the observed domain. Larger differences in the radii however occur if one abandons the assumption of identical relative abundances in planets and the host star. Such a situation can result from collisional mantel stripping as might have occurred for Mercury (e.g. Benz et al. \cite{benzanic2007}). For a 1 $\mearth$ planet, Marcus et al. (\cite{marcussasselov2010}) find a minimal plausible planetary radius achievable by maximum mantel stripping which is about 0.85 $\rearth$.}

{As a word of caution one must mention that most of the rather low uncertainties were estimated for low-mass planets. Baraffe et al. (\cite{baraffechabrier2008}) show that the uncertainties can become  significantly larger when one considers more extreme conditions, as arising in the center of a giant planet. Under such conditions, they for example find that the density of water as predicted by various equations of state can differ by up to $\sim20$\%.}

\subsubsection{Gaseous envelope}\label{sect:errorsgasenve}
\textit{All solids in the core}  {The consequences of how the heavy elements are distributed in a (giant) planet have been analyzed in details by Baraffe et al. (\cite{baraffechabrier2008}). The two limiting cases are either that all solids reside in the core, surrounded by a pure H$_{2}$/He envelope (as in our model), or that there is no core, but that the solids are uniformly mixed with the gas. For a Neptunian planet, and a heavy element mass fraction $Z=\mz/M$ of up to 20\%, putting all solids in the core versus uniform mixing typically leads to a change of the radius by  $\sim10$\% during the long-term evolution. For a Jovian planet, with $Z$=20\% or 50\%, the variation in the radius is of order 4\% and 12\%, respectively. For both planet types, putting the solids in the core leads to larger radii.}
 
 \textit{Gray atmosphere model}  {The \textit{Kepler} satellite has recently detected large numbers of planets with radii between the radius of the Earth and the ice giants, allowing to study the composition of such objects (Sect. \ref{sect:comprdistkepler}). Most of these transiting planets are inside $\sim 0.5$ AU from their host star, therefore they are subjected to significant levels of irradiation. It is well known (e.g. Guillot \& Showman \cite{guillotshowman2002}) that a strong irradiation can significantly affect the evolution of a planet, and lead to atmospheric structures more complex than the simple gray atmosphere we use (Paper I). In the current version of our model, we do not allow planets to migrate closer than 0.1 AU to the star, so that the cases undergoing the strongest effects are excluded by construction. But also at a distance of 0.1 AU, the irradiation already affects the radius of low-mass planets with H$_{2}$/He atmospheres (Rogers et al. \cite{rogersbodenheimer2011}). In order to estimate the uncertainty introduced by using the  simple gray atmosphere model, we compare in Fig. \ref{fig:mrrogerscomp} the radii of low-mass planets with H$_{2}$/He atmospheres with those found by  Rogers et al. (\cite{rogersbodenheimer2011}). These authors use a more realistic ``two-stream''  atmosphere  (Guillot \cite{guillot2010}).  The radii shown in the figure were  found not by combined formation and evolution calculations as in the rest of this study, but instead by specifying directly the composition of the planets, their semimajor axis and the intrinsic luminosity  (``equilibrium model'' in the terminology of  Rogers et al. \cite{rogersbodenheimer2011}). Then, the same method as in the evolutionary calculations was used to obtain the radii. In all models, the intrinsic luminosity per mass is  $10^{-6.5}$ erg g$^{-1}$ s$^{-1}$ as in Rogers et al. (\cite{rogersbodenheimer2011}). The left panel shows the radii for planets with an equilibrium temperature of 500 K, which corresponds for zero albedo to an orbital distance of about 0.3 AU around a solar-like star. In this and the middle panel, a composition of the core with an ice mass fraction of 67\% and 33\% ``rocky'' material was assumed. This is similar, but not exactly identical as in Rogers et al. (\cite{rogersbodenheimer2011}), because of a somewhat different composition of the silicates. The resulting radius of the core is also shown in the figure. A comparison of the thick and thin lines show that both models lead to the same global pattern (as also found in Valencia et al. \cite{valenciaikoma2010}). At a total mass of the planet of $20\mearth$, our model predicts total radii (i.e. $\tau=2/3$) which are 2\% larger than those in Rogers et al. (\cite{rogersbodenheimer2011}), independent of the envelope mass fraction $f=\mxy/M=1-Z$. This difference is partially due to an identical difference in the core radius.  When moving to smaller planetary masses, the differences get larger, in particular for planets with a high envelope mass fraction. This is expected, as such low-mass, high $f$ planets are particularly sensible to irradiation.   At $M=4 \mearth$, for $f$=0.05, we find radii which are 7\% larger, while for $f=0.5$, the difference becomes significant with about 25\%. In the combined formation-evolution calculation it is however found that the envelope mass fraction scales with the total mass of the planet (see Fig. \ref{fig:mrobs}), so that low-mass planets with very massive envelopes cannot exist. This is a natural consequence of the long KH timescales of low-mass cores.  At a total mass of $4\mearth$ for example, the mean $f$ in the synthesis is approximately 0.05, and no planet has $f>0.2$, where the overestimation of the radius due to the simple gray atmosphere is about 10\%. Preliminary calculations using also the ``two-stream'' approximation, lead to radii which are in excellent agreement with  Rogers et al. (\cite{rogersbodenheimer2011}), showing that the gray atmosphere is indeed the reason for the differences. We can thus conclude that as long as a planet is not in the regime where $R$ is clearly increasing with decreasing mass (hot, low $M$, high $f$ planets), the errors introduced by the gray atmosphere are less than $\sim$15\%.  In this light, the radii in the middle and left panel of Fig. \ref{fig:mrrogerscomp}  (equilibrium temperature of 250 K) to the right of the dotted line should not be too affected by the gray approximation.}
  
{Figure \ref{fig:mrobs} shows that  indeed, only a tiny fraction of the synthetic planets falls in the regime where the gray atmosphere introduces large differences, thanks to the fact that we model planets with $a \geq 0.1$ AU only.  We must however keep in mind that we have compared here equilibrium models only, i.e. we have not studied how a different atmospheric boundary condition affects the entire temporal evolution. This will be addressed in future work.}

\subsubsection{Summary}
{In summary we see that depending on the properties of a planet, the errors affecting the radius are of a different level. A clearly gas dominated, Jovian planet at large orbital distance is less affected than a low-mass planet with a  20\% H$_{2}$/He atmosphere at 0.1 AU, even if also for the giant planet, we cannot reach the level of refinement as in dedicated models for individual planets (e.g. Fortney et al. \cite{fortneyikoma2011}). }

{When considering the uncertainties of the radii in a bigger context, one should keep in mind the following point: The evolution of planets is by itself a complex problem, which is studied in dedicated evolutionary models. In the combined formation and evolution model shown here, however, the evolutionary part serves first of all to establish a link between observable properties of planets at late times, and the formation model, where many mechanisms are poorly understood. When used in planetary population synthesis calculations, the model must thus be sufficiently accurate that possible observational findings like correlations between radius and planet mass, host star metallicity, semimajor axis, etc., are not lost as constraints for the formation model.  This results in different requirements than when one is interested in determining as exactly as possible the composition of one specific planet.}

\subsection{Core luminosity due to radioactive decay}\label{sect:upgradelradio}
During the accretion of planetesimals, the luminosity of the core due to this process is approximatively given as 
\beq
L_{\rm core,acc}=\frac{G \mz}{\rcore} \mdotz
\eeq
where $\mz$ is the mass of the core, $\rcore$ its radius, $\mdotz$  the accretion rate of planetesimals, and $G$  the gravitational constant. This equation is applicable if the planetesimals or their debris sinks to the core (cf. Mordasini et al.  \cite{mordasinialibert2005}) and if their initial velocity is small compared to the escape velocity of the core. In our previous models, the accretion of planetesimals is the only source of luminosity of the core. This means that once the accretion of solids  ceases, the core luminosity drops to zero. However, several processes can lead to a non-zero core luminosity during the evolution of the core at constant mass (see Baraffe et al. \cite{baraffechabrier2008}): the cooling of the core due to its internal energy content $dU/dt$, its contraction $-P dV/dt$, and finally the decay of radioactive elements contained in it. Within the upgraded model, we now take into account the last two points (see Paper I for the contraction). In this section we describe how we model the last mechanism.

\subsubsection{Chondritic radiogenic heating}
Urey (\cite{urey1955}) was the first to point out that the presently measured surface heat flux of the Earth  roughly corresponds to the radiogenic heat released in the interior if the Earth has a chondritic composition. This results still holds today, even though the details depend on the exact type of chondritic material that is assumed to have formed the Earth (Hofmeister \& Criss \cite{hofmeistercriss2005}). Estimates for the total flux vary, but are of order $3 \times 10^{20}$ erg/s (Hofmeister \& Criss \cite{hofmeistercriss2005}). Due to these findings, we model the radiogenic heat production as the consequence of the decay of the three most important long-lived nuclides $^{40}$K, $^{238}$U and $^{232}$Th (Wasserburg et al. \cite{wasserburgmacdonald1964}), assuming a chondritic composition. We further assume that the radiogenic heat production rate in the interior, and the loss at the surface are in equilibrium, i.e. the luminosity at the surface is equal to the instantaneous radiogenic energy production rate inside the planet. We therefore neglect a  possible delayed secular cooling, and also possible residual heat from the impacts during the formation of the planet. This is, at least for Earth today, a good approximation (Hofmeister \& Criss \cite{hofmeistercriss2005}). 

As illustrated below, the radiogenic luminosity $L_{\rm radio}$ obtained this way is small,  but not negligible for the total luminosity of low-mass (super-Earth) planets` with a tenuous H$_{2}$/He atmosphere, i.e. $L_{\rm radio}$ provides a physically motivated minimum for the luminosity of the core, once the accretion of planetesimals has ceased. This can play a role for the evolution of the total radius, which is an observable quantity, as demonstrated in Section \ref{result:lradio}.

Valencia et al. (\cite{valenciaikoma2010}) simply use a total heat production rate of  $2\times 10^{20}$  erg s$^{-1}$ $\mearth^{-1}$ for the planet's mantel which is also assumed to have a chondritic composition. They assume that the rate is constant in time.  We improve this by considering that over timescales of billions of years, the heat production by the long-lived radionuclides will decrease according to the law of the exponential radioactive decay. 

Denoting with $Q(t)$ the energy release per second per gram of chondritic material, one can write (e.g. Prialnik et al. \cite{prialnikbarnur1987})
\beq
Q(t)=Q_0 e^{-\lambda t}
\eeq
where $Q_0$ is the heat production rate per gram of chondritic material at time $t=0$, and $\lambda=\ln 2/t_{1/2}$ is the decay constant for a radionuclide with a half-life $t_{1/2}$. The values of $Q_0$ can be found by extrapolating back from currently measured values in meteorites. In table \ref{tab:radio} we give the values of $\lambda$ and $Q_0$ derived from the data given in Lowrie (\cite{lowrie2007}) for the three nuclides we include. These values are compatible with the ones of Prialnik et al. (\cite{prialnikbarnur1987}). 

\begin{figure}
\begin{center}
\includegraphics[width=\columnwidth]{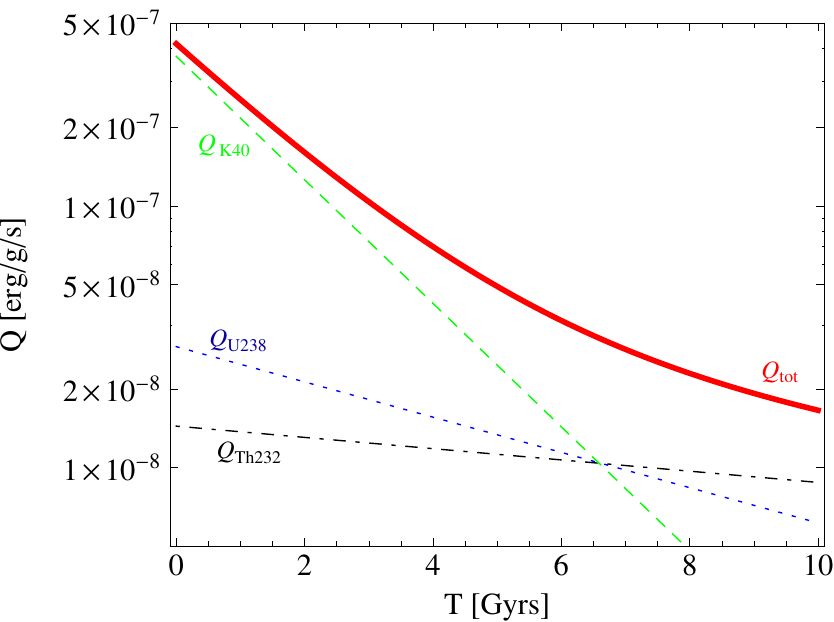}
\caption{Radiogenic heat production rate  in one gram of chondritic material as a function of time due to the decay of long living nuclides. The lines show the contribution from the different radionuclides (as labeled in the figure), as well as the total rate (red, solid line). }\label{fig:lradio}
\end{center}
\end{figure}

\begin{table}
\caption{Parameters for the radiogenic heat production. Besides the three long-lived radionuclides we also give the parameters for $^{26}$Al. }
\begin{center}
\begin{tabular}{l|cc}
Nuclide & $Q_{0}$ [erg/(g s)] & $\lambda$ [1/Gyrs]  \\ \hline
$^{40}$K & $3.723\times10^{-7}$ & 0.543 \\
$^{238}$U & $2.899\times10^{-8}$ &  0.155 \\
$^{232}$Th & $1.441\times10^{-8}$ & 0.0495 \\ \hline
$^{26}$Al & $2.13\times10^{-3}$ & 966.732 \\
\end{tabular}
\end{center}
\label{default}\label{tab:radio}
\end{table}%
The total heating rate per gram of chondritic material is then simply
\beq
Q_{\rm tot}(t)=Q_{\rm 0,K}e^{-\lambda_{\rm K} t}+Q_{\rm 0,U}e^{-\lambda_{\rm U} t}+Q_{\rm 0,Th}e^{-\lambda_{\rm Th} t}.
\eeq
In Figure \ref{fig:lradio} we plot $Q_{\rm tot}$ and the contribution of the three nuclides as a function of time. Initially, potassium provides the dominant contribution, by about an order of magnitude.  At later times, $^{238}$U and $^{232}$Th take over due to their longer half-life.

The total radioactive luminosity of the core is given as
\beq\label{eq:totallradio}
L_{\rm radio}(t)=Q_{\rm tot}(t) f_{\rm mantle} f_{\rm rocky} M_{\rm Z}
\eeq
where $M_{\rm Z}$ is the total core mass,  $f_{\rm mantle}$ the mantle fraction and $f_{\rm rocky}=1-f_{\rm ice}$ the mass fraction of  ``rocky'' material in the core. 

With this  setting, one finds for an Earth like planet ($f_{\rm mantle}$=2/3, $f_{\rm rocky}$=1, $M_{\rm Z}$=1 $\mearth$) at $t=4.5$ Gyrs a radioactive luminosity of $L_{\rm radio}=2.26\times10^{20}$ erg/s which is in good agreement with the value from Valencia et al. (2009). For $t=0$, one finds $L_{\rm radio}=1.65\times10^{21}$ erg/s, or about a factor 7.3 higher than today. This is in good agreement with Wasserburg et al. (\cite{wasserburgmacdonald1964}) who estimate a ratio of 4.5 to 8.2  depending on the chemical composition of the Earth.  {Nettelmann et al. (\cite{nettelmannfortney2011}) find in a similar approach as here a present time radiogenic heat production rate of $2.3 \times10^{20}$ erg/s, and a value of $2.7 \times10^{21}$ erg/s at $t=0$. The latter value is higher than found here, probably because we have not included $^{235}$U due to it's relatively short half-life of 0.47 Gyrs.}  The value of the radiogenic luminosity due to long-lived nuclides at $t=0$ (or up to small $t\sim10$ Myrs, Prialnik et al. \cite{prialnikbarnur1987}) has no practical meaning, as short-lived radionuclides, and heat produced by collisional growth would then dominate by many orders of magnitude. It gives however an impression by how much the heat flux due to long-lived radionuclides has varied over the lifetime of the Solar System.  

\subsubsection{Short-lived radionuclides}
We have neglected  the radiogenic heating due to short-lived radionuclides, in particular Aluminium 26. It is well known that $^{26}$Al could have been a powerful heating mechanism in the first few million years in the Solar System (e.g. Prialnik et al. \cite{prialnikbarnur1987}). According to the study of Castillo-Rogez et al. (\cite{castillorogez2009}) who assume an initial isotopic abundance ratio of $^{26}$Al/$^{27}$Al =$5\times 10^{-5}$, an aluminium concentration of 1.2 wt \% in silicates, one finds a  $Q_{\rm 0,Al}=2.13\times10^{-3}$ erg/(g s). This value is compatible with the one given in Prialnik et al. (\cite{prialnikbarnur1987}). This is many orders of magnitude more than for the long-lived nuclides. If we were to have a fully assembled rocky 1 $\mearth$ planet at $t=0$,  then $^{26}$Al would lead to a luminosity of about 2.5 $\lj$, and a surface temperature of 414 K (neglecting melting and a delayed heat transport, which is clearly not realistic in this case). 

It is difficult to derive a general model (i.e. also for other stars than the sun) including short-lived radionuclides for a number of reasons: First, it is unclear how universal the initial abundance of short-lived radionuclides is for different planetary systems. If non-local effects play an important role (like the proposed nearby supernova, Cameron \& Truran \cite{camerontruran1977}), then we probably must expect very different initial concentrations. Second, due to the short half-life of $^{26}$Al (about 0.72 Myrs), results regarding this nuclide are very sensitive to the exact time delay between CAI formation and inclusion into the planetesimals and protoplanets (e.g.  Prialnik et al. \cite{prialnikbarnur1987}). The time corresponding to $t=0$ is  not very well defined in our simulations. For example, the distribution of initial disk  masses is  based on observations of YSO in $\rho$ Ophiuchi, which is about 1 Myr old (Andrews et al. \cite{andrewswilner2009}). Third,  the inclusion of short-lived nuclides would require a detailed thermic model of the core, allowing melting and heat transport. This is  currently beyond the scope of the model. The same applies for the heating due to impacts.

In the result Section \ref{result:lradio}, we illustrate the effect of the radiogenic heat production on the luminosity of a low-mass planet with a tenuous primordial H$_{2}$/He atmosphere. 

\section{Disk Model}\label{sect:disk model}
{We next turn to three improvements regarding the model of the protoplanetary disk. The properties of the protoplanetary disk are the initial and boundary conditions for the growth and migration of the planets. The disk model yields for example the outer boundary conditions for the structure equations, the radial slopes of the temperature and surface density for the migration model, or the gas accretion rate in the runaway phase (cf. Alibert et al. \cite{alibertmordasini2005} and Paper I).}  In this computational module, we have improved the initial conditions for the disk evolution,  the description of photoevaporation, and the treatment of the outer and inner disk edge.

\subsection{Standard $\alpha$ model with irradiation and photoevaporation}\label{standardalphamodel} 
As in previous studies, we solve the standard equation for the evolution of the surface density of a viscous accretion disk (Lynden-Bell \& Pringle \cite{lyndenbellpringle1974}),  in a 1+1D model (vertical and radial structure), assuming that the viscosity can be calculated with the $\alpha$ approximation  (Shakura \& Sunyaev \cite{shakurasunyaev1973}).  

A detailed description of the disk model, in particular how the vertical structure is calculated assuming an equilibrium of viscous heating and radiation at the surface, was given in Alibert et al. (\cite{alibertmordasini2005}). In Fouchet et al. (\cite{fouchetalibert2011}) we have shown how we  take into account heating by the stellar irradiation, assuming that the flaring angle  is at the equilibrium value of 9/7 (Chiang \& Goldreich \cite{chianggoldreich1997}).

In previous studies, we  used an initial gas surface density profile inspired by the minimum mass solar nebula (Weidenschilling \cite{weidenschilling1977}; Hayashi \cite{hayashi1981}) which is given as 
\beq\label{eq:oldprofile}
\Sigma(r)=\Sigma_0\left(\frac{r}{5.2\, {\rm AU}}\right)^{-3/2}
\eeq
{at a distance $r$ from the star.} 
Here, $\Sigma_0$ is the initial gas surface density at our reference distance $R_{0}=5.2$ AU. We also used a fixed outer radius of the disk of 30 AU. In the updated model, the disk is free to spread or shrink at the outer edge, and the initial gas surface density profile takes the form 
\beq\label{eq:newprofilegeneral}
\Sigma(r,t=0)=\Sigma_0\left(\frac{r}{R_0}\right)^{-\gamma}\exp\left[-\left(\frac{r}{R_{\rm c}}\right)^{2-\gamma}\right],
\eeq
which corresponds to a power law for radii much less than $R_{\rm c}$, followed by an exponential decrease. As shown by Andrews et al. (\cite{andrewswilner2009}), this form can be derived from the similarity solution of the viscous disk problem of  Lynden-Bell \& Pringle (\cite{lyndenbellpringle1974}). We use two radii, $R_{0}$ and $R_{\rm c}$ instead of only $R_{\rm c}$. This does not mean that we introduce an additional parameter, it just means that the normalization constant $\Sigma_{0}$ takes another value, namely the one at 5.2 AU, which simplifies comparison with previous work. Andrews et al. (\cite{andrewswilner2009}, \cite{andrewswilner2010}) note that profiles with an exponential decrease provide better agreement with observations than a pure power-law  profile with a sharp outer edge.  {The observed values for $\gamma$ roughly follow a normal distribution with a mean of $\gamma=0.9$ and a standard deviation of 0.2, which is compatible with one uniform value characterizing the entire observational sample of 16 disks (Andrews et al. \cite{andrewswilner2010}). We therefore use $\gamma=0.9$ in the simulations, see Appendix \ref{result:stabilitydisk}.}

\subsection{Stability against self-gravity}\label{sect:stabilityselfgrav}
In order to check whether we specify initial conditions which make the application of our (constant) $\alpha$ model self-consistent, we study which disks with the new initial profile are stable against the development of spiral waves due to self-gravity. 

The sound speed $c_{\rm s}$ is given as 
\beq
c_{\rm s}=\sqrt{\frac{k_{\rm b} T_{\rm mid}}{\mu m_{\rm H}}}
\eeq
where $T_{\rm mid}$ is the midplane temperature (obtained from our vertical structure model),  $\mu$ the mean molecular weight (set to 2.27),   {$k_{\rm b}$ the Boltzmann constant,} and $m_{\rm H}$ the mass of a hydrogen atom. The Keplerian frequency $\Omega$ at a distance $r$ around a star of mass $\mstar$ is given as 
\beq
\Omega=\sqrt{\frac{G \mstar}{r^3}}
\eeq
and the Toomre (\cite{toomre1981}) parameter is given as
\beq\label{eq:qtoomre}
Q_{\rm Toomre}=\frac{c_{\rm s} \Omega}{\pi G \Sigma},
\eeq
assuming that the disk is  in Keplerian rotation despite its own mass.

{With a Toomre $Q_{\rm Toomre}$ smaller than unity, the disk is unstable against local axisymmetric (ring-like) gravitational disturbances.  Numerical simulations show that non-axisymmetric  disturbances (multi-armed spiral waves) start growing slightly earlier at $Q_{\rm Toomre,crit}\lesssim1.7$ (Durisen et al. \cite{dangelodurisen2010}).}

{For massive disks ($\mdisk/\mstar\gtrsim0.1$), one  has to consider global  instabilities, too. In particular the m=1 mode (corresponding to a one-armed spiral) was initially thought to be crucial (Adams et al. \cite{adamsruden1989}, Shu et al. \cite{shutremaine1990}) as it can be amplified by the SLING mechanism.  It should however be noted (cf. Bodenheimer \cite{bodenheimer1992}; Papaloizou \& Lin \cite{papaloizoulin1995})  that the type of instability described by Adams et al. (\cite{adamsruden1989}) and Shu et al. (\cite{shutremaine1990}) is sensitive to the outer boundary condition of the disk as it requires a sharp outer edge  in order to reflect incoming density waves. This is not the case for the disk profile considered here,  and the nonlinear numerical simulations of Christodoulu \& Narayan (\cite{christodoulou1992}) and Heemskerk et al. (\cite{heemskerk1992}) have not confirmed this specific type of instability.  Later numerical simulations have rather found (e.g. Nelson et al. \cite{nelsonbenz1998}; Lodato \& Rice \cite{lodatorice2004}) that with increasing disk mass, lower (global) modes do become more important than at lower disk masses, but that they have a mixed character involving various modes with m$\lesssim5$. We address the global stability in Appendix \ref{sect:globalinst}.} 

\subsection{Stability against clumping}
If $Q_{\rm Toomre}\lesssim1.7$, spiral arms develop in the disk. Another question is whether these arms can fragment into bound clumps of self-gravitating gas. This question can be addressed by looking at the cooling timescale. 

The representative  gas density $\rho$ in the disk is given as $\rho\approx\frac{\Sigma}{2 H}$ where $H$ is the vertical scale height, while the optical depth $\tau$ is approximatively  $\tau=\kappa(\rho,T_{\rm mid}) \Sigma/2 $ where $\kappa$ is the opacity taken from Bell \& Lin (\cite{belllin1994}),  {and $\Sigma$ is the local gas surface density}. The effective optical depth can be estimated as  $\tau_{\rm eff}=\tau + \frac{1}{\tau}$ (Rafikov \cite{rafikov2005}), so that the  cooling timescale $\tau_{\rm cool}$ of a gas parcel is approximatively (Kratter et al. \cite{krattermurrayclay2010})
\beq
\tau_{\rm cool}=\frac{3 \gamma_{\rm ad} \Sigma c_s^2 \tau_{\rm eff}}{32 (\gamma_{\rm ad}-1) \sigma T_{\rm mid}^4}.
\eeq
This expression is valid  both in disks where irradiation or viscous dissipation is the main heating source, modulo order unity factors. For simplicity, we have assumed that  {the adiabatic index of the gas is} $\gamma_{\rm ad}=7/5$ everywhere. Strictly speaking, this is appropriate only in the colder parts of the disk where hydrogen is molecular. 

In order to form a bound clump, and prevent it from being sheared apart, the clump must cool on a timescale comparable to the orbital timescale, or in other words (Gammie \cite{gammie2001})
\beq
\tau_{\rm cool} \Omega \lesssim \xi
\eeq
where we assume a critical $\xi=3$ (Kratter et al. 2010). In the Appendix \ref{result:stabilitydisk} we use these equations to determine the most massive stable disk, and to study the general stability of irradiated $\alpha$ disks.

\subsection{Photoevaporation}\label{sect:photoevap} 
In the first generation of the model (Alibert et al. \cite{alibertmordasini2005}) we calculated the mass loss of the disk due to photoevaporation according to
\beq
\dot{\Sigma}_{w} =\left\{ \begin{array}{ll}
0 & \textrm{for}\  r < R_g \\
\frac{\dot{M}_{\rm wind}}{2 \pi (R_{\rm max}-R_{\rm g}) r} & \textrm{otherwise}
\end{array} \right.
\eeq
where we set the gravitational radius  $R_{\rm g}$ equal 5 AU, and  $R_{\rm max}=30$ AU was the total radius in the old model. In this prescription, the mass loss is concentrated at  $R_{\rm g}$ (similar to internal photoevaporation) but the total loss rate was externally  specified, similar to external photoevaporation. 

It should be noted that the primary role of photoevaporation in the framework of planet formation models like the one here is that it acts as a controlling mechanism for the disk lifetime, which is important for the final mass of giant planets. The finer description of the photoevaporation (in specific where how much gas is removed) is in contrast not very important, as the detailed description becomes important only when the typical disk surface density (or alternatively, the accretion rate onto the star) has become small. At such surface densities, the main gas driven mechanisms (gas accretion, tidal migration) have however already fallen to a low level.

\subsection{External photoevaporation}\label{sect:externalphotoevap}
In the updated model, we split the mass loss due to photoevaporation into two components: External photoevaporation due to the presence of massive stars in the birth cluster of the host star, and internal photoevaporation by the planet's host star itself. 

For the external photoevaporation, we follow the description of FUV evaporation in Matsuyama, Johnstone \& Hartmann (\cite{matsuyamajohnstone2003}). The FUV radiation (6-13.6 eV) heats a neutral layer of dissociated hydrogen (mean molecular weight $\mu_{\rm I}=1.35$) to a temperature of order  $T_{\rm I}=10^3$ K, which launches a supersonic flow. We thus get a sound speed 
\beq
c_{s,{\rm I}}=\sqrt{\frac{k_{\rm b} T_{\rm I}}{\mu_{\rm I} m_{\rm H}}}
\eeq
and a gravitational radius 
\beq
R_{g,{\rm I}}=\frac{G M_*}{c_{s,{\rm I}}^2}
\eeq
which corresponds, for a 1 $\msun$ star to about 144 AU. Following Matsuyama et al. (\cite{matsuyamajohnstone2003}) we assume that mass is removed outside the effective gravitational radius $\beta_{\rm I} R_{g,{\rm I}}$ with $\beta_{\rm I}=0.5$, and that the total mass loss rate is
\beq
\dot{\Sigma}_{w,{\rm ext}}  =\left\{ \begin{array}{ll}
0 & \textrm{for}\  r < \beta_{\rm I} R_{g,{\rm I}} \\
\frac{\dot{M}_{\rm wind, ext}}{\pi (R_{\rm max}^2- \beta_{\rm I}^2 R_{g,{\rm I}}^2)} & \textrm{otherwise}
\end{array} \right.
\eeq
where $\dot{M}_{\rm wind, ext}$ is a parameter that gives the total mass loss rate if the disk were to have an outer radius of $R_{\rm max}=1000$ AU.  The effect of this mechanism is thus to remove mass outside a radius of about 70 AU. In practice it is found that the actual total mass loss rates are clearly smaller than the specified $\dot{M}_{\rm wind, ext}$ because the disk evolves to a state where the outer radius is given by the equilibrium of mass removal by photoevaporation, and outward viscous spreading (see also Adams et al. \cite{adamshollenbach2004}). 

In this description of external photoevaporation, we have not taken into account that if the host star is born in a large cluster with OB stars, and the distance to one of the massive stars is small ($d\lesssim0.03$ pc for the Trapezium Cluster, Johnstone et al. \cite{johnstonehollenbach1998}), ionizing EUV can lead to a very rapid dispersal of the protoplanetary disk (Matsuyama et al. \cite{matsuyamajohnstone2003}). It is however rather unlikely that such a very high radiation environment is representative for the majority of young stars, in which we are interested here (Adams et al. \cite{adamshollenbach2004}; Roccatagliata et al. \cite{roccatagliatabouwman2011} for an observational study).  We have also neglected the findings of Adams et al. (\cite{adamshollenbach2004}) that significant mass loss due to external FUV radiation might also happen well inside $R_{g,{\rm I}}$ and leave this for improvements in further work.

\begin{figure*}
\begin{center}
 \begin{minipage}[lt]{12.2cm}
\begin{minipage}[lt]{6.1cm}
\includegraphics[width=6cm,angle=0]{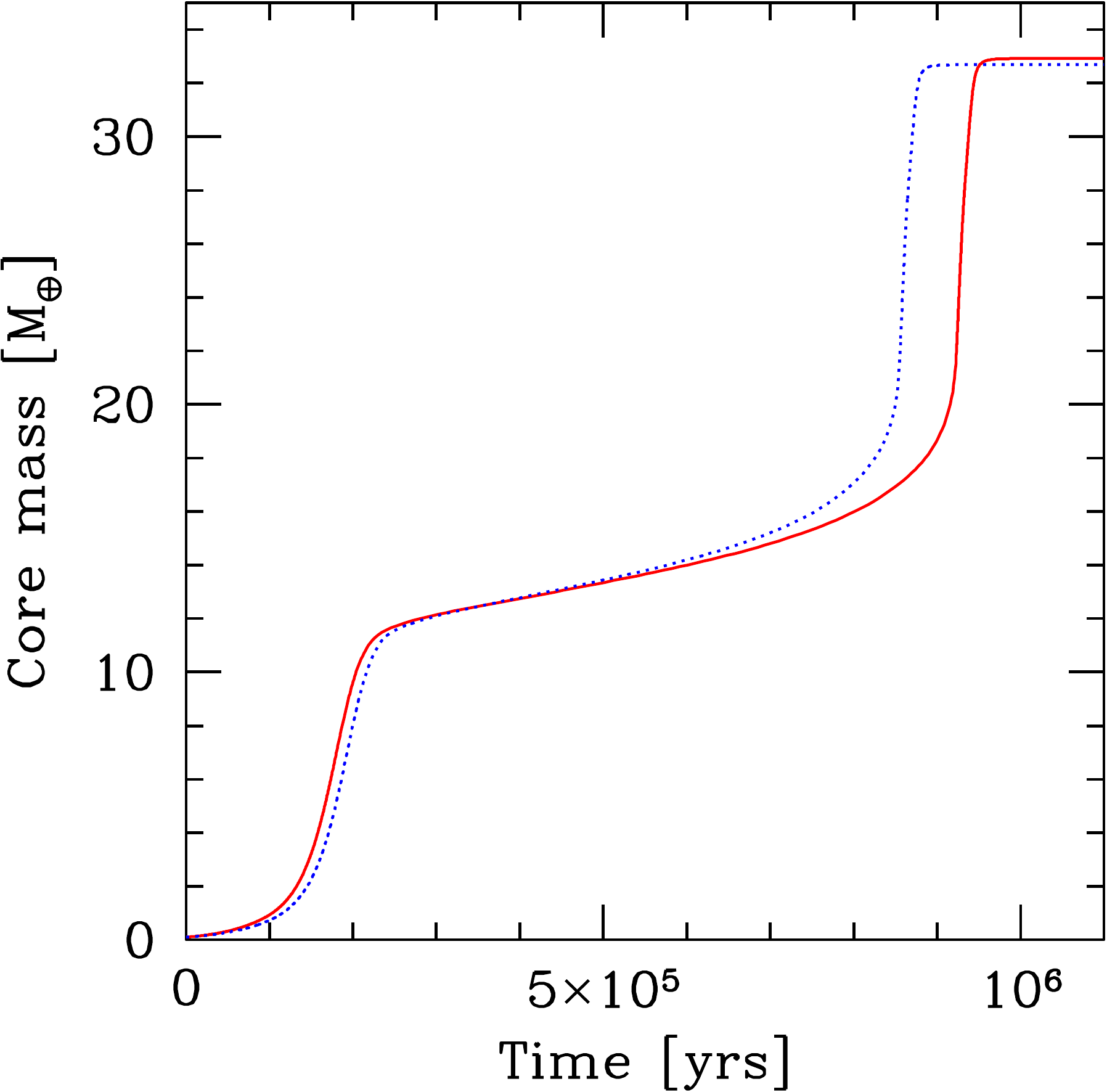}
\includegraphics[width=6cm,angle=0]{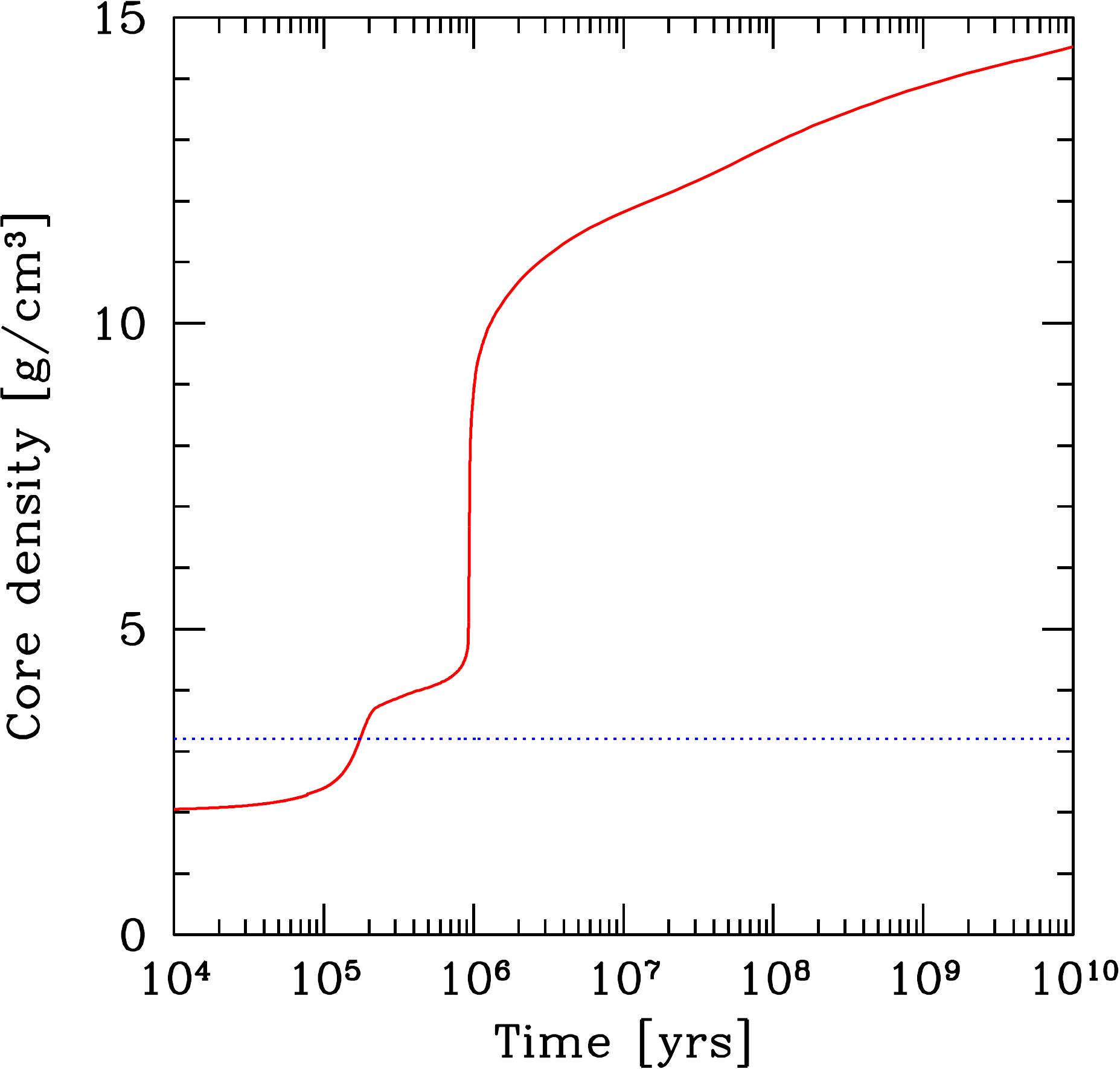}
\end{minipage}
\hfill
\begin{minipage}[lt]{6.1cm}
\includegraphics[width=6cm,angle=0]{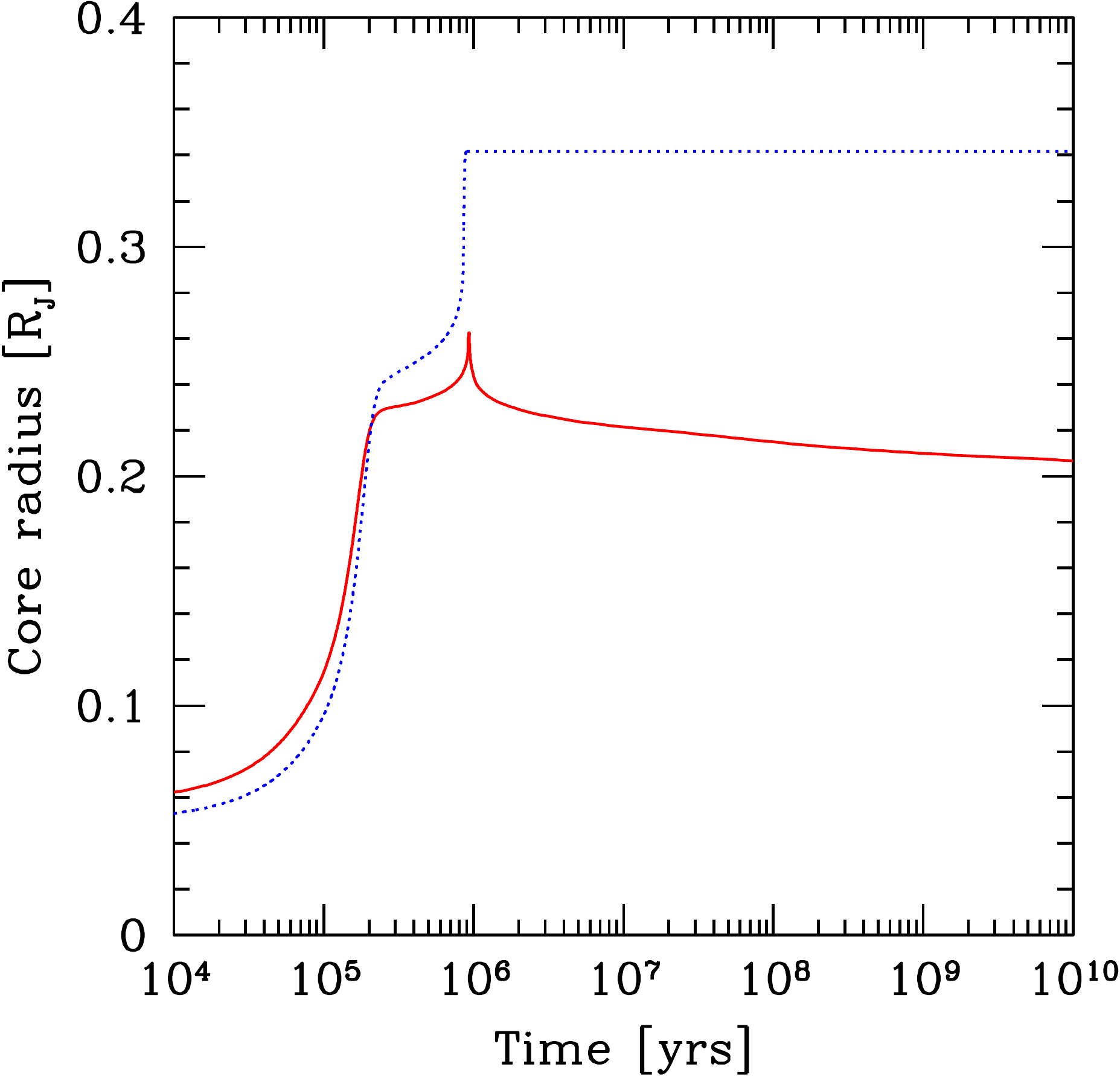}
\includegraphics[width=6cm,angle=0]{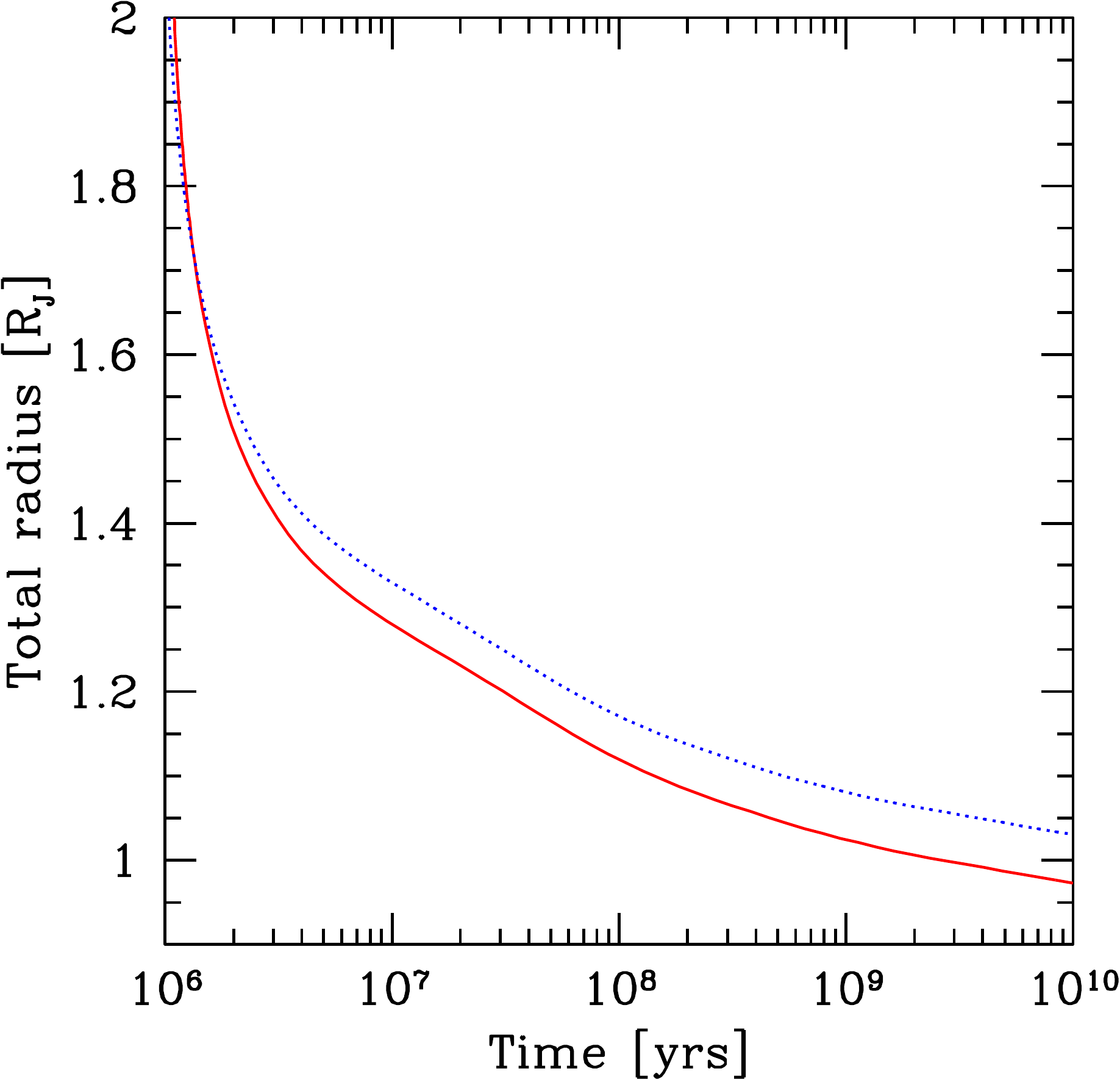}
\end{minipage}
\hfill
\end{minipage}
\begin{minipage}{6cm}
\caption{Effect of the variable core density on the in situ formation and evolution of Jupiter. In each panel, red solid lines show the case of a variable core density, while blue dotted lines correspond to a constant core density of 3.2 g/cm$^{3}$. The top left panel shows the core mass as a function of time during the formation phase, afterwards it is constant. The top right panel is the radius of the core. Note the local maximum for the variable density case near about 1 Myr. The bottom left panel is the mean core density. The bottom right panel finally shows the total radius of the planet during the long-term evolution.   }\label{fig:trhocore}
\end{minipage}
\end{center}
\end{figure*}

\subsection{Internal photoevaporation}\label{sect:internalphotoevap}
The photoevaporation due to the host star is modeled as in Clarke, Gendrin \& Sottomayor (\cite{clarkegendrin2001}) which is  based on the  ``weak stellar wind'' case studied by Hollenbach et al. (\cite{hollenbachjohnstone1994}). In this model, the EUV radiation ($>13.6$ eV) from the host star leads to an ionized layer with a temperature of order $ T_{\rm II}=10^4$ K, where the mean molecular mass is $\mu_{\rm II}=0.68$. The  sound speed is then
\beq
c_{s,{\rm II}}=\sqrt{\frac{k_{\rm b} T_{\rm II}}{\mu_{\rm II} u_{\rm H}}} \approx 10 \ {\rm km/s}
\eeq
and the gravitational radius is
\beq
R_{g,{\rm II}}=\frac{G M_*}{c_{s,{\rm II}}^2}
\eeq
which corresponds for a 1 $\msun$ star to about 7 AU. As Clarke et al. (\cite{clarkegendrin2001}) we use a scaling radius $R_{14}=\beta_{\rm II} R_{g,{\rm II}}/10^{14}\ {\rm cm}$ and assume a $\beta_{\rm II}=0.69$ which means that the effective gravitational radius is located at 5 AU as in our earlier models. The base  density can be estimated as 
\beq
n_{0}(R_{\rm 14})=k_{\rm Hol} \Phi_{\rm 41}^{1/2} R_{\rm 14}^{-3/2}
\eeq
where the constant $k_{\rm Hol}=5.7\times10^{7}$ is taken from the hydrodynamic simulations by  Hollenbach et al. (\cite{hollenbachjohnstone1994}), and $\Phi_{\rm 41}$ is the ionizing photon luminosity of the central star in units of  $10^{41}$ s$^{-1}$. We assume $\Phi_{\rm 41}=1$. The base density varies with radius as 
\beq
n_{0}(r)=n_{0}(R_{\rm 14}) \left(\frac{r}{\beta_{\rm II} R_{g,{\rm II}}}\right)^{-5/2}
\eeq
which means that most of the wind is originating close to the effective gravitational radius. The decrease of the surface density is then finally
\beq
\dot{\Sigma}_{w,{\rm int}}  =\left\{ \begin{array}{ll} 0 & \textrm{for}\  r <  \beta_{\rm II} R_{g,{\rm II}}\\
2 c_{s,{\rm II}} n_0(r) m_{\rm H} & \textrm{otherwise}.
\end{array} \right.
\eeq
This results in a total, constant mass loss rate due to internal photoevaporation of about $3\times 10^{-10}$ $\msun$/yr, a value similar to what was found by Alexander \& Armitage (\cite{alexanderarmitage2009}).

The final photoevaporation rate which enters the master equation for the evolution of the surface density (see e.g. Alibert et al. \cite{alibertmordasini2005}) is then given by the sum of the two mechanisms,
\beq
\dot{\Sigma}_{w}=\dot{\Sigma}_{w,{\rm ext}}+\dot{\Sigma}_{w,{\rm int}}.
\eeq
{The characteristic evolution of a protoplanetary disk under the influence of the  photoevaporation model presented here has essentially already been shown in  Clarke et al. (\cite{clarkegendrin2001}) and Matsuyama et al. (\cite{matsuyamajohnstone2003}). To allow comparison, and for reference, we show our results regarding the protoplanetary disk model in the appendices \ref{result:stabilitydisk} and \ref{sect:diskcharactevo}.}

\section{Results}
We use the upgrades of the model to study first the effect of the variable core density on the in situ formation and evolution of Jupiter (Sect. \ref{result:coredensity}).  In Section \ref{result:lradio} we then analyze the evolution (and formation) of a close-in super-Earth planet for which the radiogenic heating of the core influences the evolution of the radius, which is potentially relevant for transit observations. We also compare our results with previous studies.  {The most important result is shown in Section \ref{sect:formationevolutionmrr} where we study the formation and evolution of the planetary mass-radius relationship with population synthesis calculations. We compare the synthetic mass-radius relationship with the observed one, and derive an expression for the mean planetary radius as function of mass. Comparisons are also made with the radius distribution as found by the \textit{Kepler} satellite}.

\section{Jupiter formation and evolution: effect of a variable core density}\label{result:coredensity}
Figure \ref{fig:trhocore} compares the in situ formation and evolution with a constant core density of 3.2 g/cm$^{3}$ (as usually assumed in formation models, e.g. Pollack et al. \cite{pollackhubickyj1996} or Movshovitz et al. (\cite{mbpl2010}) and a variable one, using the model described above. The simulation with a variable core density is the nominal case discussed in Paper I, where all parameters used in the simulation can be found. Except for the density,  all settings  are identical in the two simulations.

\begin{figure*}
\begin{center}
\begin{minipage}{0.5\textwidth}
	      \centering
        \includegraphics[width=\textwidth]{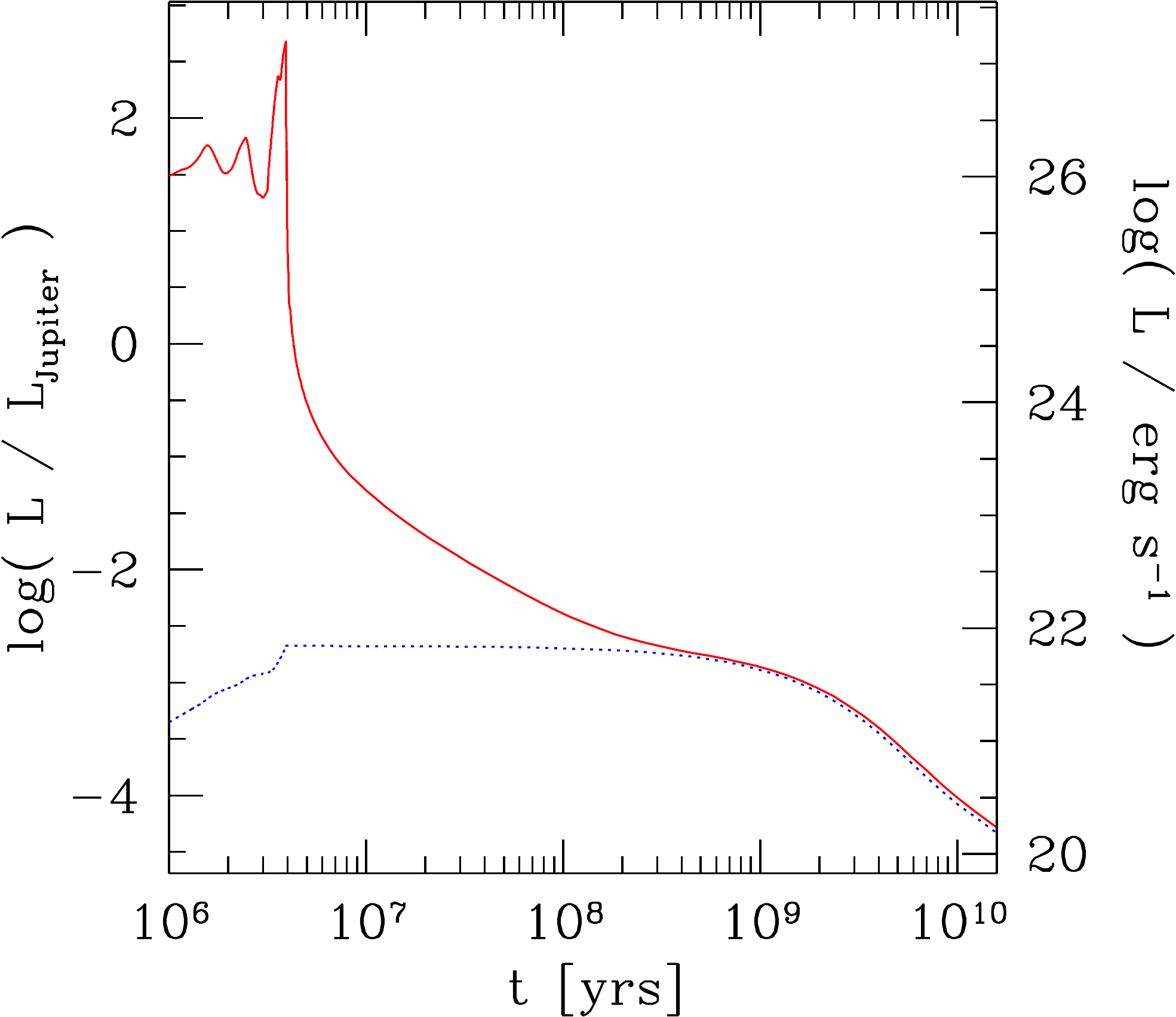}
     \end{minipage}\hfill
     \begin{minipage}{0.5\textwidth}
      \centering
       \includegraphics[width=0.85\textwidth]{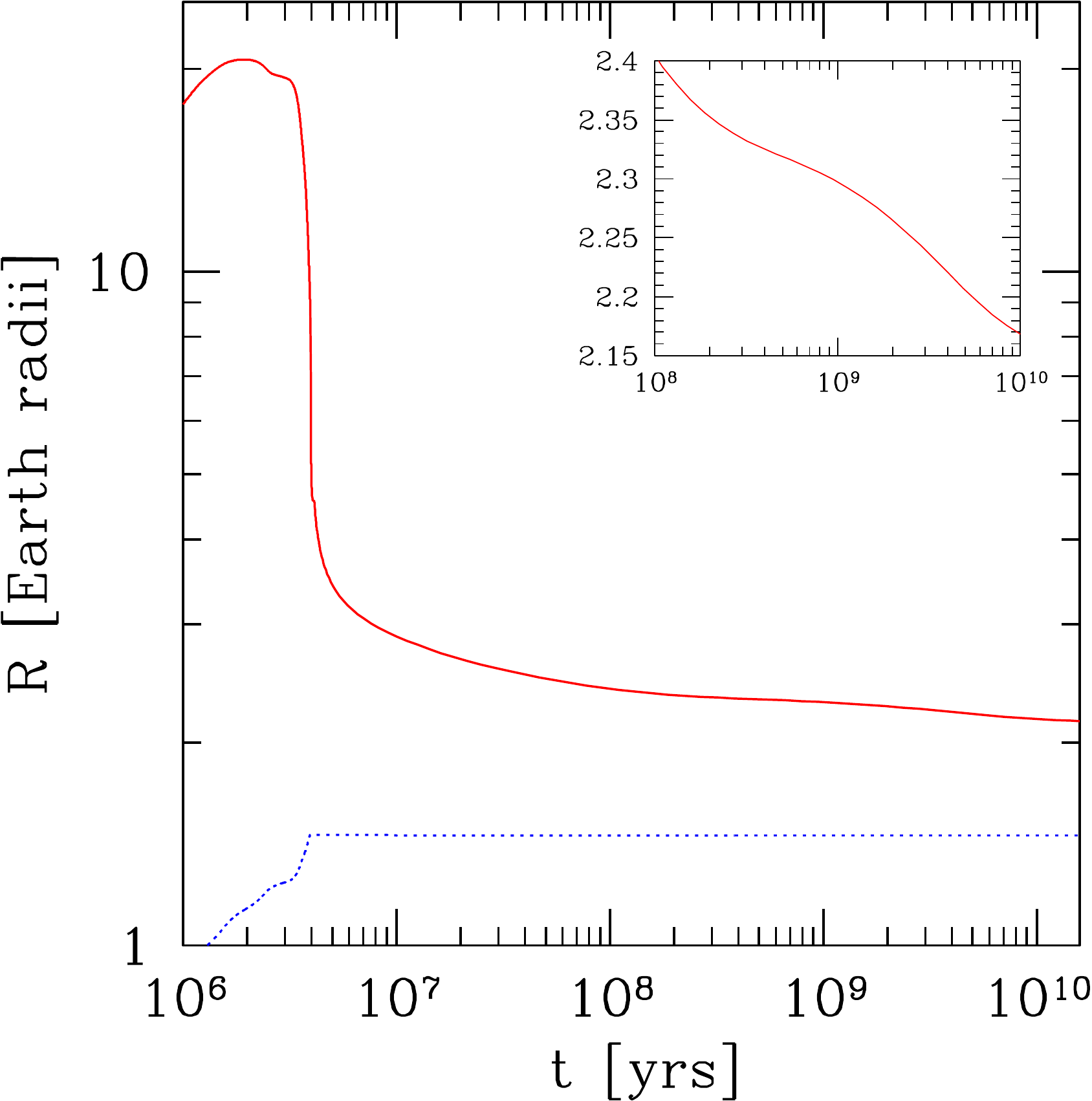}
     \end{minipage}
\caption{Temporal evolution of the luminosity and the radius of a  close-in, super-Earth planet with a tenuous  H$_{2}$/He envelope ($\mz=4.2 \mearth$, $\mxy=0.045 \mearth$ i.e. about 1\%). Left panel: intrinsic luminosity as a function of time.  The red, solid line is the total intrinsic luminosity, while the blue dotted line is $L_{\rm radio}$.  The left axis gives $L$ in units of present day intrinsic Jovian luminosities. The right panel shows the total radius (red solid line) and the radius of the solid core (blue dotted line).  {The inset figure shows the total radius at late times. Note the delayed contraction between about  {0.2} and  {2} Gyrs due to radiogenic heating.} Atmospheric mass loss, as well as outgassing are neglected  so that one directly sees the effect of $L_{\rm radio}$. These simplifications could however also mean that the evolution is in reality more complex than shown here. }\label{fig:lradioex}
\end{center}
\end{figure*}

The top left panel of the figure shows the core mass as a function of time during the formation phase. It immediately demonstrates that the formation is very similar in both cases, in agreement with Pollack et al. (\cite{pollackhubickyj1996}) who found only a weak dependence of the formation on the core density. The core grows in the simulation with the constant  density during phase I initially slightly slower, as the actual mean density is then still smaller (about 2 g/cm$^{3}$), as can be seen in the bottom left panel. This leads to a larger capture radius for planetesimals and thus to a faster growth. This is because in phase I, capture and core radius are initially identical (Paper I).

For the variable density, the growth of the core leads to an increase of the density due to the increasing pressure, therefore the two densities become equal roughly in the middle of phase I. From now on, the actual density is always larger than 3.2  g/cm$^{3}$.  The further mass accretion in phase II is similar in both cases. This is due to the fact that the capture radius for planetesimals is now between 10 to 60 times larger than $\rcore$. The capture radius is similar in both cases, because the  density structure in the envelope relevant for the capture is similar in both cases, since it is located quite far away from the core.  The mean core density in phase II is of order 4 g/cm$^{3}$, so quite close to the constant value. As the runaway gas accretion phase is approached, the density in the variable case increases, so that the evolution starts to diverge slightly, with the constant density case growing now faster.  This leads to a difference in the formation timescale of somewhat less than $10^{5}$ years, corresponding to about 10\%. 

The collapse of the planet and the rapid accretion of $\sim300$ Earth masses of gas lead to a rapid increase of the core density during the collapse/gas runaway accretion phase, due to the quickly growing external pressure exerted on the core. The density goes up quickly from about 5 to 10 g/cm$^{3}$. The core grows during the same time by another $\sim17 \mearth$ in mass.  The increase of the density  however over-compensates that, so that $\rcore$ actually shrinks after having reached a local maximum,  as visible in the top right panel. This is an interesting effect. We should however keep in mind that it is  {unclear whether} the planetesimals still penetrate  to the core in this phase (Paper I),  {therefore the moment of the maximum core radius might occur already at an earlier time than shown here where it coincides with the onset of gas runaway accretion.} 

During the long-term evolution, a core density of 3.2  g/cm$^{3}$ is clearly  {not} realistic. The actual core densities range  between 10 and 15 g/cm$^{3}$ (14.3 g/cm$^{3}$ at 4.6 Gyrs), leading at late times to a significant difference of $\rcore$ of about 0.13$\rj$ between the two cases.  

It is interesting to investigate how this difference affects the observable total radius. This is shown in the bottom right panel of Fig. \ref{fig:trhocore}.  We find that the total radius  is larger in the constant density case, too. The difference is as expected smaller than for the core radius, namely about 0.06 $\rj$ at 4.6 Gyrs. This is clearly a non-negligible difference, as it would otherwise correspond to a difference of the core mass of about $30\mearth$ (Paper I; Fortney et al. \cite{fortneymarley2007}), which is a significant quantity. For accurate comparisons of our population synthesis calculations with transit observations we must therefore include a realistic core density model,  {as can implicitly be deduced already from the results of  Fortney et al. (\cite{fortneymarley2007})}.

\section{Evolution of a super-Earth planet including radioactive luminosity}\label{result:lradio}
In Figure \ref{fig:lradioex} we illustrate the effect of $L_{\rm radio}$ on the total luminosity of a low-mass, super-Earth planet (core mass $\mz=4.2 \mearth$) which has, at the end of formation, a tenuous H$_{2}$/He envelope with a mass of  about $\mxy=0.045 \mearth$, i.e.  about 1\%.  {For comparison, Venus has an atmosphere with a mass of approximately $8\times10^{-5} \mearth$, which is a factor $\approx 560$ less. On this scale, $\mxy=0.045 \mearth$ still corresponds to a relatively massive envelope.}

The  H$_{2}$/He envelope mass of low-mass planets at the end of formation varies as it depends on a number of  factors. Important quantities are the grain opacity in the envelope (see the dedicated work of Mordasini et al. \cite{mordasiniklahr2011}), the luminosity of the core due to the accretion of planetesimals, but also, as tests have shown, the radiogenic luminosity of the core due to the decay of $^{26}$Al. Large  {radiogenic} core luminosities during the nebular phase can lead to lower envelope masses because of a stronger pressure support in the gas at a higher temperature.  {By modifying the final envelope mass, short-lived radionuclides can thus have an impact on the long-term evolution of a planet in an indirect way. The planets which are most affected by this mechanism are low-mass, core-dominated ``rocky'' planets, which undergo towards the end of their formation a phase where the accretion rate of planetesimals is low. This can for example occur if a planet is caught in a type I convergence zone, and has emptied its feeding zone. If this happens early enough compared to the half-life of the radionuclides, the radiogenic heating will reduce the amount of gas that can be accreted.}

In the light of the discoveries of the \textit{Kepler} satellite of a number of low-mass, low-density planets (e.g. the \object{Kepler-11} system, Lissauer et al. \cite{lissauerfabrycky2011}) the problem of possible primordial envelope masses in the transition regime between terrestrial and water planets on one hand (without primordial envelopes),  and mini-Neptunes (with primordial envelopes)  has become an important one.  {We show our results concerning this question in Sect. \ref{sect:comprdistkepler}.}

The left panel {of Fig. \ref{fig:lradioex}}  shows the total intrinsic and the radiogenic luminosity (neglecting $^{26}$Al) as a function of time. During the formation of the planet, the luminosity is high, mostly due to the accretion of solids. The specific shape of $L$ in this phase with some wave-like patterns is a consequence of the complex migration behavior of this planet, giving origin to phases with high and low planetesimal accretion rates, and thus core luminosities. The planet eventually migrates to $\sim0.1$ AU. At about 4 Myrs, the accretion of solids (and of gas) ceases. The luminosity is now dominated by the contraction and cooling of the gaseous envelope. At about 400 Myrs however, $L_{\rm radio}$ becomes the dominant source of luminosity for this planet, giving rise to a transient plateau phase of roughly constant $L$. Finally, beyond a few billion years, $L$ starts to decrease again, following now the decrease of $L_{\rm radio}$ according to the radioactive decay law.  {As found by Nettelmann et al. (\cite{nettelmannfortney2011}), we see that it is important to include radiogenic heating to accurately model the thermal evolution of core-dominated planets.}

\subsection{Radius evolution and comparison with Rogers et al. (\cite{rogersbodenheimer2011})}\label{sect:radevorogers}
Knowing the intrinsic luminosity (as well as the irradiation from the host star), we can calculate the radius of a (low-mass) planet as a function of time, and thus compare with results of transit observations. Figure \ref{fig:lradioex} shows the core and total radius as a function of time. During the attached phase, the planet has  a large radius (Paper I), and then contracts rapidly after accretion has stopped  {(i.e. when the protoplanetary gas disk disappears)} to a radius of initially about 3.5 $\rearth$. During the following evolution, it slowly contracts to a radius of about 2.2 $\rearth$ at 4.6 Gyrs. The core radius is only 1.46 $\rearth$, so despite the fact that the envelope   only contains about 1\% of the total mass, it contributes significantly to the radius. This is a well know effect (e.g. Valencia et al. \cite{valenciaikoma2010}; Sect. \ref{sect:errorsgasenve}).  

\begin{figure}
\begin{center}
	      \centering
        \includegraphics[width=0.9\columnwidth]{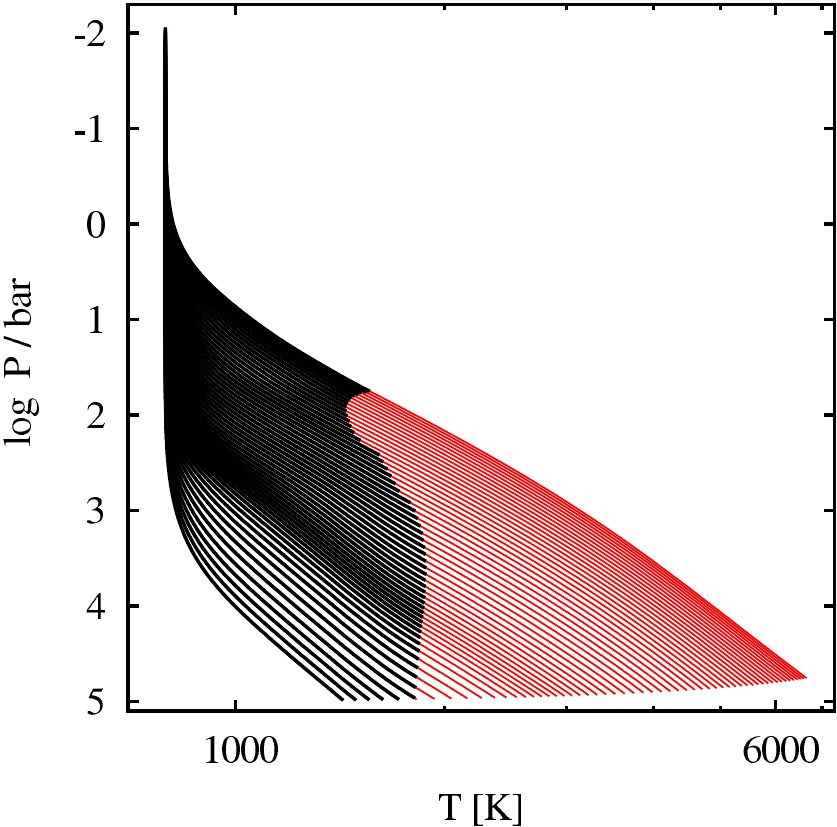}     
\caption{ {Evolutionary sequence of the $p-T$ structure of the envelope. The first structure (rightmost line) is at  5 Myrs, the last structure (leftmost line) is at a hypothetical age of 16 Gyrs. The entire internal structure is shown, which means that the lower end of a line corresponds to the core-envelope interface, while the top end corresponds to the $\tau=2/3$ surface. Red indicates convective parts, black radiative layers. Note that the actual upper part of the envelope/atmosphere will be more complex than shown here, since we only use a simple grey atmosphere in the current model.}}\label{fig:pTsuperearth}
\end{center}
\end{figure}

At an orbital distance of 0.1 AU from a solar-like star and the albedo we assume (Paper I) we get a surface temperature of the planet during the evolutionary phase of about 790 K. This temperature is virtually constant, as the contribution of the intrinsic luminosity compared to the contribution from stellar irradiation is negligible.  {The temporal variation of the stellar luminosity itself is currently not considered in the model}.   {In Fig. \ref{fig:pTsuperearth} we show the temporal evolution of the pressure-temperature profile in the planetary envelope. The entire envelope is shown, so that the lower ends of the lines correspond to the temperature and pressure at the interface with the solid core. The first profile which corresponds to the rightmost line is shown at an age of the planet of 5 Myrs, which is about one million year after the end of the formation phase. At this early time, the temperature at the core-envelope boundary is approximately 6700 K. The last profile which is the leftmost line shows the state at a hypothetical age of 16 Gyrs. The temperature at the core envelope-interface has then fallen  to $\sim1500$ K.  The intrinsic luminosity is decreasing with time, therefore the envelope becomes completely radiative at an age of about 7 Gyrs. At even later times, the envelope would become isothermal.} {Between an age of 5 Myrs and 16 Gyrs, additional $p-T$ structures are shown at irregular time intervals of a few $10^{4}$ yrs (at the beginning) to several $10^{9}$ yrs towards the end, reflecting the increase of the numerical time step due to the increase of the Kelvin-Helmholtz timescale of the planet.}

We note that our treatment of the effect of stellar irradiation on the envelope structure is very basic at the moment (Paper I) and will be improved in future (Guillot \cite{guillot2010}; Heng et al. \cite{henghayek2011}).  {It is therefore likely that the upper part of the atmosphere structure is clearly more complex than shown here.}

It is interesting to compare our results with the ones of Rogers et al. (\cite{rogersbodenheimer2011}) who use a more realistic ``two-stream'' atmosphere (Guillot \cite{guillot2010}). These authors find for a $4\mearth$ planet with a 1\% envelope a radius of about 2.7 and 3.3 $\rearth$, at an equilibrium temperature of 500 and 1000 K, respectively, and an intrinsic luminosity to mass ratio of $L/M=10^{-6.5}=3.2\times10^{-7}$ erg/(g s). These authors do not follow the temporal evolution of the internal luminosity and thus use the $L/M$ ratio as an age proxy, as was also made to calculate Fig. \ref{fig:mrrogerscomp}. 

Our planet has $L/M$ ratios of $6.6\times10^{-6}$, $5.2\times 10^{-7}$,  $1.7\times10^{-7}$ and $1.1\times10^{-8}$ erg/(g s) at 0.01, 0.1, 1 and 10 Gyrs, respectively.  {The values at 1 and 10 Gyrs simply correspond to Eq. \ref{eq:totallradio}}.  One notes the effect of the plateau phase because of radiogenic heating between roughly  {0.2} and  {2} Gyrs. This is also directly visible in the evolution of the radius (Fig. \ref{fig:lradioex}, right panel), where the contraction is  slowed down during this time interval.  At the $L/M$ ratio used by Rogers et al. (\cite{rogersbodenheimer2011}), we find a total radius of about  {2.34} $\rearth$. A lower radius than in  Rogers et al. (\cite{rogersbodenheimer2011}) is expected, as these authors assume an icy composition of the core resulting in a core radius of about 1.9 $\rearth$, while we use an Earth-like ``rocky'' composition as the planet has accreted only inside the ice line, leading to a $R_{\rm core}=1.46 \rearth$. Therefore, there is a difference of the core radii of about 0.45 $\rearth$. Taking this into account, we see that for the total radius, the models agree on a  {0.2} $\rearth$ or better level, corresponding to about $\sim${7}\%. The comparison  {here, and in Sect. \ref{sect:errorsgasenve}}, shows that the radii of low-mass super-Earth planets can be simulated for the population synthesis with  {relatively} good accuracy.  

As the evaporation of the primordial H$_{2}$/He envelope, the thermal cooling of the core as well as the outgassing of a secondary atmosphere during the long-term evolution are not included, we must  {nevertheless keep in mind} that the evolution of the planet on Gyrs could in reality be more complex than shown here. Here, our interest was to show the differential effect of $L_{\rm radio}$. To quantify the importance of subsequent $H_{2}$ outgassing, we can use the work of Rogers et al. (\cite{rogersbodenheimer2011}). These authors find that a rocky planet with a composition as we assume ({2:1 silicate-to-iron ratio}) could outgas  $H_{2}$ corresponding to about 0.2 wt\% of its  mass. This corresponds in the case here to 0.0084 $\mearth$. This is a factor $\sim5$ less than the mass of the primordial atmosphere,  {so that at least this effect should not completely change the evolution}.

\section{Formation and evolution of the planetary mass-radius relationship}\label{sect:formationevolutionmrr}
We now turn to the most important results of this paper. We use the upgraded  model to conduct planetary population synthesis calculations, and study the formation and evolution of the planetary mass-radius relationship. We also discuss the synthetic radius distribution and compare with observations. 

\subsection{Population synthesis}
The population synthesis calculations presented here are obtained with the same basic approach as introduced in Mordasini et al. (\cite{mordasinialibert2009a}), but instead of the original formation model presented first in Alibert, Mordasini \& Benz (\cite{alibertmordasini2004}) we use the extended model described in Paper I and this work.

Table \ref{tab:popsynth} {lists the most important parameters} for the calculations. Regarding the gas disk model, we use the new initial disk profile introduced in Sect. \ref{standardalphamodel} (see Eq. \ref{eq:newprofile})   with an inner border of the computational disk at 0.1 AU. We assume that the gas surface density falls to zero at this point (see Fig.  \ref{fig:sigmaevo}). The new description of photoevaporation of Sect. \ref{sect:photoevap} is applied, and for the calculation of the vertical temperature structure of the disk we now include besides viscous heating also stellar irradiation (Fouchet et al. \cite{fouchetalibert2011}). As in earlier works (Mordasini et al. \cite{mordasinialibert2009a}), we fix the $\alpha$ viscosity parameter to $7\times 10^{-3}$.

Regarding the planetary seed, we start with an embryo of 0.6 $\mearth$. This setting, and the fact that we do not calculate any planetary growth once that the gaseous disk is gone means that our model lacks an essential phase (giant impacts) in the formation of low-mass (terrestrial) planets, so that results about low-mass planets ($M\lesssim10\mearth$) must be regarded with caution (see extended discussion in Mordasini et al. \cite{mordasinialibert2009a}). For the internal structure of the core, we use the model presented in Sect. \ref{sect:variablecoredensity}.  Radiogenic heating is included as a source of core luminosity as described in Sect. \ref{sect:upgradelradio}. The fraction of ice in the core is known as we keep track of how much solids are accreted inside or outside of the iceline. The position of the iceline is given by the initial temperature structure of the nebula (Mordasini et al. \cite{mordasinialibert2009a}). 

Primordial H$_{2}$/He envelopes only are considered in the model. For the treatment of the accretion shock occurring during gas runaway accretion, we assume that the shock  radiates all  liberated energy into outer space i.e. we assume a ``cold start'' scenario (Paper I). The gas accretion rate in the runaway phase is calculated as described in Paper I allowing for non-equilibrium fluxes in the disk. We use the simplification of a constant luminosity within the envelope, due to the reasons explained in Paper I.  We assume that the grain opacity in the gaseous envelope is 0.003 times the value in the interstellar medium.  This value has been determined in Mordasini et al. (\cite{mordasiniklahr2011}) by comparison with detailed grain evolution calculations in protoplanetary envelopes (Movshovitz et al. \cite{mbpl2010}). For the opacity due to the grain-free gas, the tables of Freedman et al. (\cite{freedmanmarley2008}) are used. Two effects related to the planetary envelope which can be relevant during the long-term evolution of the planet are currently not included. First, we do not include deuterium burning in the low number of planets that grow beyond the D-burning limit. This will affect the radius of objects more massive than $\sim$13$\mj$ at early times (e.g. Spiegel et al. \cite{spiegelburrows2011}). Second, we do not take into account that planets close to the star can loose parts of their gaseous envelope due to evaporation (e.g. Lammer et al. \cite{lammerodert2009}; Murray-Clay et al. \cite{murrayclaychiang2009}). 

\subsubsection{Type I migration model}\label{typeImodel}
Regarding orbital migration, we have made a major upgrade relative to the previously used type I migration model. Based on the results of  Paardekooper \& Mellema (\cite{paardekoopermellema2006}), Baruteau \& Masset (\cite{baruteaumasset2007}), Kley \& Crida (\cite{kleycrida2008}), Kley et al. (\cite{kleybitsch2009}) and Paardekooper et al. (\cite{paardekooperbaruteau2010}), we now distinguish several sub-types of type I migration which lead to different migration directions and rates. We distinguish isothermal and adiabatic regimes as well as regimes where the corotation torque saturates. We also take into account the decrease of the surface density due to beginning gap formation. For the transition to type II migration, we now use the criterion derived by Crida et al. (\cite{cridamorbidelli2006}).  An important effect of the non-isothermal type I migration is that it leads to the existence of convergence zones, i.e. to places in the disk towards which the planets migrate both from in- and outside (Lyra et al. \cite{lyrapaardekooper2010}; Mordasini et al. \cite{mordasinidittkrist2011}; Kretke \& Lin \cite{kretkelin2012}). Once a low-mass planet has migrated into such a convergence zone, it follows the viscous evolution of the disk. This means that it migrates on a viscous timescale, which is typically much longer than the migration timescale in the isothermal type I approximation (Tanaka et al. \cite{tanakatakeuchi2002}). The rapid, and always inward migration  in the isothermal approximation as assumed in our previous studies made arbitrary reduction factors necessary. In the new migration model, no efficient factors are used any more. A short overview of the migration model is presented in Mordasini et al. (\cite{mordasinidittkrist2011}) while a detailed description will be given in Dittkrist et al. (in prep.). 

\begin{table}
\caption{Parameters and settings for the population synthesis. }\label{tab:popsynth}
\begin{center}
\begin{tabular}{l c}
\hline \hline
Quantity & value  \\ \hline
Initial disk profile          & Eq. \ref{eq:newprofile}\\
Disk viscosity parameter $\alpha$ & $7\times 10^{-3}$\\
Inner radius of computational disk & 0.1 AU \\
Outer radius of computational disk & free \\
Gas surface density at inner radius & falls to 0 \\
New photoevaporation model &  included \\
Irradiation for disk temperature profile & included \\
Rockline  & not included \\
Iceline & included \\  \hline
Embryo starting mass & 0.6 $\mearth$ \\
Core density & variable \\
Radioactive luminosity & included\\ \hline
Envelope type & primordial H$_{2}$/He\\
Accretion shock luminosity treatment & cold accretion \\
$dl/dr$ in the envelope & zero \\
Grain opacity reduction factor & 0.003 \\
Deuterium burning & not included  \\ 
Envelope evaporation & not included  \\ \hline
Type I migration & non-isothermal \\
Type I migration reduction factor & none \\
Transition criterion type I to type II & Crida et al. \cite{cridamorbidelli2006}\\ \hline
Stellar mass & 1 $\msun$\\
Simulation duration & 10 Gyrs \\ 
Number of embryos per disk & 1 \\\hline
\end{tabular}
\end{center}
\end{table}%

We fix the stellar mass to 1 $\msun$ in all calculations and follow the evolution of the planets up to an age of 10 Gyrs. The calculations presented here are  made in the one-embryo-per-disk approximation.  The effects induced by the concurrent formation of many protoplanets  (see Thommes et al. \cite{thommesmatsumura2008}; Ida \& Lin \cite{idalin2010}; Hellary \& Nelson \cite{hellarynelson2012}) like for example capture into mean-motion resonances, competition for  gas and solids, or effects of the eccentricity onto the migration,  cannot be described in the current model. The results from HARPS indicate that low-mass planets are usually found in multiple systems (Mayor et al. \cite{mayormarmier2011}). This simplification is therefore an important limitation which should be kept in mind when comparing the model outcomes for low-mass planets with observations. The model will reach its full predictive power only when the improvements presented here will be combined with the calculation of the concurrent formation  of many embryos per disk (Carron et al. in prep.).  Giant planets on the other hand are more frequently found as single planets (e.g. Latham et al. \cite{lathamrowe2011}). For this class of planet, our predictions should be more solid. 

\begin{figure*}
\begin{center}
\begin{minipage}{0.5\textwidth}
	      \centering
        \includegraphics[width=0.9\textwidth]{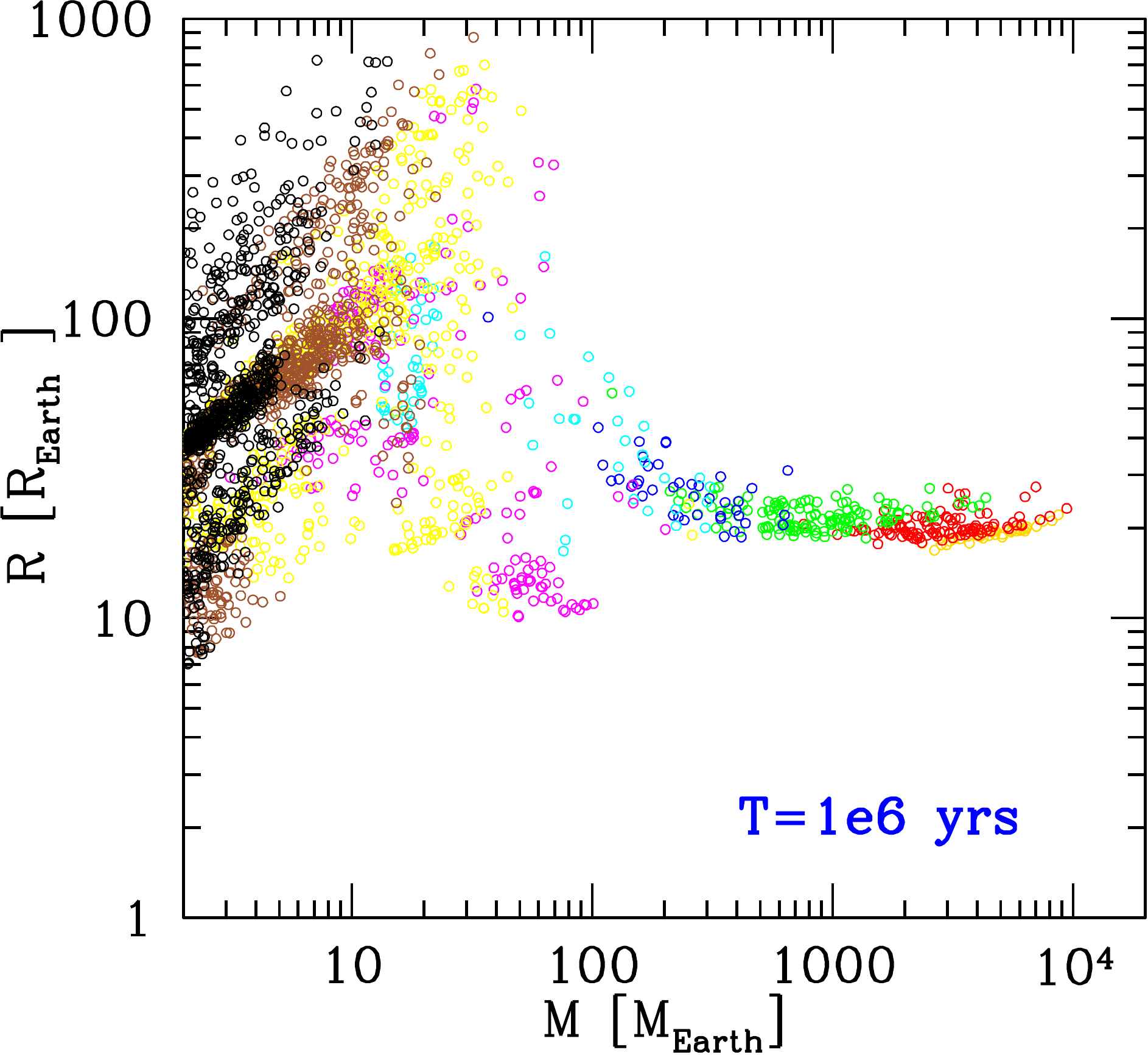}
        \includegraphics[width=0.9\textwidth]{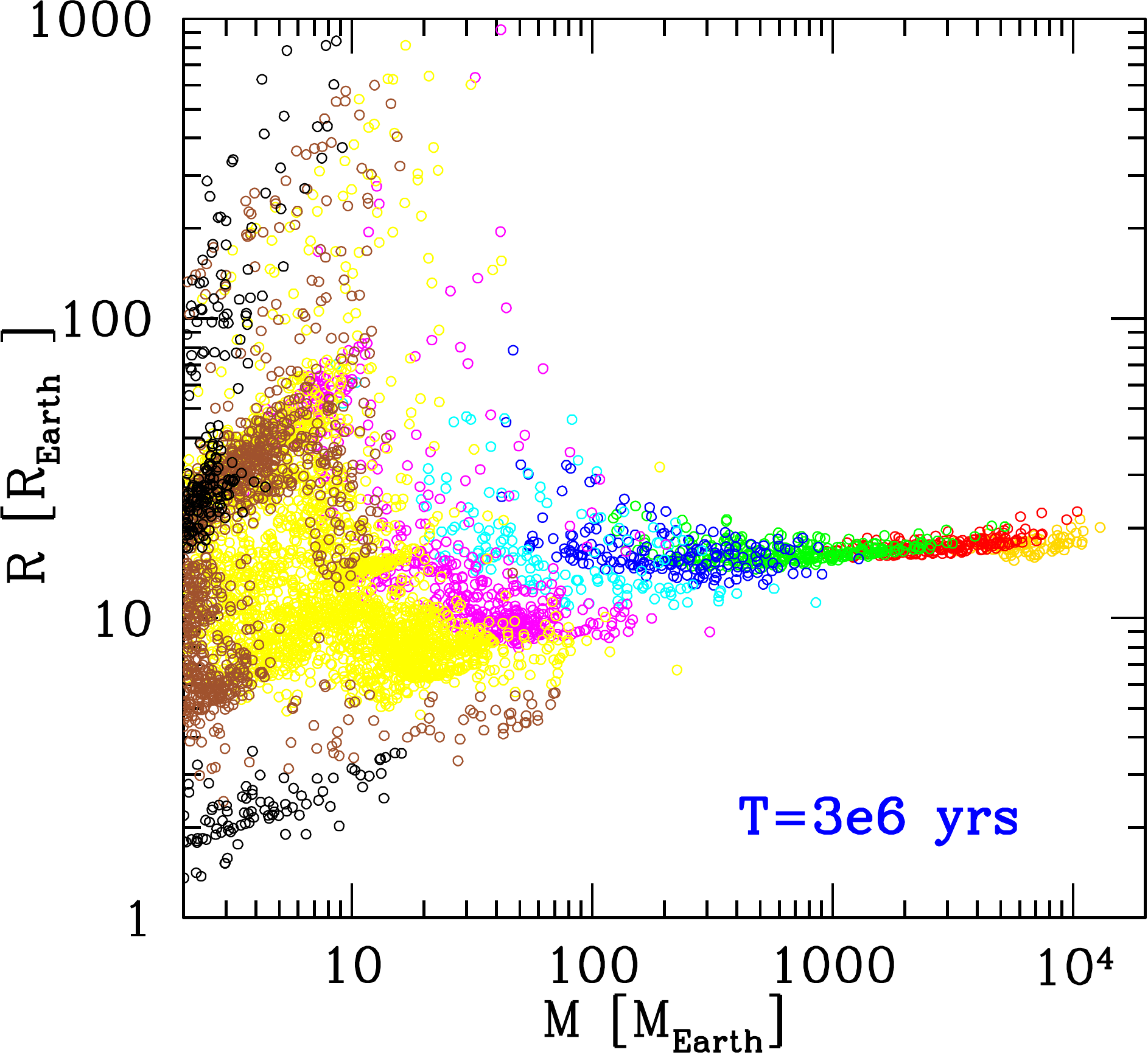}
     \end{minipage}\hfill
     \begin{minipage}{0.5\textwidth}
      \centering
       \includegraphics[width=0.9\textwidth]{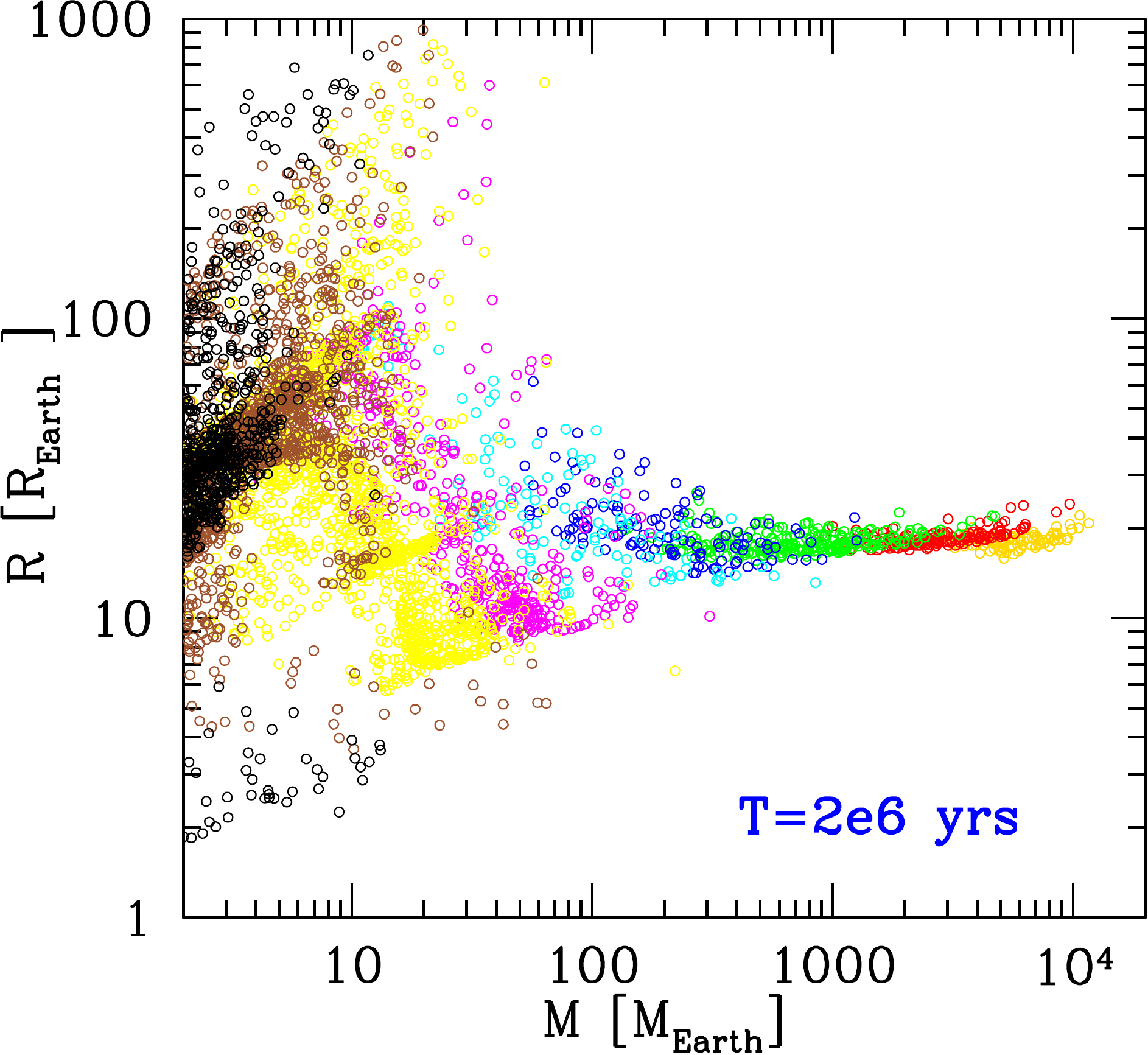}
       \includegraphics[width=0.9\textwidth]{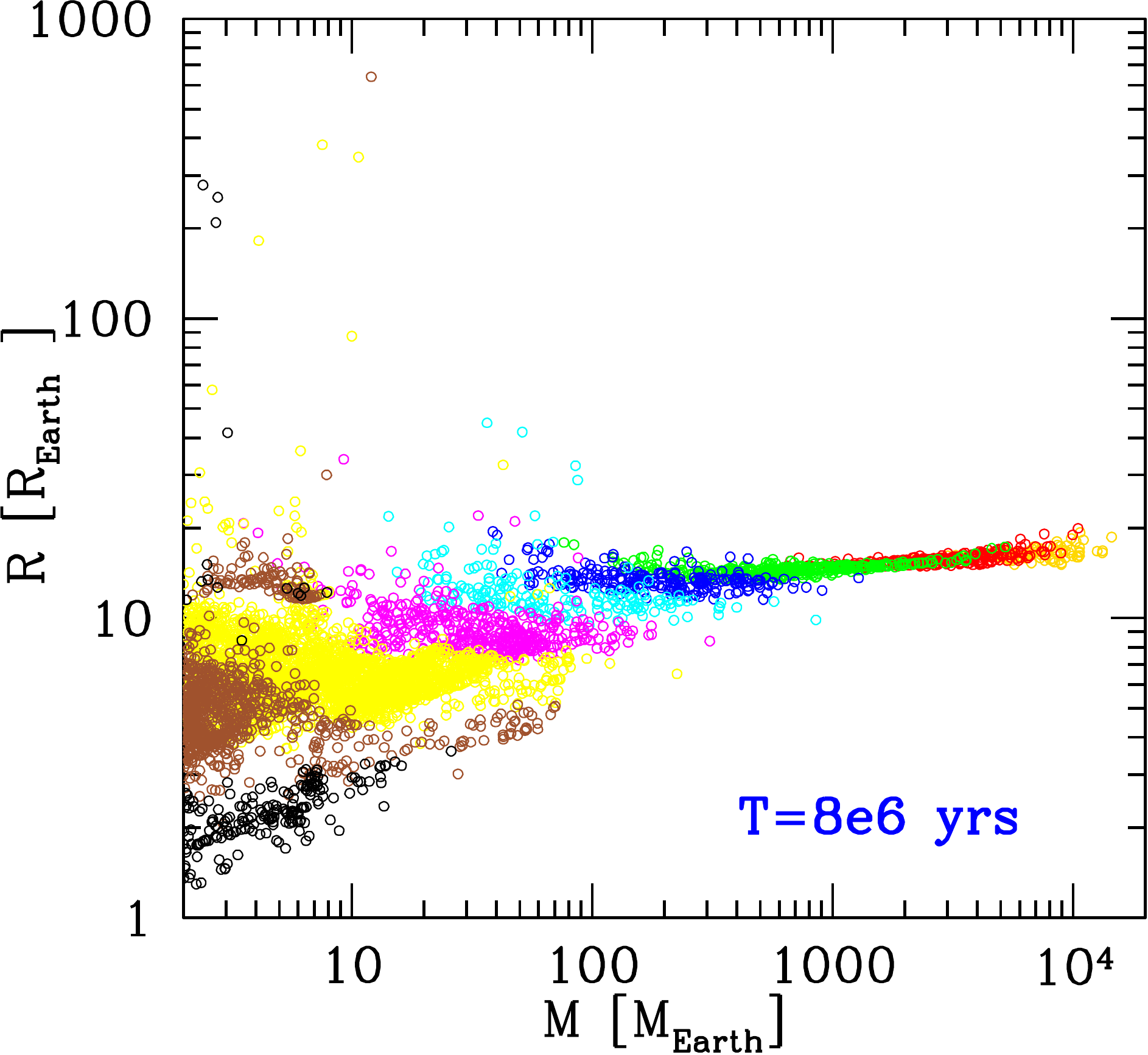}

     \end{minipage}
\caption{Temporal evolution of the planetary mass-radius diagram during the formation phase.  The colors show the fraction of heavy elements $Z=\mz/M$ in a planet. Orange: $Z\leq1\%$. Red: $1<Z\leq5\%$. Green: $5<Z\leq20$\%. Blue: $20<Z\leq40$\%. Cyan: $40<Z\leq60$\%. Magenta: $60<Z\leq80$\%.  Yellow: $80<Z\leq95$\%.  Brown: $95<Z\leq99$\%. Black: $Z>99$\%. }\label{fig:mrform}
\end{center}
\end{figure*} 

\subsubsection{Probability distributions}
As in earlier studies, we use four Monte-Carlo variables to represent the variations of the initial conditions of planet formation. For the disk metallicity and the initial semimajor axis of the embryo we use the same probability distributions as in Mordasini et al. (\cite{mordasinialibert2009a}). For the distribution of the initial gas mass in the protoplanetary disks we now use the distribution found by Andrews et al. (\cite{andrewswilner2010}) for the Ophiuchus star forming region. This replaces the much older observational data of  Beckwith \& Sargent (\cite{beckwithsargent1996}). The new distribution is characterized by a mean and standard deviation in $\log(\mdisk/\msun)$ of  -1.66 and 0.56, respectively. We have adjusted the external photoevaporation rates in order to get a distribution of lifetimes of the synthetic disks which is in good agreement with the observed distribution (Fedele et al. \cite{fedeleancker2010}).

\subsection{Mass-radius relationship during formation}\label{sect:mrrformation}
Figure \ref{fig:mrform} shows the synthetic mass-radius diagram at four moments in time during the formation phase when planets  grow in mass, i.e. in the first few million years when the protoplanetary disk is present. The color coding indicates the fraction of heavy elements in the planet $Z=\mz/M$. We assume for simplicity that all heavy elements sink to the core, so that $\mz=\mcore$. This might not be the case in reality since planetesimals can get destroyed in protoplanetary atmospheres, cf. Mordasini et al. (\cite{mordasinialibert2005}). This means that our results should rather be associated with the total amount of heavy elements in a planet, and not its actual core mass. {Observations and modeling of the interiors of planets both in the Solar System and of extrasolar planets (cf. Nettelmann et al. \cite{nettelmannfortney2011} for GJ1214b) indicate that the assumption of a few chemically homogenous layers is an idealization. Interior models of Uranus and Neptune that consist of pure rocky, icy and H/He layers seem not to be consistent with the gravity field data (Fortney et al. \cite{fortneybaraffe2010}). The gravitational moments can instead be reproduced with models without hard density jumps (Helled et al. \cite{helledanderson2011}). Such a more complex distribution of the heavy elements would affect the planets in several ways: During the formation phase, the efficiency of gas accretion is higher for strongly enriched envelopes, and the critical mass for runaway is reduced (Hori et al. \cite{horiikoma2011}). During the evolutionary phase, the evolution of the radius of the planet is affected as discussed in  Sect. \ref{sect:errorsgasenve}, and the assumption of an effective convective energy transport may break down due to composition gradients  (e.g. Stevenson \cite{stevenson1985}; Leconte \& Chabrier \cite{lecontechabrier2012}). Finally, also the structure of the (remaining) solid core would differ, since its mass would be lower compared to the sinking approximation we currently use. For these reasons, it is important to take these effects into account in future models.}   

\subsubsection{Basic structure}
The  basic structure of the plots can be understood when considering the evolution of the radius as seen in the Jupiter in situ formation and evolution simulation presented in Paper I. When looking at the top left hand panel showing the $M-R$ relation at 1 Myr, we can identify different regimes:

At low masses ($M\lesssim 30 \mearth$) planets are in the attached phase and sub-critical for gas runaway accretion. Their radius is approximately equal to the Hill sphere radius $\rhill=(M/( 3 \mstar))^{1/3}a$ (or their Bondi radius in hotter parts of the nebula, Paper I). Depending on the semimajor axis of planets, the radius can therefore be very large, between 100 to 1000 $\rearth$. We see that in this regime, the outer radius increases with mass, as expected from the functional form of  $\rhill$. Planets thus move diagonally upwards in the $M-R$ plane as they grow. The substructures which are seen in the $M-R$ diagram at low masses come from different semimajor axes and different migration behaviors.  Planets in this mass regime are dominated by solids and contain less than $\sim20$\% gas (for $M\lesssim10\mearth$). The lower the planet mass, the lower the envelope mass fraction. This is because the  timescale of the envelope accretion increases nonlinearly with decreasing core mass (e.g. Ikoma et al. \cite{ikomanakazawa2000}). 

At high masses ($M\gtrsim 100 \mearth$), planets have in contrast a small radius between roughly 20 and 30 $\rearth$ ($\sim$2-3 $\rj$). This are giant planets which are undergoing gas runaway accretion. They are in the detached phase, and their radius has  collapsed (Paper I). As they grow in mass, they move nearly horizontally towards the right in the $M-R$ plane. In terms of internal bulk composition, these planets contain $\sim50$\% gas  or more.

The transition from the first to the second  group occurs typically at a total mass of about 40 to 60$\mearth$. At this moment the gas accretion rate found by calculating the Kelvin-Helmholtz timescale of the planet becomes so high that the disk can no longer deliver the necessary amount of gas to keep the envelope in contact with the nebula, so that the radius detaches and collapses. These planets have hence  just started runaway gas accretion.  In this stage, planets move nearly vertically downwards in the $M-R$ plane, as the collapse is a rapid process. The colors show that the core and the envelope have at this moment roughly the same mass, again as seen in the Jupiter models. 

\subsubsection{Development for $t>2$ Myrs}
The top right panel shows the situation at 2 Myrs. The global structure is similar as at 1 Myr, but there is one exception. At 1 Myr, all low-mass planets are still in the attached phase, and thus have a relatively large radius. As an example, for planets with a mass less than 10 $\mearth$, the smallest radius at 1 Myr is about 7 $\rearth$.  This is because our synthetic disks all have a lifetime of at least 1 Myr. At 2 Myrs, in contrast, some gaseous disk have disappeared. As we have seen in Sect. \ref{sect:radevorogers}, this triggers for low-mass planets the rapid contraction of the envelope. Therefore, at 2 Myrs, there are now some low-mass planets which are already in the evolutionary phase and have a radius of 2 to 3 $\rearth$.  

At 3 Myrs, this group has grown in number, as more disks have disappeared. The global structure is otherwise similar. One also notes that some extremely massive planets have formed ($M\gtrsim 10\,000 \mearth\approx31 \mj$) which did not yet exist at 1 Myr. This is because the formation of very massive planets is only possible in disks which have a long lifetime (Mordasini et al. \cite{mordasinialibert2012}). It is clear that such massive planets can only form in the simulation because we consider the limiting case  that gap formation does not reduce the planetary gas accretion rate. Such a situation arises if the eccentric instability (Papaloizou et al. \cite{papaloizounelson2001}; Kley \& Dirksen \cite{kleydirksen2006}) allows the planets to efficiently access disk material even after a gap has formed. For circular orbits, gap formation would in contrast strongly reduce the gas accretion rate (Lubow et al. \cite{lubowseibert1999}, Bryden et al. \cite{brydenchen1999}), and limit planetary masses to $\lesssim 10 \mj$. 

At 8 Myrs, nearly all low-mass planets with a large radius (i.e. which are in the attached phase) have disappeared. This is simply because almost all protoplanetary gas disks have now disappeared.

\subsection{Mass-radius relationship during evolution}\label{sect:mrduringevo}
\begin{figure*}
\begin{center}
\begin{minipage}{0.5\textwidth}
	      \centering
        \includegraphics[width=0.9\textwidth]{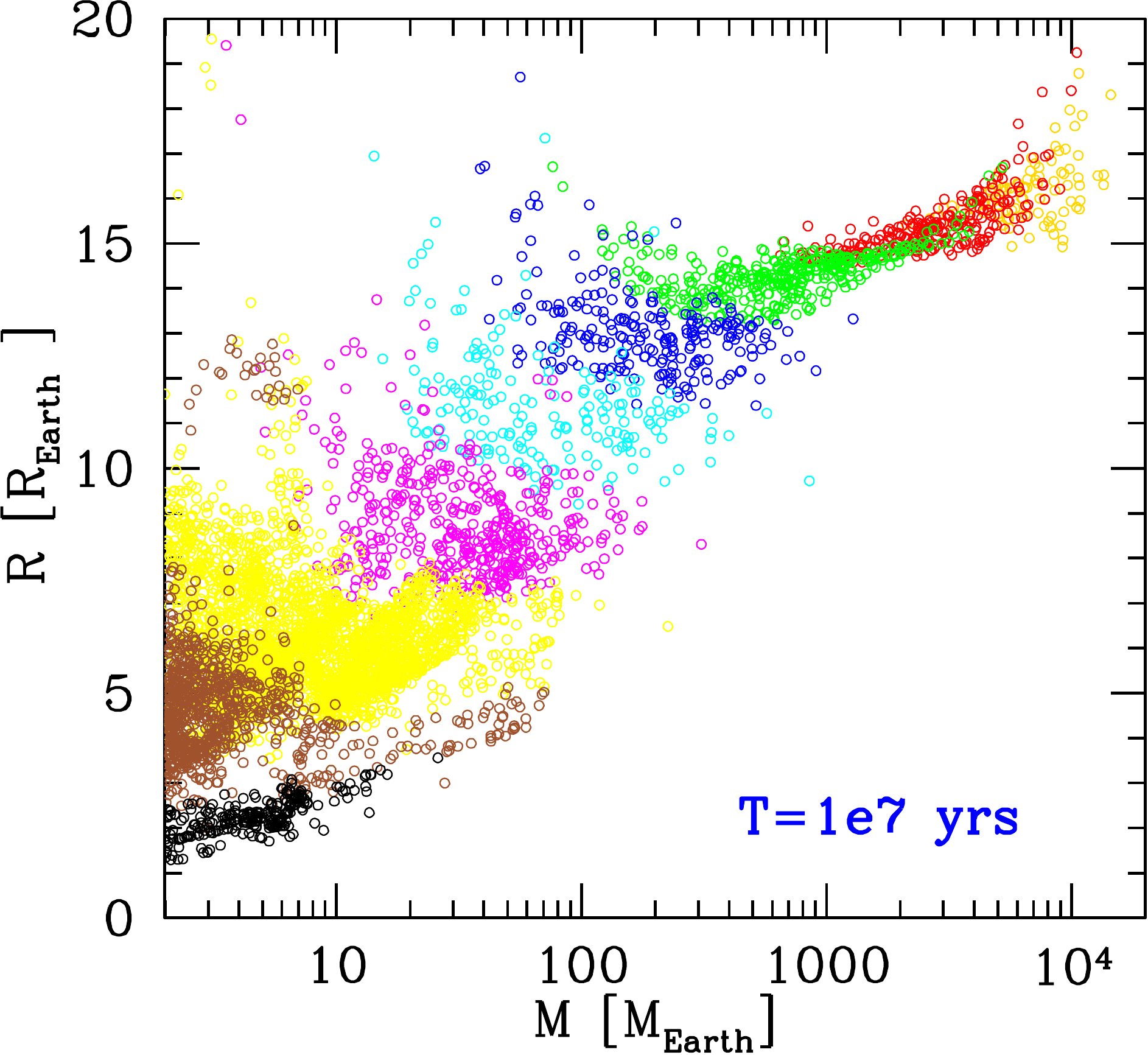}
        \includegraphics[width=0.9\textwidth]{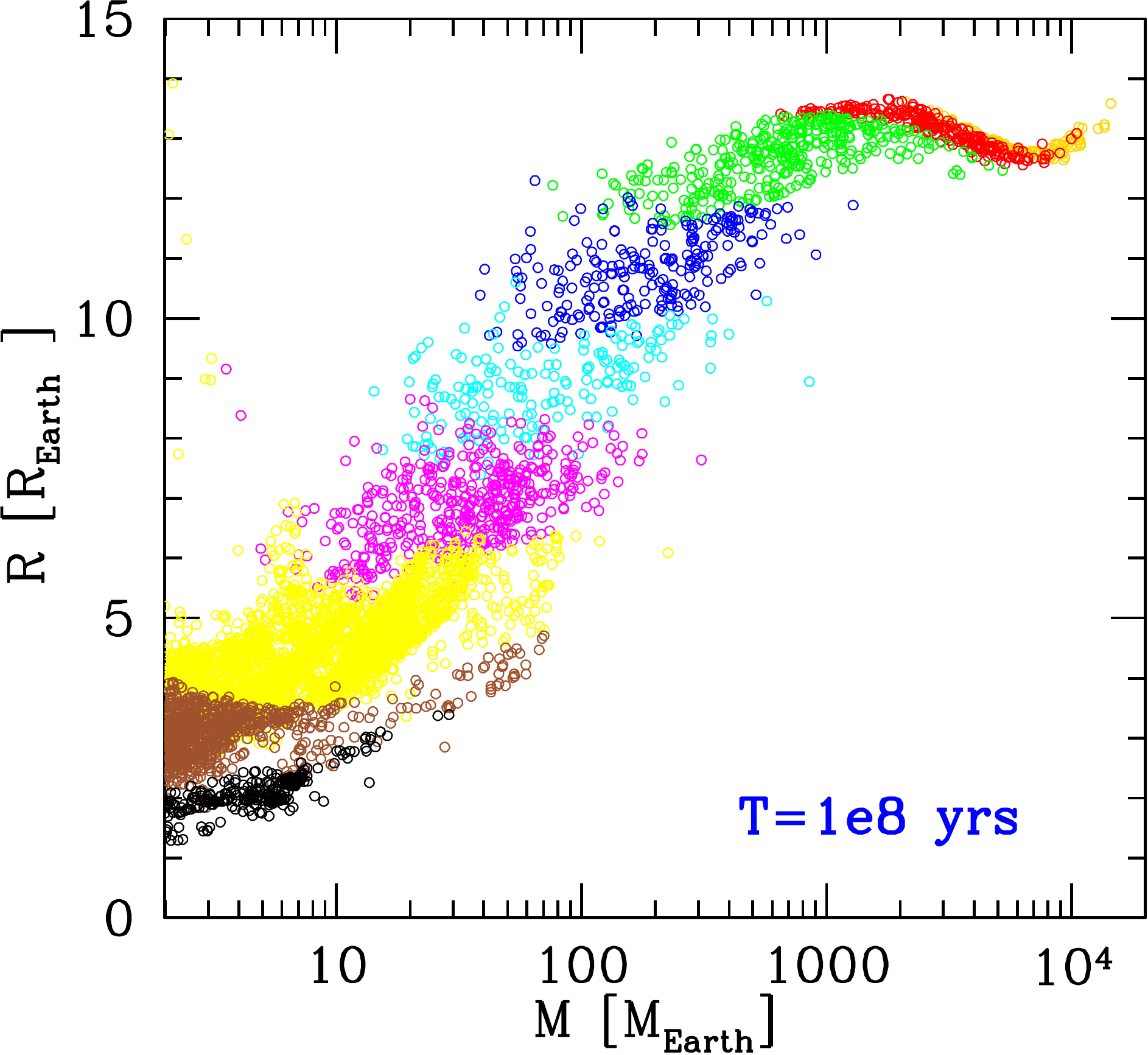}
        \includegraphics[width=0.9\textwidth]{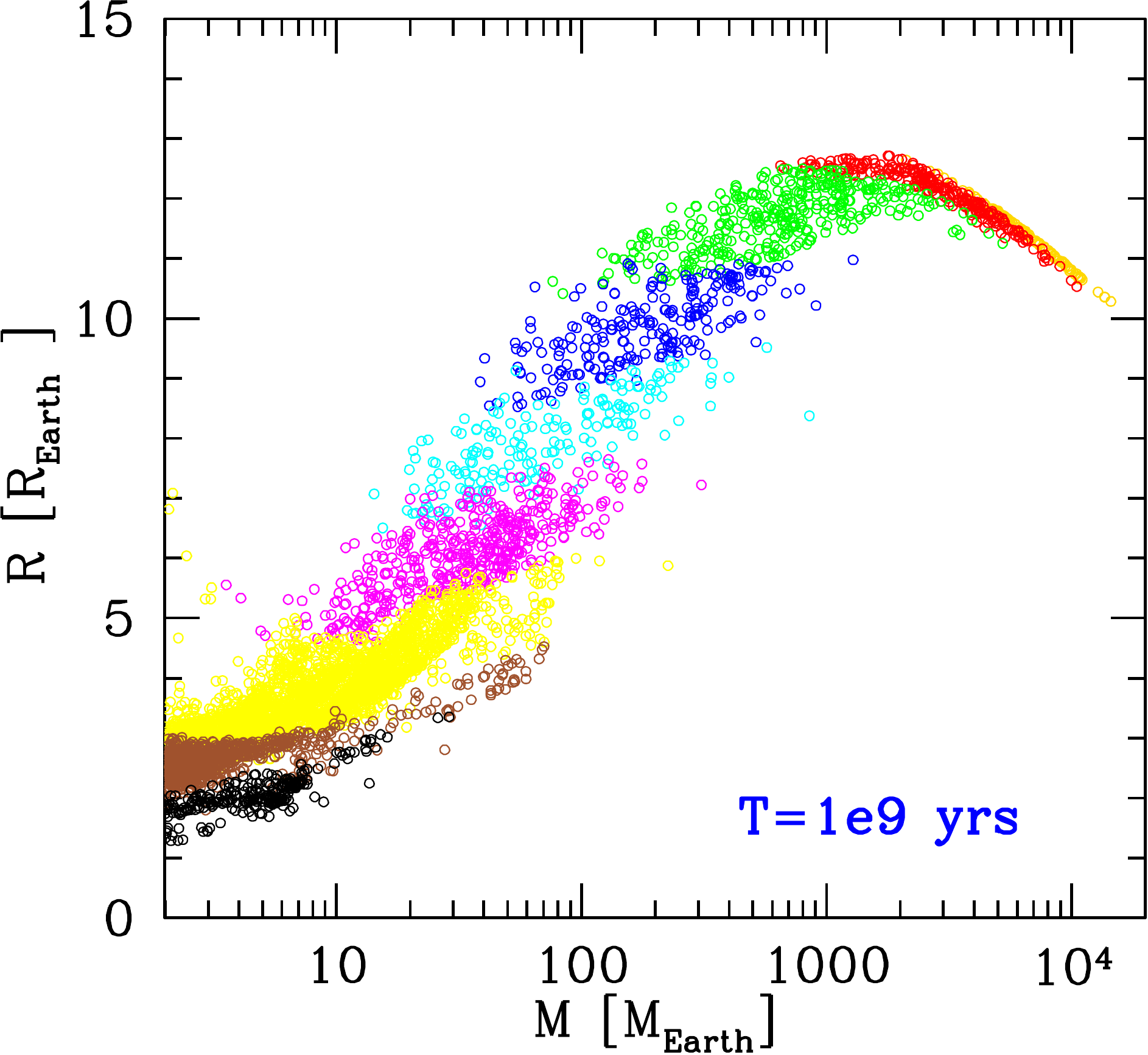}
     \end{minipage}\hfill
     \begin{minipage}{0.5\textwidth}
      \centering
       \includegraphics[width=0.9\textwidth]{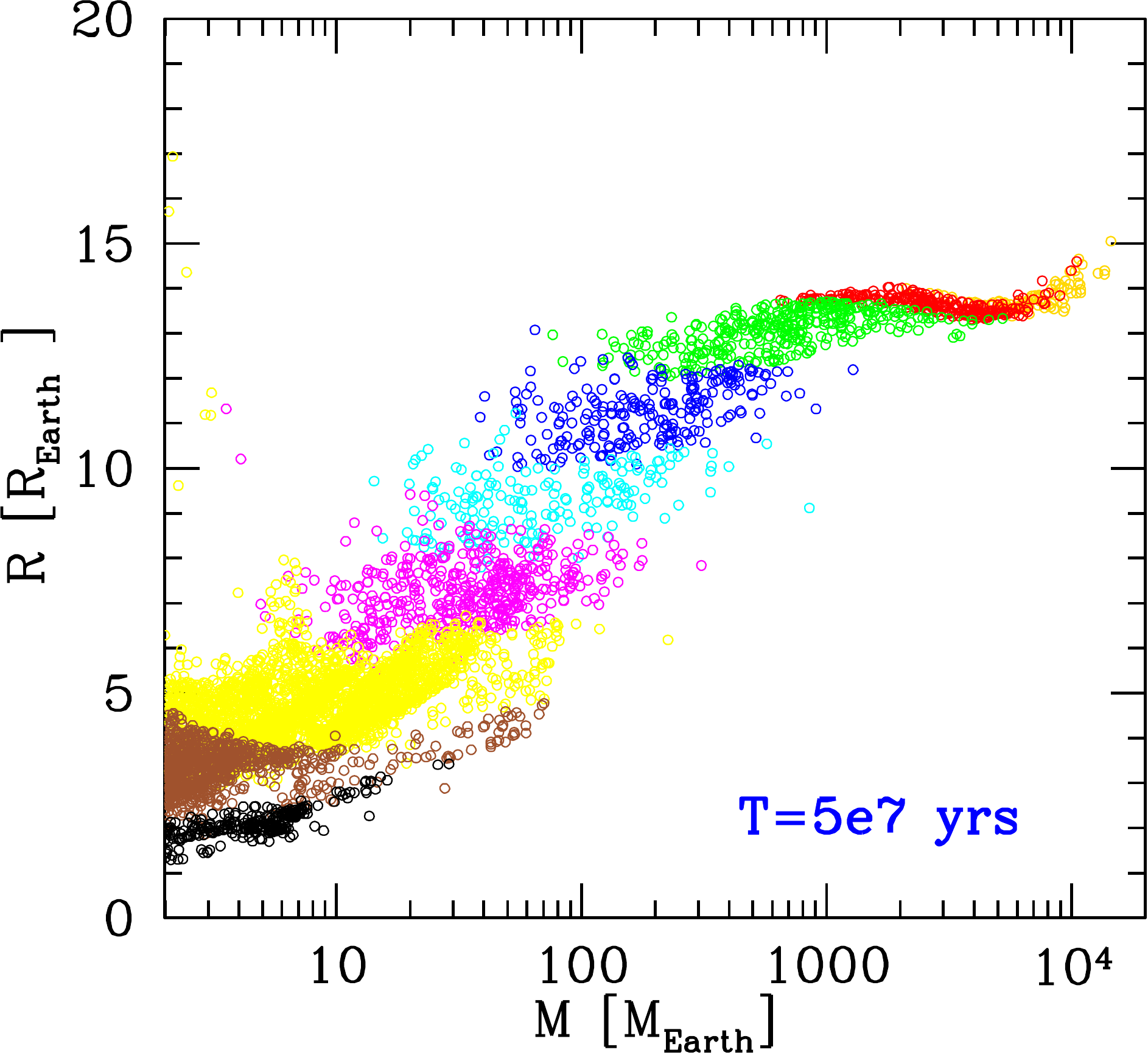}
       \includegraphics[width=0.9\textwidth]{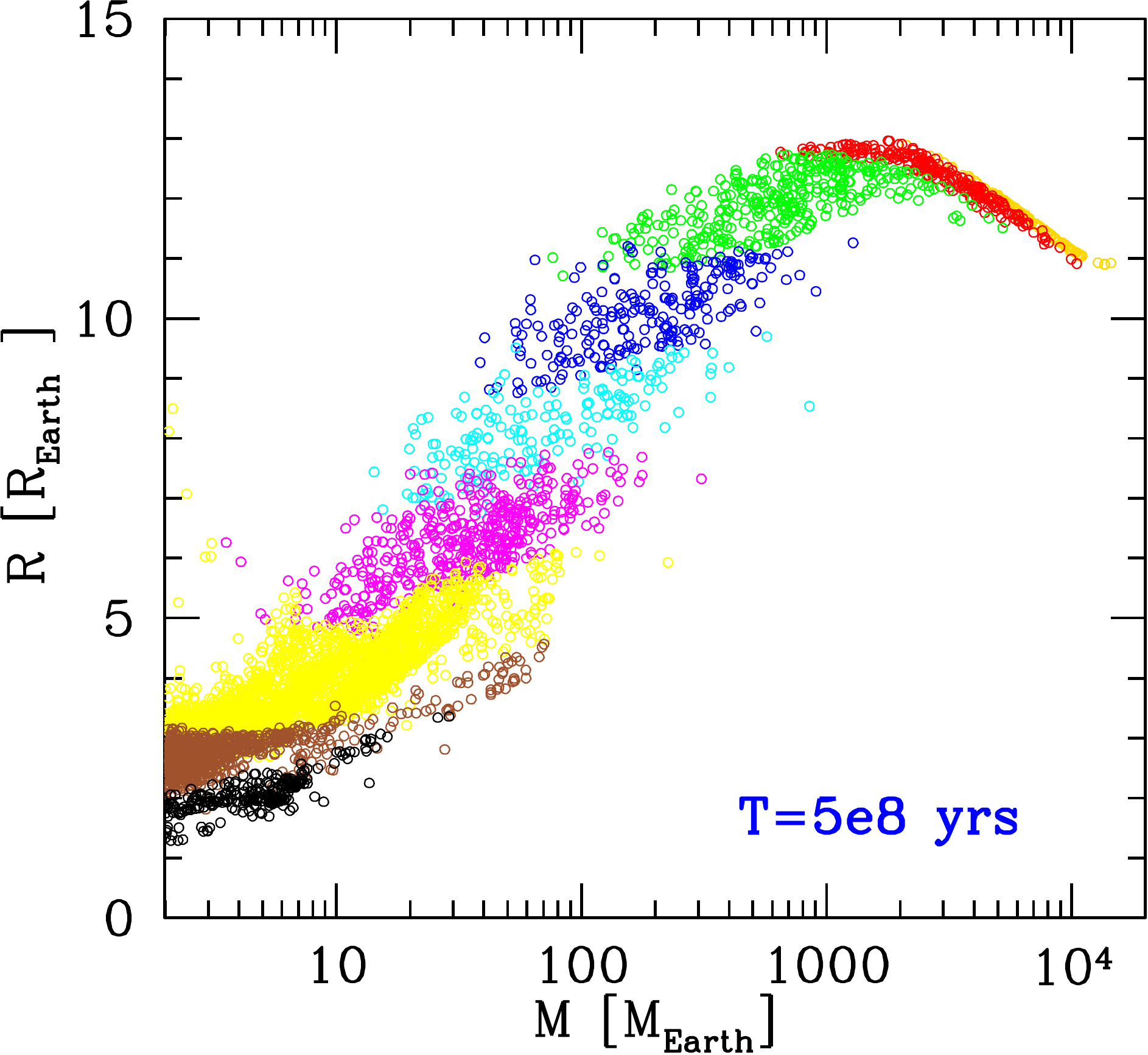}
       \includegraphics[width=0.9\textwidth]{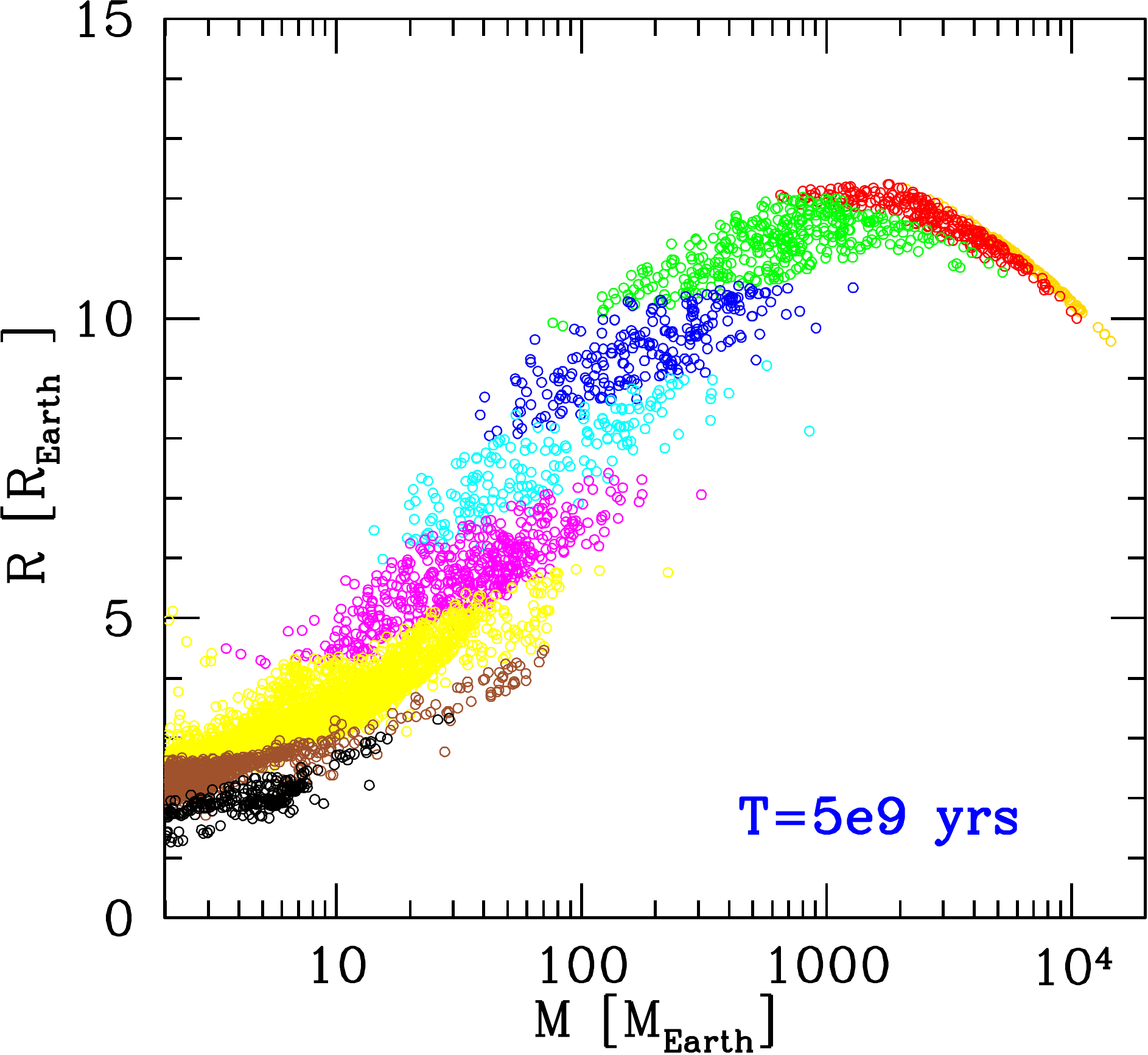}

     \end{minipage}
\caption{Temporal evolution of the planetary mass-radius diagram after the final masses have been reached.  The colors show the fraction of heavy elements $Z=\mz/M$ in a planet. Orange: $Z\leq1\%$. Red: $1<Z\leq5\%$. Green: $5<Z\leq20$\%. Blue: $20<Z\leq40$\%. Cyan: $40<Z\leq60$\%. Magenta: $60<Z\leq80$\%.  Yellow: $80<Z\leq95$\%.  Brown: $95<Z\leq99$\%. Black: $Z>99$\%. }\label{fig:mrevo}
\end{center}
\end{figure*}

Figure \ref{fig:mrevo} shows the $M-R$ diagram during the evolution of the planets at constant mass. Since we do not include atmospheric mass loss, outgassing, deuterium burning or any special inflation mechanisms (see e.g. Batygin et al. \cite{batyginstevenson2011} and reference therein), all planets  only move vertically downwards in the $M-R$ plane.

\subsubsection{Evolution at early times} 
At an age of 10 Myrs (top left panel), some planets still have very large radii.  The mass-radius relationship is at this moment still considerably different from the shape  at later times.  For massive giant planets we see that for one mass, there is a significant spread in associated radii. This is a visible imprint of different formation histories, namely of the initial entropy of these planets. The initial entropy for giant planets depends on the way the gas and the solids have been accreted during the formation phase. High gas accretion rates in the runway phase for example make the ``cold start'' scenario become more similar to the ``hot start'' scenario {(see Spiegel \& Burrows \cite{spiegelburrows2012})}. This is because at a high gas accretion rate, a planet accretes  a lot of matter when its radius is still relatively big. This means that the accretion shock is weak, leading to a relatively high entropy of the accreted gas. This in turn leads to a large radius. The entropy of young planets will be studied in a dedicated work on the planetary luminosities (Mordasini et al. in prep). 

At an age of  50 Myrs, we start to recognize the characteristic shape of the planetary mass-radius relation. It first consists of  small radii for low-mass planets, followed by an increase of the radius with mass up to a local maximum of the radius at about 4 Jovian masses, and finally a decrease of the radius with mass in the Super Jupiter domain. The latter is a consequence of the well known fact that the degree of degeneracy in the interior  increases (e.g. Chabrier et al. \cite{chabrierbaraffe2009}). This gives the planetary mass-radius relationship an overall shape reminiscent of an elongated ``S''.

We currently do not include  deuterium burning. This means that the radii of objects more massive than 13 $\mj$ at times when D burning occurs ($t\lesssim$ 0.1 - 1  Gyr) will be too small. 

\subsubsection{Evolution at late times} 
The relative shape of the $M-R$ relation hardly changes between an age of 500 Myrs and 5 Gyrs. The planets slowly contract on Gyr timescales, and thus move vertically downwards, keeping their relative position. The vertical movement  can be seen by looking e.g. at the local radius maximum in the giant planet domain. 

{At these late times, the spread in radius at a given mass no longer comes directly from their different accretion histories, but instead from their different heavy element contents.} The fact that for a given mass, a planet with a higher heavy element content has a smaller radius gives rise to the diagonal pattern seen in the color coding, as illustrated by the following example for t=5 Gyrs. When we move at a fixed radius of, say, 7 $\rearth$ from left to right across the envelope of points, we first encounter planets with a mass of about 20 $\mearth$ and a heavy element content of 40\% to 60\% (cyan). When moving further right to higher masses, in order to keep the same radius of  7 $\rearth$, $Z$ must increase. Therefore, at a mass of about 100 $\mearth$, the color  changes from cyan to magenta, indicating that the planets now have a heavy element fraction of 60 to 80\%. In other words, when we  fix the heavy element content, and increase the mass, then the radius increases also, leading to diagonal bands of identical color. Obviously, this is only true in the domain where the radius is an increasing function of mass ($M\lesssim 4 \mj$).

Also the distance from the star plays a certain role for the radius, with objects closer to the star having a larger radius (e.g. Fortney et al. \cite{fortneymarley2007}). As the minimal allowed radius is 0.1 AU, the effect is, at least for giant planets, not so large, and of order 0.5 to 1 $\rearth$. This is different for the few close-in, low-mass ($M<10 \mearth$), gas-rich  (up to 20\% gas) planets that are visible close to the left-hand border of the plot. Such planets have, as expected (Fig. \ref{fig:mrrogerscomp}; Rogers et al. \cite{rogersbodenheimer2011}) large radii (up to 5 or 10 $\rearth$, depending on age) which are increasing with decreasing mass. Note that the radius of these planets is affected by the simple gray atmosphere model as discussed in Sect. \ref{sect:errorsgasenve}. It is also clear that such objects are sensible to envelope evaporation, especially at early times just after formation when they have very large radii, sometimes exceeding 15 $\rearth$. The effects of envelope loss will be included in future work.

The calculation of the radius for a given age, composition and orbital distance can be done in a purely evolutionary model, too. But only a combined formation and evolution model in a population synthesis calculation gives a prediction for the spread of the radii for a given mass domain, and the distribution of the radii. When looking at the vertical width in radius that the envelope of points covers, we see that depending on mass, there are regimes where all planets nearly have the same radius (very massive giant planets). In other mass domains, there is a very significant spread in possible radii. In Sect. \ref{sect:rdistmassranges} below we show histograms of the radius distribution for different mass domains.  For example, at $t=5$ Gyrs and a mass of about 40 $\mearth$,  we find planets with radii between about 3 and 9 $\rearth$  i.e. a variation of a factor 3. The reason is that in this mass domain, both clearly solid dominated planets ($95<Z\leq99$\%) as well as small gaseous planets ($20<Z\leq40$\%) exist. Not all types are however found at all semimajor axes (see Sect. \ref{sect:impacta}). 

\subsubsection{General shape of the mass-radius relation}\label{sect:generalshapemrr}
Concerning the overall shape of the planetary mass-radius relation, we see that low-mass planets have small radii, as they consist mostly of solids, while high-mass planets have large radii as they consist mostly of gas. This is clearly visible  from the color coding showing the decrease of the heavy element fraction $Z=\mz/M$ with mass. This is a direct consequence of the fact that during formation, low-mass cores cannot accrete large amounts of gas as their Kelvin-Helmholtz timescale is too long. Such planets never undergo gas runaway accretion. In a symmetrically way must massive, super-critical cores always accrete significant amounts of gas (at least if they form during the lifetime of the  gaseous disk) as they necessarily must trigger runaway gas accretion. Neither a (nearly) core-free e.g. 10$\mearth$ planet (which would have a radius of $\sim1\rj$ at 1 AU, Fortney et al. \cite{fortneymarley2007}), nor a pure $3\mj$ ice planet (which would have a radius $R\sim0.4\rj$,  Sect. \ref{sect:rcoremass}) can form in the core accretion scenario. This leads, combined with the fundamental properties of matter as expressed in the EOSs, to the two ``forbidden'' regions in the planetary $M-R$ plane in the top left and bottom right, resulting in a shape of the mass-radius relationship similar to an elongated ``S''. We see that the synthetic $M-R$ relationship contains the imprint of the basic concepts of the core accretion formation theory. 

\subsection{Comparison with the observed mass-radius relation}\label{sect:compobsmrr}
\begin{figure*}
\begin{center}
\includegraphics[width=1.9\columnwidth]{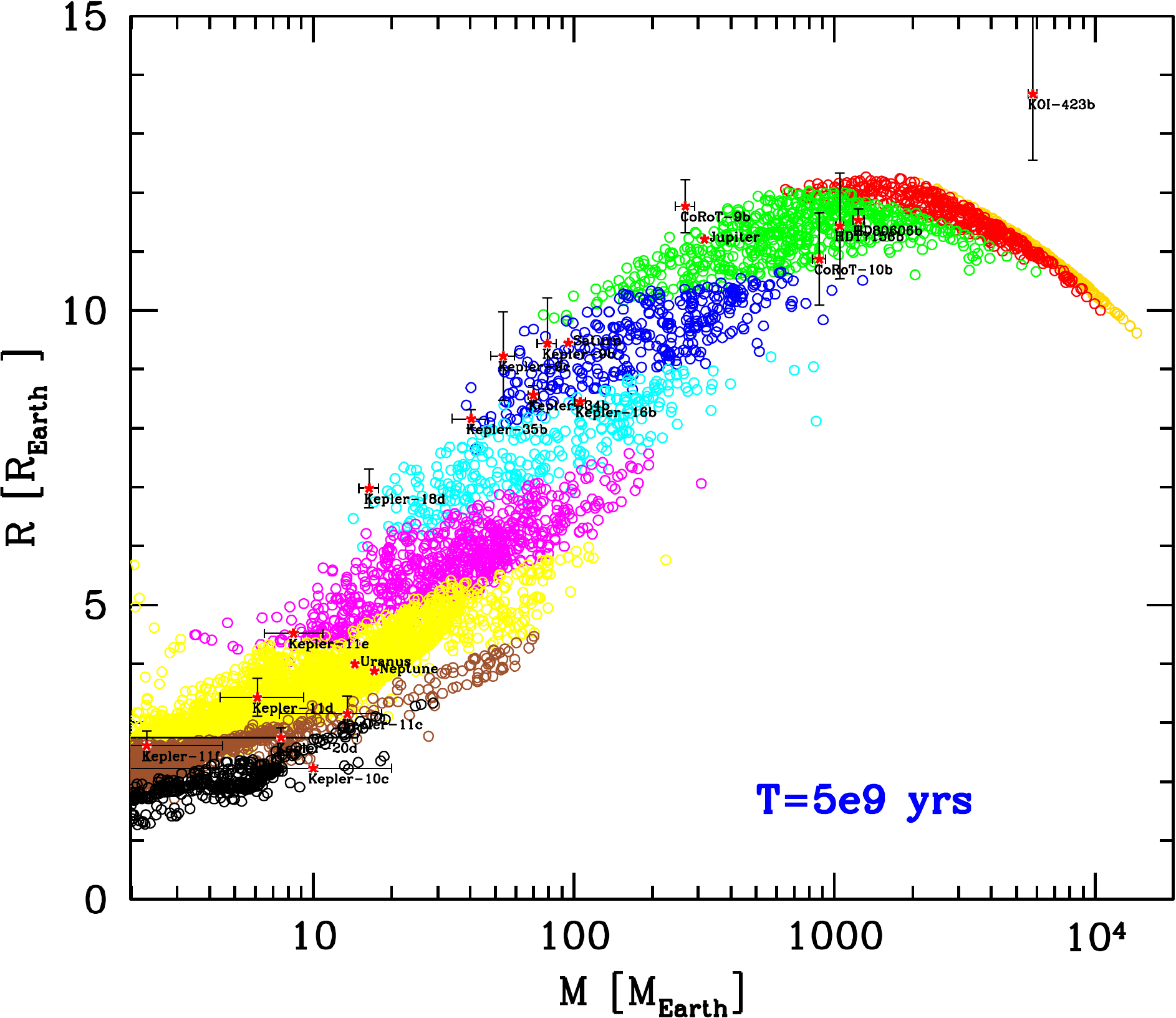}
\caption{Comparison of the synthetic and the observed mass-radius relationship. The synthetic population, and all actual planets with known mass and radius (both in the Solar System and around other stars)  and a semimajor axis of at least 0.1 AU are shown. For extrasolar planets, error bars are given. The color code for the internal composition of the synthetic planets is the same as in Fig. \ref{fig:mrevo}. The synthetic population only contains planets with primordial H$_{2}$/He atmospheres, thus at low masses, discoveries of planets lying below the synthetic population are expected. The two ``forbidden'' regions in the top left and bottom right are a consequence of the core accretion formation process (and the EOS). This leads to a characteristic, elongated ``S'' shape of the planetary mass-radius relationship.}\label{fig:mrobs}
\end{center}
\end{figure*}
Figure \ref{fig:mrobs} shows again the synthetic population at 5 Gyrs, but now together with all actual planets that have a semimajor axis larger than 0.1 AU (as in the model), and a well constrained mass and radius. We additionally plot \object{Kepler-20d} (Gautier et al. \cite{gautiercharbonneau2012}) and \object{Kepler-10c} ({Fressin et al. \cite{fressintorres2011}), for which only upper mass limits are known, as it is unfortunately the case for many other interesting \textit{Kepler} planets like those around \object{Kepler-30}  or \object{Kepler-11g} (Fabrycky et al. \cite{fabryckyford2012}, Lissauer et al. \cite{lissauerfabrycky2011}). The observational data was taken from exoplanets.org (Wright et al. \cite{wrightfakhouri2011}) on 23.3.2012. The host stars of the planets  are roughly solar-like with masses between about 0.7 and 1.3 $\msun$. In the synthesis, the stellar mass is fixed to 1 $\msun$. For simplicity, we also include the circumbinary planets \object{Kepler-34b}, \object{Kepler-35b} and \object{Kepler-16b} (Welsh et al. \cite{welshorosz2012}, Doyle et al. \cite{doylecarter2011}). 

It is clear that the age of the observed planets is not exactly 5 Gyrs, therefore the comparison can only be approximate. For a more exact comparison, each planet should be compared individually with the synthetic population at its specific age.  A comparison of the general structure of the $M-R$ diagram can however still be made, for two reasons. First, all host stars shown here are thought to be old main sequence stars, causing typical uncertainties in the age of an individual star of several Gigayears, often allowing for an actual age of 5 Gyrs. Second, the evolution of the radius at such late time is very slow.  The radius of a 1 $\mj$ planet decreases for example between an age of 1 and 5 Gyrs only by about 0.4 $\rearth$ (Paper I). 

Looking at the envelope of the actual and synthetic planets, we note that the observed planets all lie in the envelope covered by the synthetic population if we take into account the error bars, for the clear exception of KOI-423b (Bouchy et al. \cite{bouchybonomo2011}). Except for this planet, there is thus a good agreement between the model and the observation. 

For  planets with a mass similar to Jupiter or higher, we see that the observed planets seem to be distributed over the entire vertical width of the envelope of the synthetic planets. For planets with a mass similar to Saturn or less, one notes that the observed planets cluster at the upper boundary of the envelope covered by the synthetic planets. As discussed in Sect. \ref{sect:impacta} below, an explanation for this could be an imprint of the semimajor axis. It is recalled that  synthetic planets at all semimajor axes are shown  here (i.e. between 0.1 AU and  $\sim20$ AU), while the observed exoplanets are close to the lower limit of this range. For planets at even lower masses, in the Neptunian mass region, observed planets of a similar mass can have very different radii, as illustrated by \object{Kepler-18d} (Cochran et al. \cite{cochranfabrycky2011}) which has a mass similar to Neptune, but which is about 1.8 times as large in radius. Clearly,  this reflects that the spread in possible heavy element contents is large. Note that we assume that the atmospheric composition is always solar. It is clear that different chemical compositions would lead to different opacities, which in turn lead to different cooling histories and radii (cf. Burrows et al. \cite{burrowsheng2011} for the brown dwarf regime). This would further increase the vertical width of the envelope. 

\subsubsection{Individual objects}\label{sect:individobj}
{The internal composition of the actual extrasolar and Solar System planets shown in Fig. \ref{fig:mrobs} has been analyzed in numerous works with more detailed physics than used here. It is therefore important to understand to what extent the results of these studies can be reproduced with our simpler model.}

In Table \ref{tab:compoplanets} we list the inferred fraction of heavy elements $Z$ in the individual planets shown in  Fig. \ref{fig:mrobs}, assuming again an age of 5 Gyrs. To derive these fractions, only the position in the $M-R$ plane was used. Additional constraints like the semimajor axis or the metallicity of the host star were not considered here.  The third column of the table shows the composition of the synthetic planet which lies closest to the observed values in the $M-R$ plane. A $\sim$ symbol means that either no reasonably close synthetic planet was found in $M-R$ plane, or that for the planet, observationally, only upper limits of the mass are known.  The fourth column shows possible $Z$ found in the rectangle in the $M-R$ plane covered by the error bars of the observations.  Values for Solar System objects are assumed to be exact. 

\begin{table}
\caption{Derived fraction of heavy elements $Z=\mz/M$ in planets with $a>0.1$ AU. The value for the synthetic planet lying closest in the $M-R$ plane  (closest $Z$) and the domain covered by the error bars ($Z$ range) are shown. An age of 5 Gyrs is assumed.  The planets are approximatively listed by increasing mass as indicated in the second column.}\label{tab:compoplanets}
\begin{center}
\begin{tabular}{l c c c }
\hline \hline
Name         &  $M$ [$\mearth$] & closest $Z$ 	& $Z$ range  \\ \hline
Kepler-11f	& 2$^{+2.2}_{-1.2}$	&$\sim$0.95		&	0.90-0.97 \\
Kepler-11d	& 6$^{+3.2}_{-1.7}$	&0.88		&	0.84-0.94 \\
Kepler-20d	& 8$\pm$7.2	&$\sim$0.96		&	$>$0.89	  \\
Kepler-11e	&	8$^{+2.5}_{-1.9}$ &$\sim$0.78 		&	0.76-0.79	  \\
Kepler-10c	&	10$\pm$10&$\sim$0.99		&	$>$0.95		  \\
Kepler-11c	&	13$^{+4.8}_{-6.1}$ &0.98		&	$>$0.85	  \\
Uranus		& 14.5	&0.88		&	-	  \\
Kepler-18d	& 16.4$\pm$1.4	&$\sim$0.50		&	0.43-0.52 \\
Neptune	& 17.1	&0.90		&	- 	  \\
Kepler-35b	&	40$\pm$6.4 &$\sim$0.37		&	0.28-0.37 \\
Kepler-9c	&54$\pm$5.5	&0.25		&	0.24-0.30 \\
Kepler-34b	& 70$^{+3.5}_{-3.2}$	&0.35		&	0.32-0.39 \\
Kepler-9b	&	79$\pm$6.7 &0.30		&	0.16-0.36 \\
Saturn		& 95.2	&0.27		&	-	  \\
Kepler-16b	& 106$\pm$5.1 	&$\sim$0.41		&	0.40-0.44 \\
CoRoT-9b	& 270$\pm$23 	&$\sim$0.11			&	0.09-0.11 \\
Jupiter       &  317.8  &0.10 & - \\    
CoRoT-10b	& 880$\pm$51 	&0.17		&	0.08-0.18 \\
HD17156b	& 1050$\pm$36	&$\sim$0.08		&	0.04-0.11	  \\
HD80606b	&	1240$\pm$60  &0.09		&	0.07-0.10 \\
KOI-423b	& 5800$\pm$228	&	-		&	($<$0.05)   \\ \hline
\end{tabular}
\end{center}
\end{table}%

The general trend that the heavy element fraction $Z=\mz/M$ decreases with increasing mass is obvious. It is worth looking at some extrasolar planets individually. Starting at the smallest masses ($\lesssim10\mearth$), we see that the low-mass planets around \object{Kepler-10} ({Fressin et al. \cite{fressintorres2011}), \object{Kepler-11} (Lissauer et al. \cite{lissauerfabrycky2011}) and \object{Kepler-20} (Gautier et al. \cite{gautiercharbonneau2012}) can be reproduced with envelopes masses of  $\sim$1 to $\sim$20 \% of the total mass of the planets, similar as found by Lissauer et al. (\cite{lissauerfabrycky2011}).

It is expected that at such low masses, and $a\geq0.1$ AU, planets will be found in future with radii smaller than in the synthetic population. Such objects have already been found at very small orbital distances like for example \object{Kepler-10b}, with $M=4.6\mearth$, $R=1.4\rearth$ and a  semimajor axis of 0.017 AU (Batalha et al. \cite{batalhaborucki2011}). This is much closer than the domain we currently model in our simulations. Objects with an Earth-like composition without H$_{2}$/He are however also expected to exist at larger semimajor axis, as demonstrated by the terrestrial planets in the Solar System. From the comparison of such objects with the domain covered in the synthesis, it will become possible to constrain the transition from terrestrial and water planets to mini-Neptunes (with primordial H$_{2}$/He envelopes). This helps to understand various processes of primordial envelope accretion and loss, which is difficult from the Solar System alone with its large  gap in mass between the Earth and the ice giants. 

\object{Kepler-18d} (Cochran et al. \cite{cochranfabrycky2011}) is a remarkable planet  with a mass ($16.4\pm1.4 \mearth$) approximately between the one of Neptune and Uranus, but a radius that is 80\% larger than Neptune.  It thus seems to have a clearly lower heavy element fraction than the two Solar System planets. As can be seen in Fig. \ref{fig:mrobs}, this planet lies in the $M-R$ diagram at the upper border of the $M-R$ region covered with synthetic planets. We derive a heavy element fraction of 0.43 to 0.52 ($\mz$= 7.1 to 8.5 $\mearth$), similar as in Cochran et al. (\cite{cochranfabrycky2011}) who find a $\mz=10.4\pm1.4\mearth$.  Such a composition of about half gas, and half solids does not exist in the Solar System. As described in Mordasini et al. (\cite{mordasiniklahr2011}), we find that planets of such a composition can only form if the grain opacity in the envelope during the formation phase is much smaller than the interstellar grain opacity.  This object is thus important to constraint uncertain parameters of the core accretion model. A complication arises from the fact that at a semimajor axis of about 0.12 AU, this planet might still get affected by the irradiation of the host star in a more complex way than modeled here. Then it could have a relative composition closer to the Solar System ice giants.

The system of the two Saturn-mass planets \object{Kepler-9b} and \object{Kepler-9c} (Holman et al. \cite{holmanfabrycky2010}) has been analyzed in details in Havel et al. (\cite{havelguillot2011}). Comparing our simpler analysis with their more sophisticated calculations still indicates a relatively good agreement: Havel et al. (\cite{havelguillot2011}) derive for Kepler-9b and an age of 5 Gyrs as in our calculation a possible range of  $Z$ between 0.19 and 0.56, with a nominal value of 0.37. For Kepler-9c, also at 5 Gyr, they derive a $Z$ between 0.17 to 0.58, with a nominal value of 0.37. Our simulations indicate a heavy element fraction for Kepler-9b of 0.24 to 0.30 with a nominal value of 0.25,  while for Kepler-9c values of 0.16 to 0.36 with a nominal value of 0.30  are found. The allowed ranges thus agree in the two analyses. The smaller nominal values in our analysis are not surprising, as the effect of additional heating in the envelope of the planet due to stellar irradiation  is not taken into account in our simulations. Therefore, lower core masses are sufficient to explain the observed radius. In Havel et al. (\cite{havelguillot2011}), the effect of irradiation leads to an uncertainty of about 30\% for the value of $Z$.   The error bars in their analysis are larger (and more realistic) because  more sources of uncertainties like atmospheric opacities are taken into account. 

For Jupiter, we find that the closest synthetic planet contains about 31 $\mearth$ of heavy elements, similar as in Paper I. This is in good agreement with the allowed range for the total amount of heavy elements found by Guillot (\cite{guillot1999}) of 11 to 41 $\mearth$. For Saturn, the derived heavy element mass is 25.6 $\mearth$, also in good agreement with Guillot (\cite{guillot1999}) who finds masses between 19 and 31 $\mearth$. For the ice giants, we find that they contain about 10 to 12\% of gas, similar as e.g. Hubbard et al. (\cite{hubbardnellis1991}).

\object{CoRoT-10b} (Bonomo et al. \cite{bonomosanterne2010}) has a mass of about 880$\pm$51 $\mearth$ and is interesting because it seems to contain a high fraction of heavy elements despite its high mass, as already noted in Bonomo et al. (\cite{bonomosanterne2010}). These authors infer that 120 to 240 $\mearth$ of heavy elements should be contained in the planet. We find that this planet should contain about 150 $\mearth$ of metals, assuming that the star is 5 Gyrs old, compatible with their result.  In the $M-R$ plane, the planet lies towards the lower boundary of the domain covered with synthetic planets, but still well within it. Planets with such high solid masses can form by core accretion also without collision in metal-rich disks (high [Fe/H] and high gas mass) at large semimajor axes of $\sim10$ AU. An example of the formation of such a planet is shown in Mordasini et al. (\cite{mordasinialibert2009a}).  Compared to the formation of Jupiter, the isolation mass is much larger under such circumstances, allowing for the accretion of many more planetesimals so that the plateau phase II (Pollack et al. \cite{pollackhubickyj1996}) is skipped. It should however be noted that such a formation is probably only feasible if the random velocities of the planetesimals remain low while the planet grows (``monarchical'' growth, Weidenschilling \cite{weidenschilling2005}).  

The most massive object in the comparison, \object{KOI-423b} (Bouchy et al. \cite{bouchybonomo2011}) with a mass of approximately 5800$\mearth$ ($\approx 18 \mj$), cannot be explained with our model which assumes standard cooling sequences with solar composition opacities. The same conclusion was reached by Bouchy et al. (\cite{bouchybonomo2011}).  These authors show that large, ad-hoc increases in the atmospheric opacities are necessary to reconcile the measured mass and radius with cooling models given the high age of the host star, estimated to be between 3.6 and 6.6 Gyrs. With the assumptions of our model (solar opacities, no bloating mechanisms), we can only reproduce the large radius if we assume an age of the planet of less than 100 to 500 Myrs. This object is  currently difficult to understand from a theoretical point of view, as stated already by Bouchy et al. (\cite{bouchybonomo2011}).

{In summary we see that our simple model leads to estimated heavy element contents which agree relatively well with the results from more detailed models. However, an important factor for this agreement is unfortunately that despite great efforts, also the latter studies can deduce the internal composition, including the one of the Solar System planets, only with significant error bars.  This is due e.g. to the uncertainties in the equations of state,  the distribution of heavy elements (mixed vs. layered structure), the efficiency of convection, and the currently limited amount of observational data.  The reader is referred to Fortney \& Nettelmann (\cite{fortneynettelmann2010}) or  Baraffe et al. (\cite{baraffechabrier2010}) for recent reviews. }

\subsubsection{Constraints from a multi-dimensional parameter space}
{Due to the still  limited number of extrasolar planets with an accurately known mass and radius at large orbital distances}, the analysis shown here  represents only the very beginning of the possible constraints that can be derived in future from the planetary mass-radius relationship. 

Once large numbers of planets with accurately measured mass, radius, semimajor axis, host star mass and metallicity will be known, it will be possible {to compare observation and theory with Kolmogorov-Smirnov tests, and} to search for correlations in this multi-dimensional parameter space. An already known example is the correlation between stellar [Fe/H] and planetary heavy element content (Guillot et al. \cite{guillot2006}; Burrows et al. \cite{burrowshubeny2007}; Miller \& Fortney \cite{millerfortney2011}). This correlation can be reproduced with planet population synthesis calculations (Mordasini et al. \cite{mordasinialibert2009b,mordasiniohp2011}).  

Below, in Sect. \ref{sect:impacta} we will show a correlation between semimajor axis and radius found among synthetic planets. It is particularly important to have many well characterized planets with semimajor axes clearly larger than 0.1 AU. This should reduce the impact of  several   only poorly understood mechanism influencing close-in planets during their formation (e.g. the exact disk structure close to the inner rim) and evolution (strong irradiation), and thus allow to see the correlations more clearly.

\subsection{Bimodal planetary radius distribution}\label{sect:planetaryrdist}
A prediction of the planetary radius distribution is one of the most important results of a  population synthesis calculation, and similar in its importance as a prediction of the planetary mass function (Ida \& Lin \cite{idalin2008}, Mordasini et al. \cite{mordasinialibert2009b}). 
Figure \ref{fig:distall} shows the distribution of the radii of all synthetic planets with a radius larger than 2 $\rearth$ at an age of 5 Gyrs. Planets at all semimajor axes are included. The distribution is normalized i.e. the sum of the fractions in all bins is unity. Table \ref{tab:rdist} lists the corresponding fractions  at ages of 1, 5 and 10 Gyrs. 

\begin{figure}
\begin{center}
\includegraphics[width=\columnwidth]{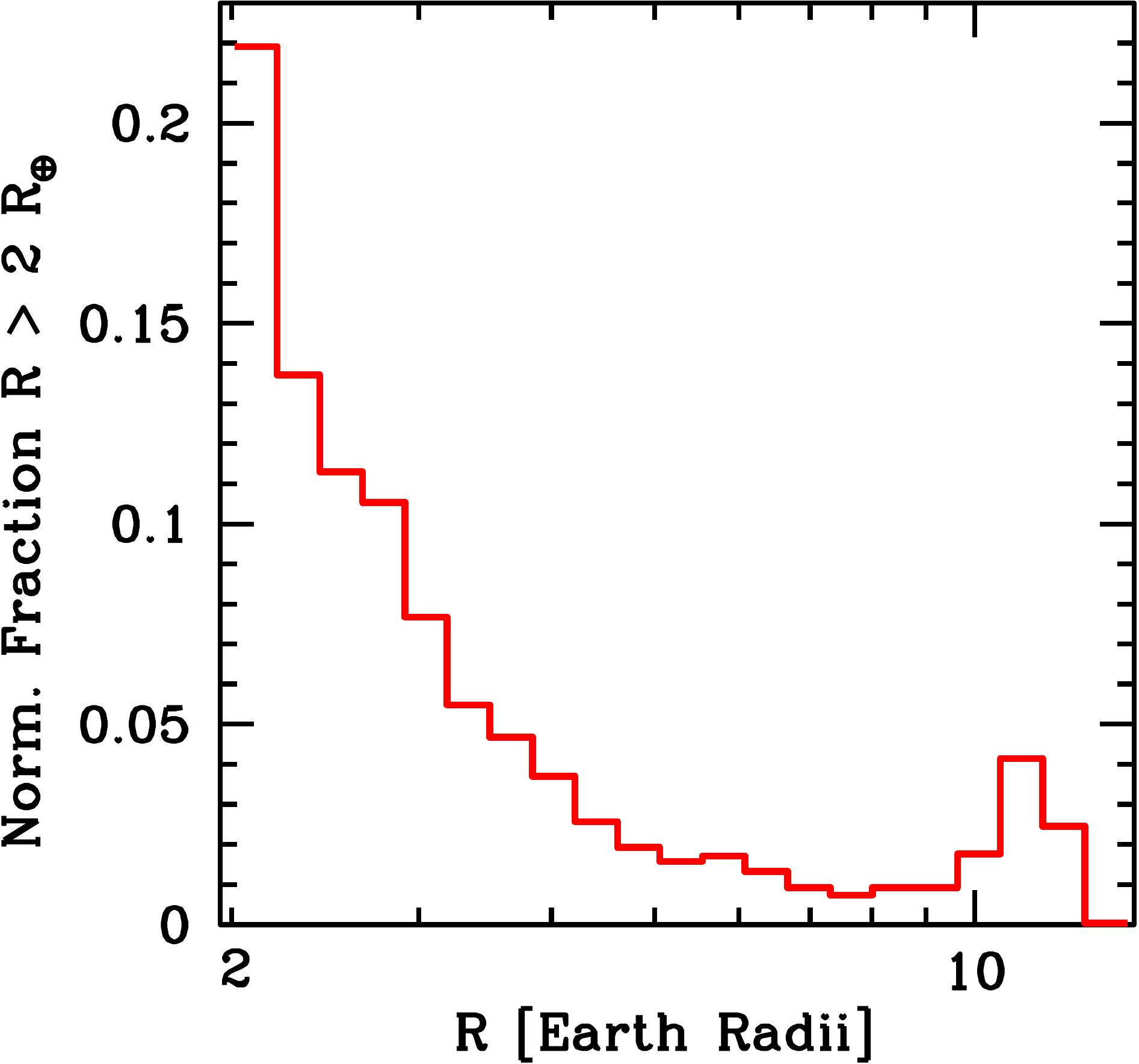}
\caption{Predicted radius distribution for planets with primordial H$_{2}$/He atmospheres and a radius $R>2\rearth$. Synthetic planets at all semimajor axes have been included. The age of the population is 5 Gyrs.}\label{fig:distall}  
\end{center}
 \end{figure}
 
 \begin{table}
\caption{Radius distribution for planets with a primordial H$_{2}$/He atmosphere and $R>2\rearth$. The first two columns are the radius bins, while the remaining three columns are the corresponding fractions  at ages of 1, 5, and 10 Gyrs.}\label{tab:rdist}
\begin{center}
\begin{tabular}{l l c c c }
\hline \hline
$R/\rearth$ & $R/\rj$ & 1 Gyr & 5 Gyrs  & 10 Gyrs\\ \hline
      2.11 &       0.19 &      0.134 &      0.219 &      0.202 \\ 
       2.31 &       0.21 &      0.157 &      0.137 &      0.134 \\ 
       2.54 &       0.23 &      0.134 &      0.113 &      0.135 \\ 
       2.78 &       0.25 &      0.101 &      0.105 &      0.088 \\ 
       3.05 &       0.27 &      0.082 &      0.077 &      0.060 \\ 
       3.34 &       0.30 &      0.078 &      0.055 &      0.053 \\ 
       3.66 &       0.33 &      0.059 &      0.047 &      0.052 \\ 
       4.02 &       0.36 &      0.050 &      0.037 &      0.039 \\ 
       4.41 &       0.39 &      0.037 &      0.026 &      0.027 \\ 
       4.83 &       0.43 &      0.023 &      0.019 &      0.022 \\ 
       5.30 &       0.47 &      0.017 &      0.016 &      0.019 \\ 
       5.81 &       0.52 &      0.014 &      0.017 &      0.020 \\ 
       6.37 &       0.57 &      0.014 &      0.013 &      0.013 \\ 
       6.98 &       0.62 &      0.009 &      0.009 &      0.010 \\ 
       7.66 &       0.68 &      0.008 &      0.007 &      0.008 \\ 
       8.39 &       0.75 &      0.007 &      0.009 &      0.011 \\ 
       9.20 &       0.82 &      0.008 &      0.009 &      0.012 \\ 
      10.09 &       0.90 &      0.009 &      0.018 &      0.022 \\ 
      11.07 &       0.99 &      0.022 &      0.041 &      0.056 \\ 
      12.13 &       1.08 &      0.039 &      0.024 &      0.017 \\ 
      13.30 &       1.19 &      0.000 &      0.000 &      0.000 \\    \hline	       
\end{tabular}
\end{center}
\end{table}%

The distribution has a  characteristic, bimodal shape (cf. Schlaufman et al. \cite{schlaufmanlin2010}; Wuchterl \cite{wuchterl2011}): A global maximum at the smallest radii, and a second, lower local maximum at a radius of about 1 $\rj$. The increase towards small radii is  due to the increase of the underlying mass distribution towards small masses, combined with the fact that with decreasing mass, the fraction of heavy elements increases (Sect. \ref{sect:mrduringevo}). This means that low-mass planets also have small radii. Note that it is well possible that the increase towards small radii may be even stronger in reality than predicted by the model. This is due to the fact that we only include (relatively large) primordial H$_{2}$/He envelopes and one embryo per disk.

The second maximum at about a Jovian radius has a fundamental reason, too. It is due to the fact that in the giant planet domain ($M\gtrsim 100 \mearth$), planets all have approximately the same radius, independent of their mass. This is in turn due to the  property of matter to become degenerate for such massive objects (e.g. Chabrier et al. \cite{chabrierbaraffe2009}).  This property of the EOS makes that a high number of planets from a wide range of different  masses all fall into the same radius bin (radii between $\sim$0.9 and 1.1 $\rj$), causing the maximum in the distribution. The local minimum of the distribution occurs at a radius of 7 to 8 $\rearth$. As can be deduced from Fig. \ref{fig:mrobs}, this corresponds to masses between $\sim20$ to $\sim200 \mearth$, with a typical mass of $\sim  70 \mearth$. This coincides roughly to the mass domain of the ``planetary desert'' where several planet formation models based on the core accretion theory (e.g. Ida \& Lin \cite{idalin2008}, Mordasini et al. \cite{mordasinialibert2009b}) predict a lower abundance of planets. This additional effect makes the second maximum even  more prominent. 
 
The figure shows the radius distribution at the specific age of 5 Gyrs. In reality, stars of a given observational sample will have a distribution of ages. The evolution of the radii at late times ($t\gtrsim 1$ Gyr) is, however,  slow. The distribution of the radii in an age range between 1 to 10 Gyrs indeed only changes  slightly. As expected, there is still a slow general contraction, which makes for instance that at an age of 1 Gyr instead of 5 Gyrs, the local maximum in the giant planet domain is shifted by one bin to the right (i.e. by about 0.1 $\rj$). But the general shape remains very similar, as can also be seen from Table \ref{tab:rdist}.
 
Note that our results become increasingly uncertain with decreasing radius. This is due to the fact that we do not model several effects which are important for planets with small radii, like for example the concurrent formation of many  planets in one disk, or the loss of the gaseous envelope due to evaporation. This, and the fact that we only include primordial H$_{2}$/He envelopes whereas in reality, there are also planets with other types of atmospheres (or without atmosphere) means that the actual radius distribution could be substantially different from the synthetic one for smaller radii. The comparison of the synthetic and the actual distribution then helps to quantify these effects.

\subsection{Comparison with \textit{Kepler} results}\label{sect:comprdistkepler}
It is interesting to compare the radius distribution as found in the synthesis with the one observed by the \textit{Kepler} satellite (Borucki et al. \cite{boruckikoch2011}). Howard et al. (\cite{howardmarcy2011}) have analyzed the radius distribution of \textit{Kepler} planet candidates with high S/N transits for a subsample of bright, GK dwarfs. They corrected for the observational bias to derive the underlying radius distribution for planets with $R>2 \rearth$ and an orbital period of less than 50 days. In Fig.  \ref{fig:Kepler} we compare their observational result with the synthetic radius distribution.
 
\begin{figure*}
      \centering
      \begin{minipage}[lt]{8.9cm}
        \includegraphics[width=\textwidth]{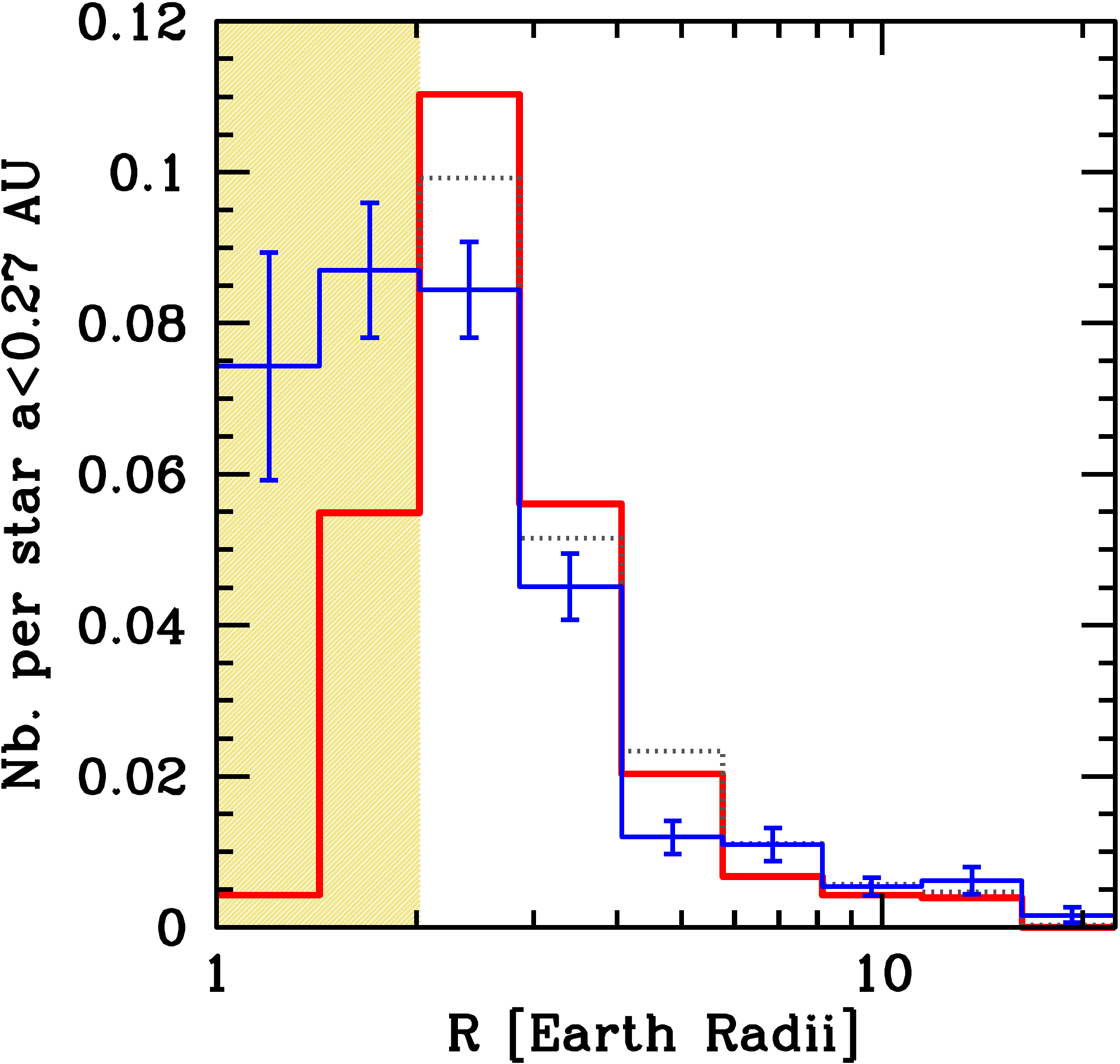}
        \end{minipage}
        \hfill
             \begin{minipage}[lt]{8.9cm}
             \includegraphics[width=\textwidth]{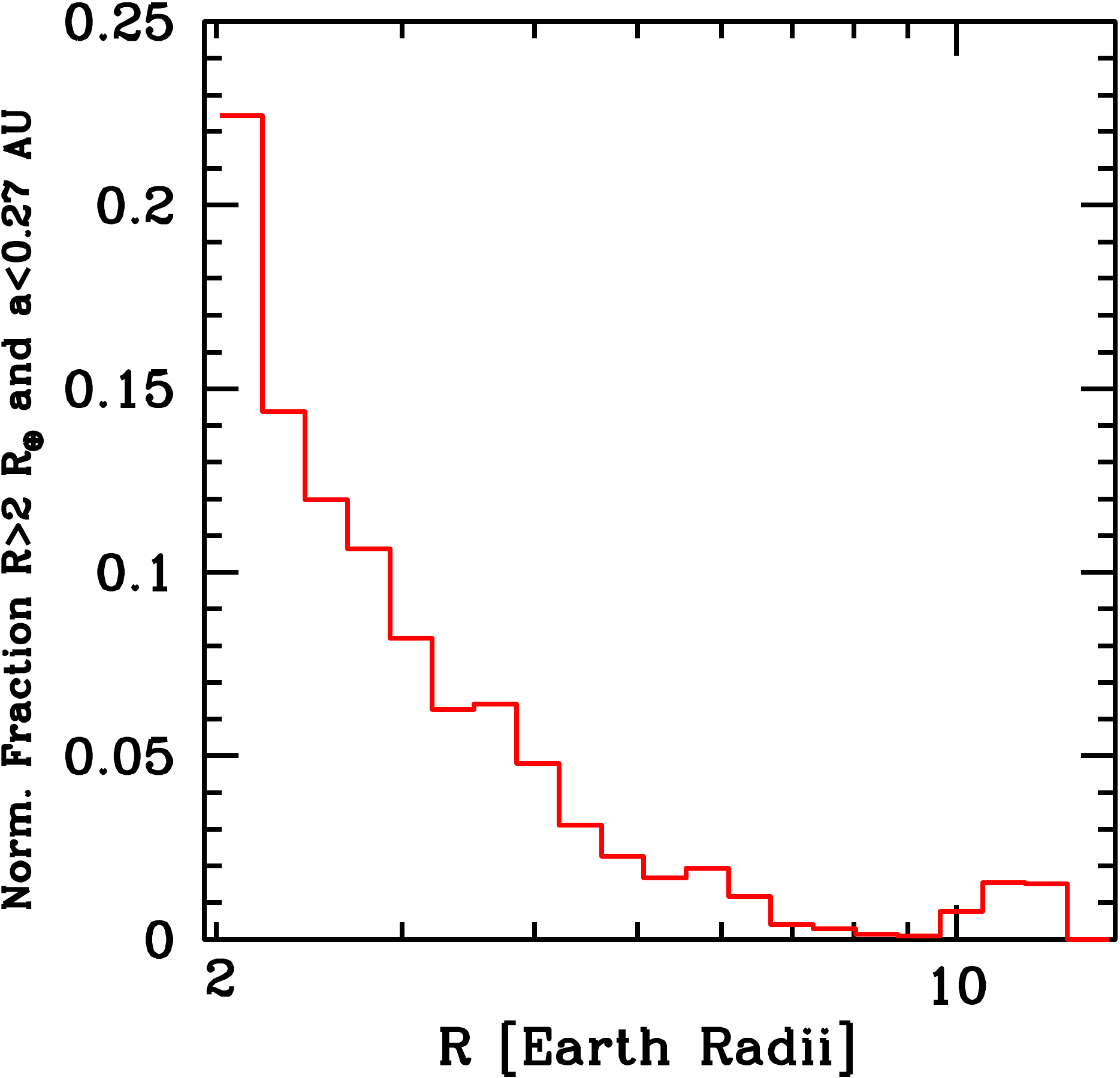}
        \end{minipage}
        \hfill        
      \caption{Left panel: Number of planets per star in a given radius bin and an orbital period $P<50$ d ($a<0.27$ AU for a 1$\msun$ star). The thick red line is the synthetic population. The blue line with error bars is from Howard et al. (\cite{howardmarcy2011}) for the \textit{Kepler} data of Feb. 2011 (Borucki et al. \cite{boruckikoch2011}). The black dotted line is the preliminary analysis of Howard et al. (\cite{howardmarcy2012}) for the data of Sep. 2011 (Batalha et al. \cite{batalharowe2012}). The observational data in the hatched region with $R<2\rearth$ is incomplete.  Right panel: Normalized fraction of radii of the same synthetic planets as on the left, but with finer bins, and including only planets with $R>2\rearth$.}
       \label{fig:Kepler}
\end{figure*}

In the left panel we show as Howard et al. (\cite{howardmarcy2011}) the number of planets per star in a given radius bin and a semimajor axis $<0.27$ AU. Synthetic planets which migrate to the inner border of the computational disk at 0.1 AU are evolved at this orbital distance. Note that the  majority of planets analyzed by Howard et al. (\cite{howardmarcy2011}) are found outside of 0.1 AU, as the planetary frequency is increasing with distance, especially for companions which are not giant planets. This means that the comparison of our synthetic planets with $0.1\leq$a/AU$\leq$0.27 with the sample of Howard et al. (\cite{howardmarcy2011}) which includes all planets inside $\sim$0.27 AU, can still be made, at least in an approximate way.

\subsubsection{Agreement for $R\gtrsim2\rearth$}
The number of planets per star can be directly calculated from the synthesis as the outcome for all simulated planets is known\footnote{The total number of simulated synthetic planets is about 35\,000, which requires about one week of computational time on $\sim100$ CPUs.}. We see that the synthetic and the observed result are in good agreement, especially for the new, preliminary  analysis of Howard et al. (\cite{howardmarcy2012}) for the updated \textit{Kepler} data set released in Sep. 2011 (Batalha et al. \cite{batalharowe2012}). We see that both data sets are characterized by an increase of the frequency with decreasing radius. The slope of the increase is similar in both cases, with the synthesis predicting a somewhat higher number of planets with a small radius. But not only the relative increase, also the absolute number of planets is similar in the model and observation. This could indicate that the inward migration of planets (especially at low masses, i.e. in type I) is quite efficient. The total fraction of stars with a planet with $P<50$ d and $R>2\rearth$ in the synthesis is 20.1\%. The value derived from \textit{Kepler} is 16.5\% to 20\% (Howard et al. \cite{howardmarcy2011,howardmarcy2012}).

We must however note that the very good agreement also in absolute terms must be regarded with caution. This is in particular because we are using the one-embryo-per-disk approximation, so that we cannot catch the fact that the occurrence rate calculated by Howard et al. (\cite{howardmarcy2011}) includes multiple-planet systems and gives the number of planets per star, which can be higher than one, while it is exactly one in our simulations by construction when considering all synthetic planets. Including only one planet per star would reduce the number of \textit{Kepler} candidates by about 20\% (Latham et al. \cite{lathamrowe2011}; Wolfgang \& Laughlin \cite{wolfganglaughlin2012}), and mainly lead to a reduction of planets with small radii. Planets with small radii are more frequently found in multiple systems, while giant planets are more often single (Latham et al. \cite{lathamrowe2011}). But the basic result that both the model and the synthesis find a strong increase towards small radii, and that this increase follows a similar slope, remains.

\subsubsection{Divergence for $R\lesssim2\rearth$}
Figure  \ref{fig:Kepler} shows that the synthetic radius distribution  decreases for radii smaller than $\sim2$ $\rearth$. The reason for this is that low-mass planets with a H$_{2}$/He atmosphere, even when it is only tenuous (0.1 to 1\% of the core mass) have radii of at least  $\sim2\rearth$ at the relevant equilibrium temperatures for $a<0.27$ AU  and ages (i.e. intrinsic luminosities), see  Fig. \ref{fig:mrrogerscomp}. This is in agreement with the findings of Rogers et al. (\cite{rogersbodenheimer2011}). As we only model this type of atmospheres, and even low-mass cores accrete such amounts of gas during the formation phase, there are not many synthetic planets with a radius smaller than $\sim 2 \rearth$.  It is clear that our quantitative results for the distribution of the radii of  very low-mass planets are strongly affected by the various  simplifications in the model like for example  {the assumption of chemically homogeneous layers (in particular of a pure H$_{2}$/He envelope) or the fact that we neglect the evaporation of the envelope.} On the other hand, the basic mechanism responsible for the decrease of the radius distribution, which is the mass-radius relationship of close-in, low-mass planets with H$_{2}$/He atmosphere, should not be affected by this. We have checked that the decrease remains similar if we start the simulations with embryos with a mass of 0.1 instead of 0.6 $\mearth$. 

This decrease is in contrast to the (however incomplete) observational result (Howard et al. \cite{howardmarcy2011})  where the fraction remains at least constant for $R<2 \rearth$ (see also Youdin \cite{youdin2011}). The good agreement of the synthetic and the actual radius distribution for $R>2 \rearth$, and the divergence at smaller radii is interesting. It could indicate that for planets larger than $\sim2 \rearth$, H$_{2}$/He atmospheres are dominant (mini-Neptune type of planets), while for smaller radii, other types of planets (silicate-metal terrestrial planets or ocean planets without primordial H$_{2}$/He) dominate. While our quantitative results are as mentioned to be taken with great caution, this  result seems to be in agreement with detailed analyses of the combined constraints coming from the \textit{Kepler} data for the radius, and from radial velocity for the mass (Howard et al. \cite{howardmarcy2011}; Wolfgang \& Laughlin \cite{wolfganglaughlin2012}; Gaidos et al. \cite{gaidosfischer2012}). These studies show that with decreasing radius, a transition from low planetary densities to higher densities occurs. Wolfgang \& Laughlin (\cite{wolfganglaughlin2012}) in particular find that when they use a mixture of planets with H$_{2}$/He atmospheres and rocky planets, the later type of planets becomes dominant for radii $R\lesssim2.5\rearth$ for their simulation that best reproduces the combined constraints from HARPS and \textit{Kepler}. This is similar as the result found here  from theoretical calculations. The results on the radius distribution of planets with primordial atmospheres presented here, and below in Sect. \ref{sect:meanrasfctofmass}, should allow for an even better understanding of the dividing line between these types of planets.

\subsubsection{Local maximum at $\sim 1\rj$}
The left panel of Fig. \ref{fig:Kepler} shows the synthetic planets in the same bins as the observed data. No peak in the giant planet regime can be seen due to the broadness of the bins. The right panel shows the same synthetic planets, but now using finer bins (and $R>2\rearth$). The bimodal shape of the radius distribution becomes visible again.  The local maximum at about one Jovian radius is, however, about 2 to 3 times smaller than for the distribution that includes synthetic planets at all semimajor axes (Fig. \ref{fig:distall}). This is due to the fact that giant planets are less frequent inside 0.27 AU than further out.

We thus see that at the resolution of the radius distribution in the current observational data, and the restriction that $P<50$ d, it is difficult to see the bimodality. If we use finer bins (which will be possible with a higher number of transiting planets), or if we keep the same broader bins, but include planets at all semimajor axes, the peak at about 1 Jovian radius should become visible. This is an important prediction which is in agreement with other models of planet formation and evolution (Schlaufman et al. \cite{schlaufmanlin2010}; Wuchterl \cite{wuchterl2011}). It will be tested in the next years with the \textit{Kepler} mission as an increasing number of planets at large semimajor axes will be found. Concerning the local maximum at about 1 $\rj$, we must distinguish two aspects of it: First the local minimum at a radius of about 8 $\rearth$, and second the sharp absence of planets larger than $\sim1.2 \rj$. The second aspect is more fundamental, as it is a direct consequence of the equation of state.   When making the comparison, we must keep in mind a number of  possible complications which could cause differences between the synthetic and the actual radius distribution. 

1) First of all it is clear that the results shown here can only be compared with planets which are not affected by the various proposed inflation mechanism (e.g. Batygin et al. \cite{batyginstevenson2011}). An inflation to larger radii naturally blurs the maximum at one specific radius as found in standard cooling calculations as conducted here. In this context it is interesting to note that Demory \& Seager (\cite{demoryseager2011}) find that at irradiation fluxes of less than about $2\times10^{8}$ erg cm$^{-2}$ s$^{-1}$, giant planets seem not be  significantly affected any more by inflation.  For a solar-like star, the critical irradiation level corresponds to a semimajor axis of about 0.07 AU. Instead all giant planets in their sample are characterized by  a constant radius of $\sim 0.87\pm0.12\rj$. This is  compatible with our results for the location of the local maximum in the radius distribution.  

2) It is possible to construct a mass distribution which has no maximum at about one Jovian radius. First, such a mass distribution must contain a higher number of planets with intermediate radii  compared to the synthetic mass distribution, so that the local minimum at about 8 $\rearth$ disappears. This would mean that the ``planetary desert''  (Ida \& Lin \cite{idalin2004}); dedicated discussion in Mordasini et al. \cite{mordasinialibert2009a}) is more populated than found in the model. In this case, the radius distribution between $\sim6$ and 12 $\rearth$ would simply be flat, or decreasing with $R$. Second, the mass distribution could contain a lower number of giant planets.  This would naturally lower the number of planets in the second maximum. When comparing the  fraction of planets with radii close to 1 $\rj$ in the model and in the data of the \textit{Kepler} satellite (left panel of Fig. \ref{fig:Kepler}) one notes that the number of synthetic giant planets is in good agreement with the observed fraction. This would mean that this second possibility is in fact not feasible. 

3) Demory \& Seager (\cite{demoryseager2011})  find that the majority of \textit{Kepler} candidates of their sample which receive only a modest irradiation flux, but still have large radii $\gtrsim1.1\rj$ have to be classified as ``ambiguous'' candidates i.e. as cases where the secondary eclipse signature is consistent with both a planetary or stellar companion, and not as relatively clear planetary companions as it is the case for most candidates with smaller radii. This could indicate that a part of the \textit{Kepler} candidates which are difficult to understand from a theoretical point of view (radius clearly larger than 15 to 20 $\rearth$, orbital period longer than 10 to 20 days, high age)  are false positives. False positives would also blur the upper radius limit of planets. A finding of Traub (\cite{traub2012}) could  also point in this direction: He finds that in the \textit{Kepler} data, for large planets  with radii of about 8 to 40 $\rearth$, there are significantly more candidates around faint stars than around brighter host stars. This is not expected from astrophysical reasons,  therefore Traub (\cite{traub2012}) puts forward the explanation that it is more difficult to distinguish background eclipsing binaries from planetary transits when the stars are faint. This would mean that this radius class is more affected by false positives. As more data is accumulated by \textit{Kepler}, it should become possible to better assess such a reason. It is further interesting that when \textit{Kepler} candidates in multiple systems only are considered (Latham et al. \cite{lathamrowe2011}), then the observed radius distribution  is bimodal. A minimum at about 1 $\rj$, and a local maximum at about 1.2 $\rj$ is seen, which is at least qualitatively similar as found in the simulations here.  The number of planets in the relevant bins is however very small, so that it is currently unclear whether this is a statistically significant result. 

4) The synthetic radius distribution shows planets which all have exactly the same age. While also in the \textit{Kepler} data, the majority of the host stars are on or close to the main sequence (Batalha et al. \cite{batalharowe2012}; Latham et al. \cite{lathamrowe2011}) the spread in ages will still blur to some extent the maximum at about 1 $\rj$, especially its upper edge, which is in contrast very sharp in the synthetic population at one single age.  A similar effect is expected from our simplification that all planets   have identical atmospheric opacities. 

5) Finally, {objects} like  KOI-423b (Bouchy et al. \cite{bouchybonomo2011})  which is quite far from the theoretical expectations,  could indicate that our theoretical understanding of the planetary radius evolution does not take into account all relevant mechanism which affect at least some planets. Possible mechanism could for example be tidal heating for very eccentric planets (Dong et al. \cite{dongkatz2012}) or giant impacts (Anic et al. \cite{anicalibert2007}).

\subsection{Impact of the semimajor axis}\label{sect:impacta}
  \begin{figure*}
\begin{center}
\includegraphics[width=0.44\textwidth,,trim = 0mm -3.9mm 0mm 0mm, clip]{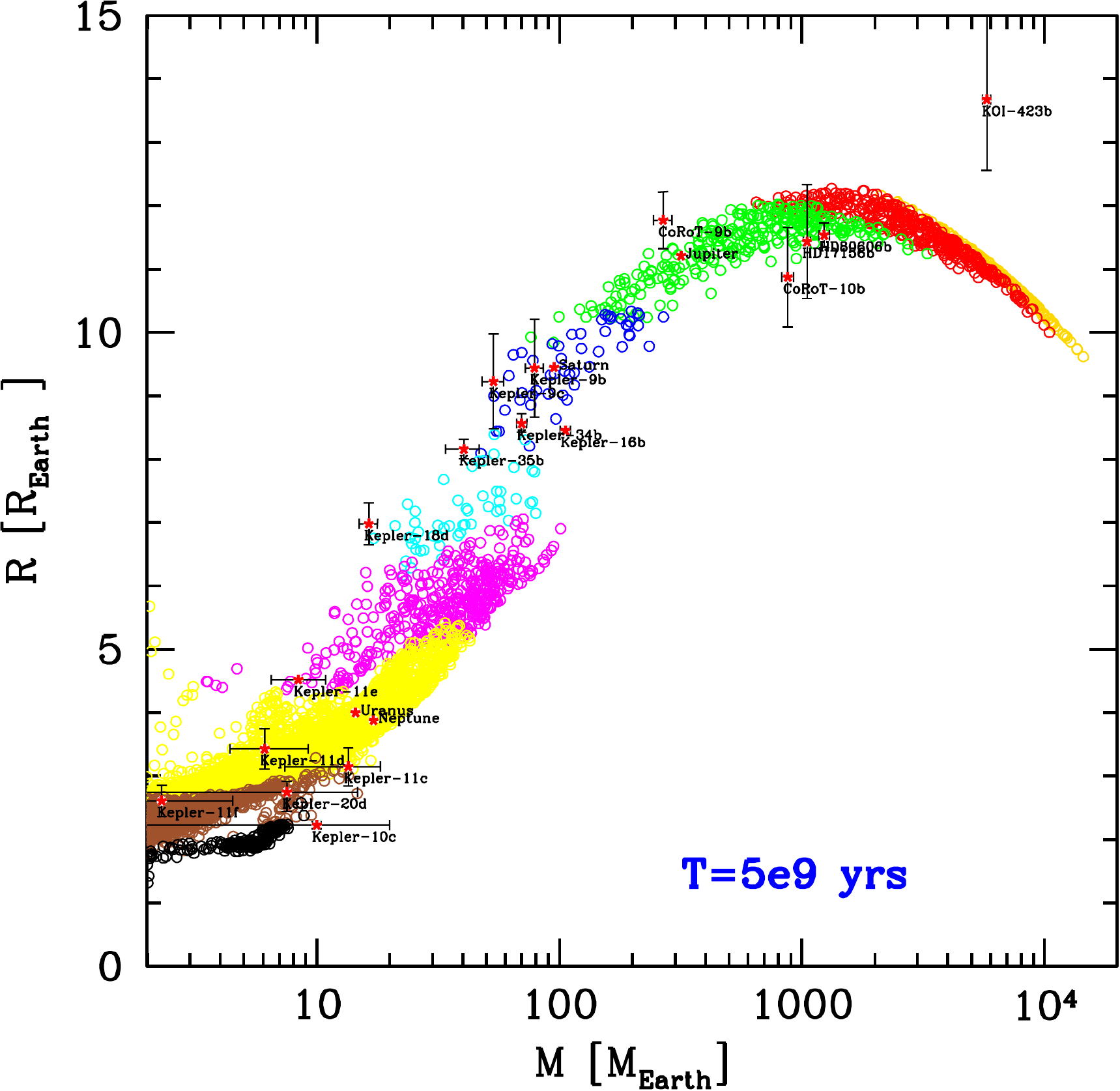}
\hspace{0.5cm}
\includegraphics[width=0.45\textwidth,height=7.95cm,trim = 0mm 1mm 0mm 0mm, clip]{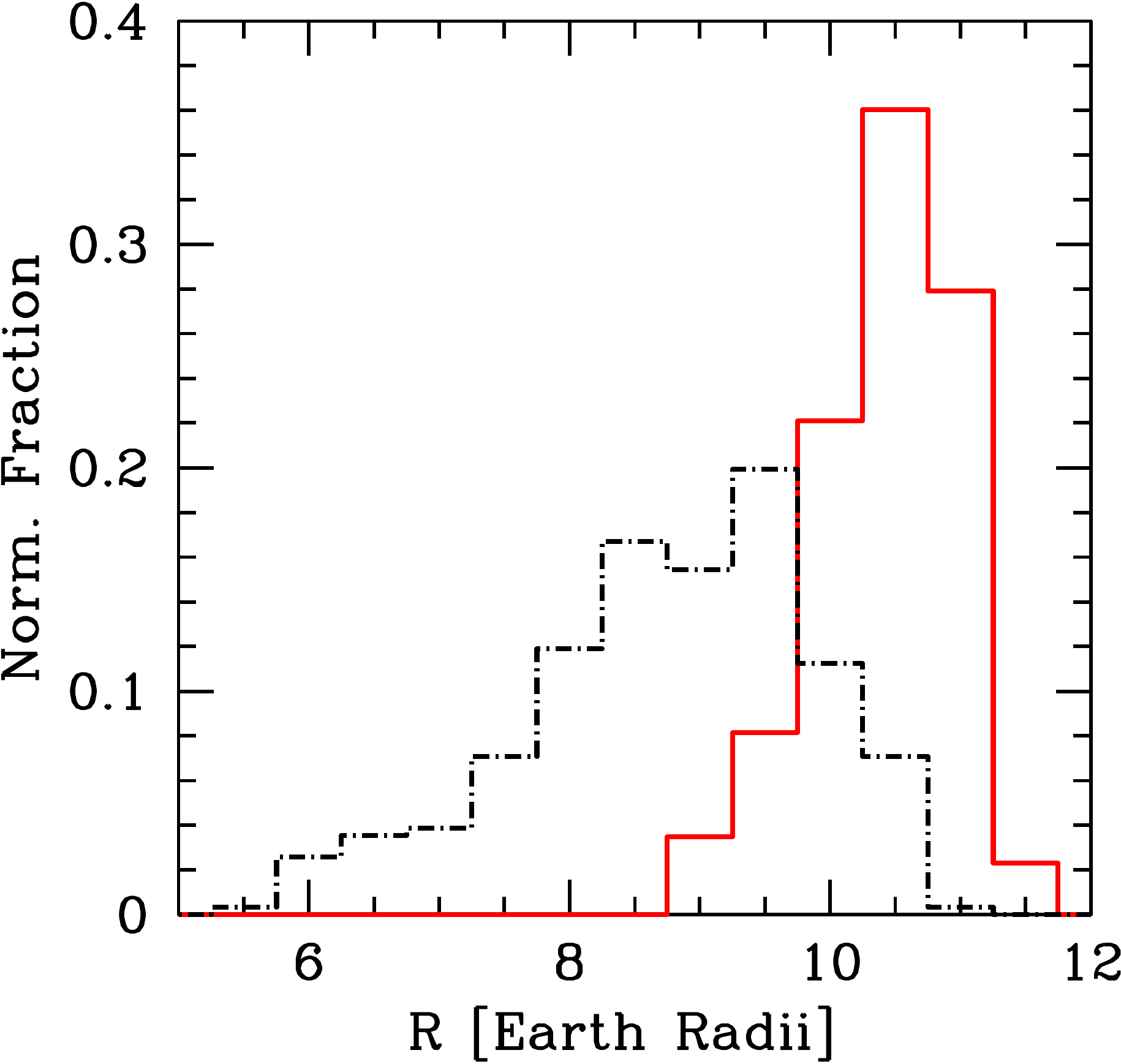}
\caption{Left panel: Synthetic mass-radius relationship as in Fig. \ref{fig:mrobs},  but including only synthetic planets with a final semimajor axis smaller than 1 AU. The color coding and the observational comparison sample is the same as in Fig. \ref{fig:mrobs}. Right panel: Distribution of radii of planets with masses between 100 and 300 $\mearth$. The red solid line  is for planets with a semimajor axis less than 1 AU, while the black dashed-dotted line shows planets outside of this distance. Planets reaching the inner border of the computational disk have been excluded. The age of the population is 5 Gyrs.}\label{fig:mrobs1au}
\end{center}
\end{figure*}

Figure \ref{fig:mrobs} shows the $M-R$ relationship for synthetic planets at all semimajor axes, including thus in particular also planets at large orbital distances. Observationally, most of the known transiting planets have in contrast small semimajor axes. To see the impact of the semimajor axis on the radii, we plot in Fig. \ref{fig:mrobs1au}, left panel again the $M-R$ relationship, but now including only synthetic planets with a semimajor axis between 0.1 and 1 AU.  The extrasolar planets in the observational comparison sample  all have semimajor axes less than 1.1 AU (the planets of the Solar System are found at larger distances, of course).   

\subsubsection{Semimajor axis - radius anticorrelation} 
Comparing the synthetic planets at smaller distances with the entire population shows that there is a clear difference. At all masses, the lower part of the envelope of $M-R$ points corresponding to those planets with the smallest radii for a given mass are not found if only the planets inside 1 AU are considered. This reduces the vertical width of the area populated by $M-R$ combinations, making that Kepler-10c and Corot-10b now lie at the lower boundary of the region populated by synthetic planets. When all semimajor axes are included, they are  more embedded in the cloud of points.  
  
The reason for this finding is that the maximum mass fraction of heavy elements $Z$ for a given total mass is lower for planets inside 1 AU than for those further out. For example, planets with a total mass of  100 $\mearth$ can contain up to $\sim90$ \% solids (in rare cases, though) when considering all semimajor axes (Fig. \ref{fig:mrobs}). When limiting to $a\leq1$ AU, the heavy element fraction is at most $\sim60$\%, as can be seen in Fig. \ref{fig:mrobs1au}. 

This is a consequence of the following. The reservoir of solids available for a planet to accrete in a disk with a dust-to-gas ratio $\fpg$ scales with distance $r$ as $\fpg \Sigma(r,t=0) r^{2}$, where $\Sigma(r,t=0)$ is the initial surface density of gas. This quantity is given by equation \ref{eq:newprofile}, and thus scales in the inner, relevant parts of the disk as $r^{-0.9}$. This means that the amount of mass contained in an annulus increases with semimajor axis. This trend gets amplified by the increase of the solid surface density at the iceline. It is clear that this radial dependence gets blurred by orbital migration, but as the simulations show, the general pattern remains conserved. 

It is interesting to see that the actual known extrasolar planets of the comparison sample (which have semimajor axes between 0.1 and  1.1 AU) mostly cluster close to the upper edge of the synthetic population at least for intermediate masses between $\sim50$ to 300 $\mearth$, and thus seem more similar to the synthetic population inside 1 AU than to synthetic planets further out. No planet with a mass of e.g. 100 $\mearth$ and a radius of only 6 $\rearth$ has been detected to date, despite such metal-rich objects {existing} in the synthetic population at large distances (Fig. \ref{fig:mrobs}). The right panel of Fig. \ref{fig:mrobs1au} compares the  distribution of radii for a similar mass domain  for planets in- or outside 1 AU. The difference is very obvious. The better match of the observed planets with the synthetic population inside 1 AU could indicate that also in nature, close-in giant planets contain a smaller fraction of heavy elements than their counterparts further out which formed in a larger reservoir of solids. Kepler-18d has a mass similar as the ice giants in the Solar System, but is much poorer in solids (Sect. \ref{sect:individobj}). This could point into the same direction. 

This allows to make the following two statements: First, for the ongoing high precision transit searches, it is predicted that they will find intermediate and giant planets with a higher fraction of heavy elements (i.e. smaller radius) as they detect planets at larger semimajor axes.  Second, this anti-correlation of semimajor axis and radius via $Z$ might offer a possibility to distinguish between disk migration and other mechanisms like scattering (Malmberg et al. \cite{malmbergdavies2011}) or Kozai migration (Fabrycky \& Tremaine \cite{fabryckytremaine2007}) as  formation mechanisms of close-in hot Jupiters. This is because it is at least not obvious that the latter two mechanism would also lead to a semimajor axis - $Z$ correlation as found here for disk migration. 

\subsubsection{Possible complications}
Note that a number of complications arise: First, it seems reasonable to assume that a higher bulk content in metals goes along with a higher atmospheric opacity. Higher opacities are well know to delay the contraction of giant planets (e.g. Burrows et al. \cite{burrowsheng2011}; Paper I).  In this work, in contrast, we use opacities for solar composition gas in all cases.  The impact of higher opacities for more metal rich planets would counteract the semimajor axis-radius anti-correlation.  Second, the evolution of the radius of a giant planet is influenced also directly by its semimajor axis, via the intensity of the stellar irradiation. When the irradiation flux is not deposited deep in the interior of the planet, this has not a very drastic influence on the radius for $a>0.1$ AU. For instance, a Jovian mass planet at 0.1 AU has at 4.5 Gyrs a radius that is about 6\% larger than a counterpart at 10 AU (Fortney et al. \cite{fortneymarley2007}).  At least for planets inside 0.1 AU, there are however special inflation mechanisms  which affect the radius much stronger, as illustrated by the large number of transiting planets with anomalously large ``bloated'' radii. It is currently not clear out to which distance these inflation mechanisms are effective deepening on e.g. the planetary mass or composition.  The inflation also leads to an anti-correlation of semimajor axis and radius, i.e.  goes in the same direction as the correlation describe here due to the composition, and thus complicates the picture.  
  
In summary we see that the combined constraints of planet mass,  radius, semimajor axis and metallicity provide powerful observational constraints for planet formation and evolution theory, in particular when they are coupled to population synthesis calculations which allow to see the correlations between these quantities. A step which has however to be taken in future for more accurate analyses is a more self-consistent coupling of bulk interior composition and atmospheric properties in terms of  composition and opacity.

\subsection{Mean radius as function of mass}\label{sect:meanrasfctofmass} 
\begin{figure*}
\begin{center}
\includegraphics[width=0.48\textwidth]{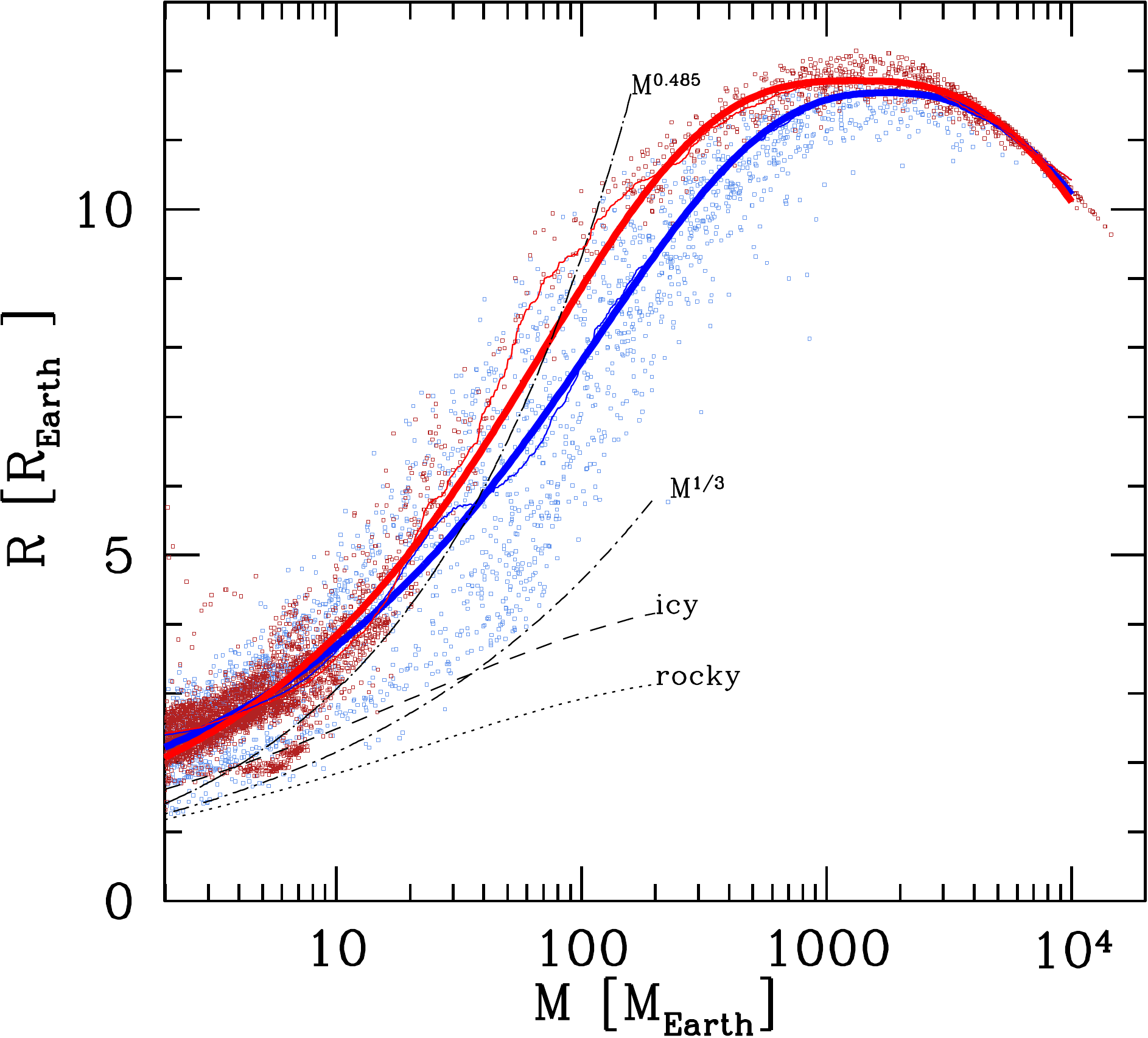}
\hfill
\includegraphics[width=0.48\textwidth]{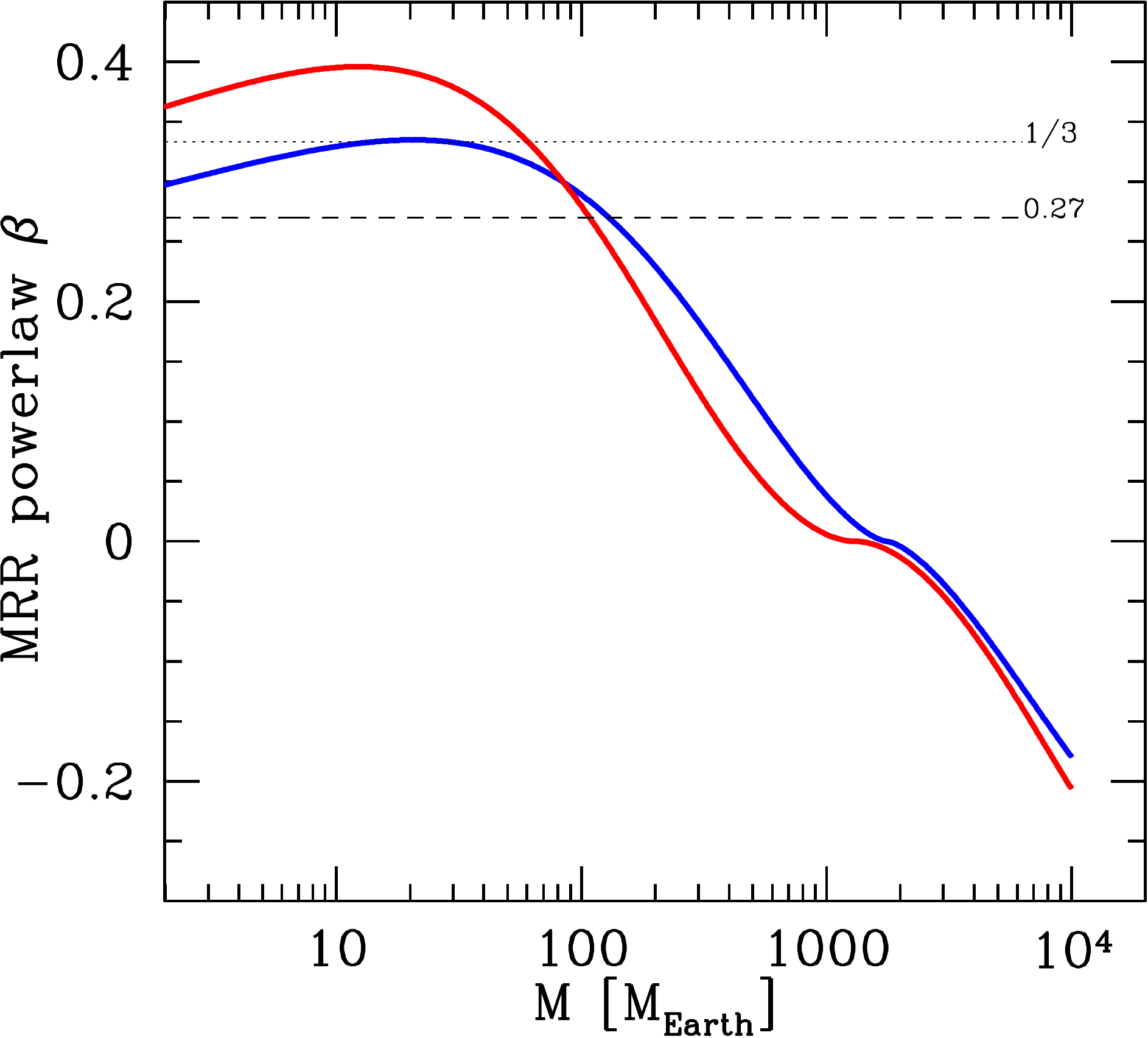}
\caption{Left panel: Synthetic mass-radius relationship  with analytical approximations for the mean radius $\bar{R}(M)$. Blue dots: synthetic planets with 0.1$<$a/AU. Thick blue line: analytical fit. Thin blue line: running mean. Red dots: synthetic planets with 0.1$<$a/AU$\leq1$. Thick red line: analytical fit. Thin red line: running mean. Black lines: mass-radius relationship for $\beta=1/3$ and 0.485, for solid planets with an Earth-like composition, and for planets with a composition as expected beyond the ice line. Right panel: Power law exponent $\beta$ for the $M-\bar{R}$ relationship as function of mass.}\label{fig:mrmean}
\end{center}
\end{figure*}

Several studies (e.g. Howard et al. \cite{howardmarcy2011};  Wolfgang \& Laughlin \cite{wolfganglaughlin2012}; Gaidos et al. \cite{gaidosfischer2012}) have analyzed combined RV and \textit{Kepler} datasets as this allows to get insight  into the mean density of the planets. An important quantity in this context is the power law exponent $\beta$ which relates planetary mass and radius  
\beq
\frac{R}{\rearth}=k \left(\frac{M}{\mearth}\right)^{\beta}.
\eeq
It is clear that in reality, there cannot be a single value of $\beta$ (and $k$) for all planetary masses and compositions. Still it remains a useful quantity when comparing theory and observations. 

Observationally, a fit to the terrestrial planets of the Solar System yields  $\beta=0.33$ as for constant density spheres (Laughlin \& Wolfgang \cite{wolfganglaughlin2012}), while a fit including the Earth and Saturn (Lissauer et al. \cite{lissauerragozzine2011}) gives $\beta=0.48$, with $k=1$ in both cases. From theoretical calculations, one expects for scaled versions of the Earth in the 1-10 $\mearth$ regime $\beta\approx0.27$ (Valencia et al. \cite{valenciaoconnell2006}) while for massive gaseous planets of about 10 $\mj$, $\beta\approx-1/8$ (Chabrier et al. \cite{chabrierbaraffe2009}), as expected from polytropic models. The negative values show that in this domain, the radius decreases with mass. At the maximum of the mass-radius relationship at about 4 $\mj$ (Fig. \ref{fig:mrobs}), $\beta$ must be equal to zero. This means that with increasing mass, a decrease of $\beta$ occurs. This is a consequence of the combined effects of a change in the interior composition of the planets with mass (the gas fraction increases), and increasing degeneracy of the matter. The specific dependence of $\beta$ on mass is a proxy for the bulk interior composition of the planets. 

\subsubsection{Analytical expression for the mean radius}
To calculate $\beta$, we first have to find an analytical expression for the mean radius as function of mass $\bar{R}(M)$ for which the best fitting parameters for the synthetic population are determined. We employ  the  function proposed by Traub (\cite{traub2011}) which has four parameters $b, M_{0}, w, p$ and is given as
\beq\label{eq:eqMR}
\bar{R}(M)=\frac{b}{1+\left|\frac{\log(M/M_{0})}{w}\right|^{p}}, 
\eeq
where $\log$ denotes the logarithm to the base 10.  The units in the fit are Earth masses and radii.  Compared to the original function of Traub (\cite{traub2011}), we have omitted an additional additive constant which is found to be unnecessary. In the equation, the parameter $p$ controls how abruptly  the radius increases when going from terrestrial planets to the maximum in the giant planet regime. It also controls how flat the maximum in the giant planet regime is. The parameter $w$ controls the width of the peak in the radius function.  The parameter $M_{0}$ is the mass (in units of Earth masses) corresponding to the maximum radius (i.e. 4 to 5 Jovian masses), while $b$ provides a  vertical scaling. The value of $b$ corresponds to the radius (in units of Earth radii) at $M=M_{0}$.

We have fitted our nominal population at 5 Gyrs with this function using a nonlinear least-squares Marquardt-Levenberg algorithm to find the parameters. We only include planets with a mass larger than 2$\mearth$. The result is shown in Tab. \ref{tab:mrfit} and Figure \ref{fig:mrmean}. The best fitting parameters are given for all planets which have not reached the inner border of the computational disk (i.e. $0.1<a/AU$), and for a selected semimajor axis range $0.1<a/AU\leq1$ to study again the effect of the semimajor axis.    {Planets that reach the inner border of the computational disk are excluded from the analysis since their fate is more uncertain.} 

\begin{table}
\caption{Fitting parameters for the $M-\bar{R}$ relationship.}\label{tab:mrfit}
\begin{center}
\begin{tabular}{l c c  }
\hline \hline
Parameter  & $0.1<a/AU$ & $0.1<a/AU\leq1$  \\ \hline
$b$ [$\rearth$]    & 11.684 & 11.858          \\
$M_{0}$  [$\mearth$] & 1756.7 & 1308.7 \\    		      
$w$ & 1.646 &  1.635 \\  
$p$ & 2.489  &  2.849 \\ \hline  		      
\end{tabular}
\end{center}
\end{table}%

The left panel of Fig. \ref{fig:mrmean} shows that the fit (thick lines) yields a good representation of the mean synthetic mass-radius relationship. In the plot, the thin colored lines show the mean radius found directly with a running mean with a window 0.15 dex wide in $\log(M/\mearth)$. The fitting function and the running mean lie in general close to each other, expect for a few special, fine features that the fitting function cannot reproduce. As expected from the discussion in the last section, the line including only planets inside 1 AU lies at larger radii for intermediate mass planets. Note that with the two sets of parameters, one finds a radius of about 1.8 and 1.6 $\rearth$ for a 1 $\mearth$ planet. Clearly, this is a consequence of  only simulating planets with primordial H$_{2}$/He atmospheres. 

\subsubsection{Mass-radius power law exponent}
The power law exponent $\beta$ is given as $d\ln \bar{R} / d \ln M$. The power law exponent for Traub's (\cite{traub2011}) fit is given as ($\ln$ is the natural logarithm) 
\beq\label{eq:expMRs}
\beta(M)=\frac{p\left(\frac{|\log(M/M_{0})|}{w}\right)^{p}}{\ln(M_{0}/M) \left(1+\left(\frac{|\log(M/M_{0})|}{w}\right)^{p} \right) }
\eeq

Three different regimes can be distinguished  for $\beta$. The mean planetary density decreases with increasing mass for $\beta>1/3$,  is independent of mass for $\beta=1/3$, while values of $\beta$ smaller than 1/3 indicate that the density is increasing with mass.  The right panel of Fig. \ref{fig:mrmean} shows the value of $\beta$ as a function of mass, again for the two different semimajor axis ranges. Note that the exact numerical values of $\beta$ partially depend on the specific fitting function. Using for instance polynomial functions instead of eq. \ref{eq:eqMR}, leads to variations of $\beta$ by of order 0.05 to 0.1, but preserves  the general pattern.  At small masses, the values of $\beta$ are relatively large. For $M\lesssim 30 \mearth$ and when all planets with $a>0.1$ AU are included (blue line), then $\beta$ lies between 0.3 and 0.33, close to the constant density case. This is the result of two counteracting effects: The increase of the gas mass fraction in the planets with increasing mass tends to lower $\beta$, while the growing effect of gravitational self-compression of the more compressible gas tends to increase $\beta$. When only planets inside 1 AU are included (red line), then $\beta$ is  higher (0.35 to 0.4) showing that for these planets, the density is decreasing with increasing mass, due to the smaller heavy element content (Sect. \ref{sect:impacta}). At a mass of roughly 100 $\mearth$, $\beta$ starts to decrease, as the effect of self-gravitational compression takes over, reaching $\beta=0$ at a few Jovian masses, as expected.

Towards the lowest masses (2 $\mearth$), $\beta$ decreases slightly with decreasing mass. This is a consequence of the following: at sufficient irradiation from the host star, the radius of low-mass planets with a significant H$_{2}$/He atmosphere increases with decreasing total mass when the heavy element fraction $Z$ is hold constant (see Fig. \ref{fig:mrrogerscomp}). This would correspond to negative values of $\beta$. The effect that for the synthetic planets, $Z$ is itself increasing with decreasing mass however overcompensates this. Due to this imprint of formation (lower mass cores can only bind smaller envelopes during the formation phase), very low values of $\beta$ do not occur at low planetary masses, especially as we have excluded planets with $a\leq0.1$ AU. The fact that the power law exponent for low-mass planets with H$_{2}$/He atmospheres is found to be similar as for solid planets without atmospheres could make the observational distinction more difficult. In terms of absolute sizes, the two planetary types are however clearly distinct. 

We stress that it is \textit{not} expected that the fit provides a good representation of the planetary mass-radius relationship at low masses in the terrestrial and super-Earth regime. The terrestrial planets of the Solar System, as well as some low-mass extrasolar planets like Corot-7b  or Kepler-10b  do not posses primordial H$_{2}$/He envelopes. But the comparison of the predicted and the actual mass-radius relationship  allows to gain insight about the actual nature of such objects, in particular when studying  the fraction of silicate-iron or water planets versus mini-Neptunes (Wolfgang \& Laughlin \cite{wolfganglaughlin2012}, Gaidos et al. \cite{gaidosfischer2012}). 

\subsection{Radius distribution for given mass intervals}\label{sect:rdistmassranges} 
In the previous section, the mean radius as function of mass was studied. The mean radius can however not catch that for every mass, planets with different internal compositions and thus different radii form in the synthesis. The importance of such a multi-valued mass-radius relationship for the interpretation of \textit{Kepler} and radial velocity data has been stressed by Wolfgang \& Laughlin (\cite{wolfganglaughlin2012}). These authors consider also terrestrial planets without any atmosphere, while we only consider planets with primordial H$_{2}$/He atmospheres. But already for this single type of planets, there is a large diversity in mass-radius relations. We therefore plot in Fig. \ref{fig:histosr}  the distribution of the radii for nine mass intervals, ranging from super-Earth planets to objects in the brown dwarf regime. In each panel, the radii of planets with semimajor axes of $0.1<a/$AU $\leq3$ are included. Planets reaching the feeding limit are not included here, as their evolution is  most affected by mechanism we do not model (e.g. evaporation). The outer border is chosen so that the distributions do not become too  dominated by planets at even larger distances, facilitating comparison with observational data. 

\begin{figure*}
\begin{center}
\begin{minipage}{0.33\textwidth}
\includegraphics[width=1.0\textwidth]{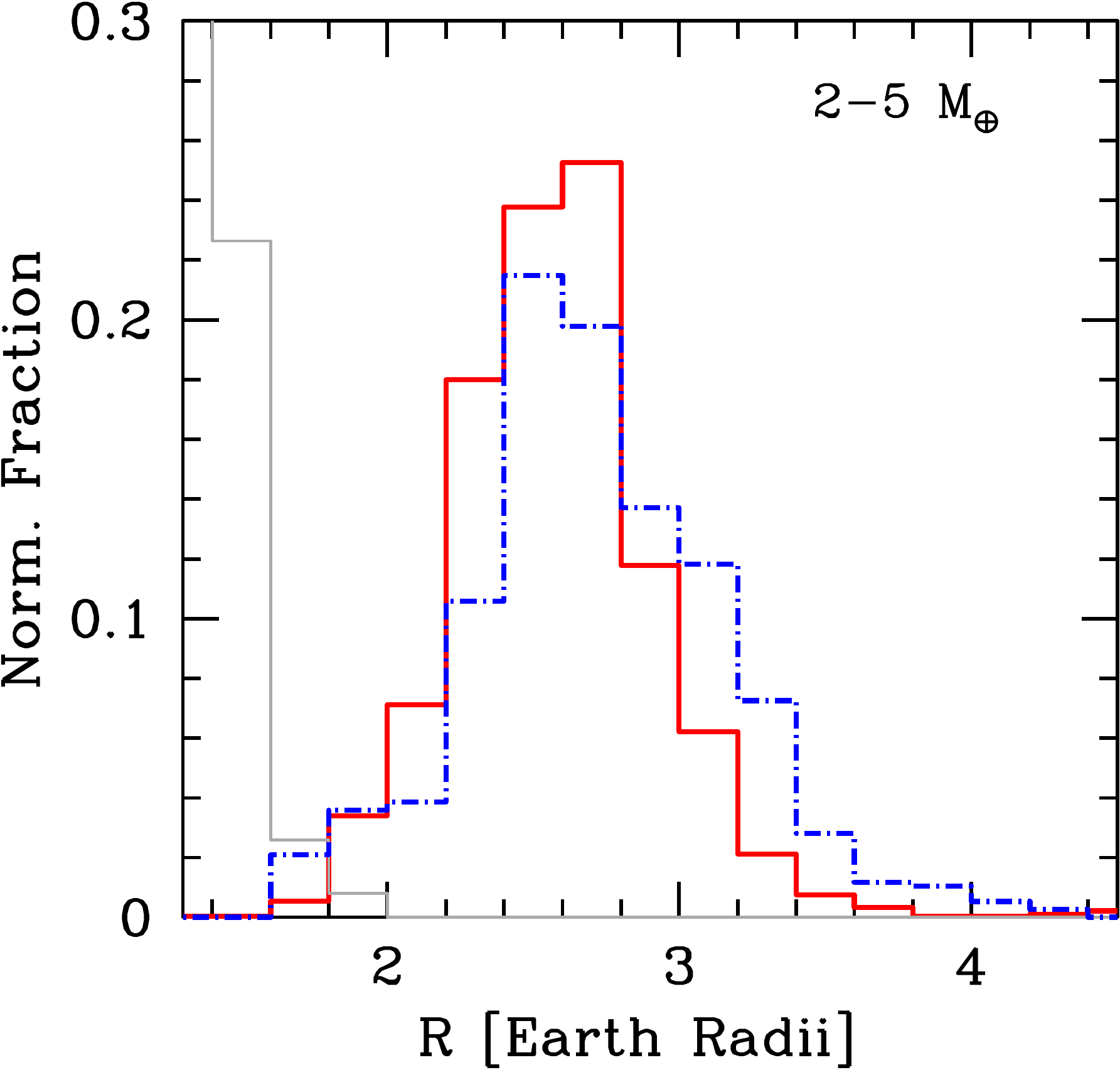}
\end{minipage}\hfill
\begin{minipage}{0.33\textwidth}
\includegraphics[width=1.0\textwidth]{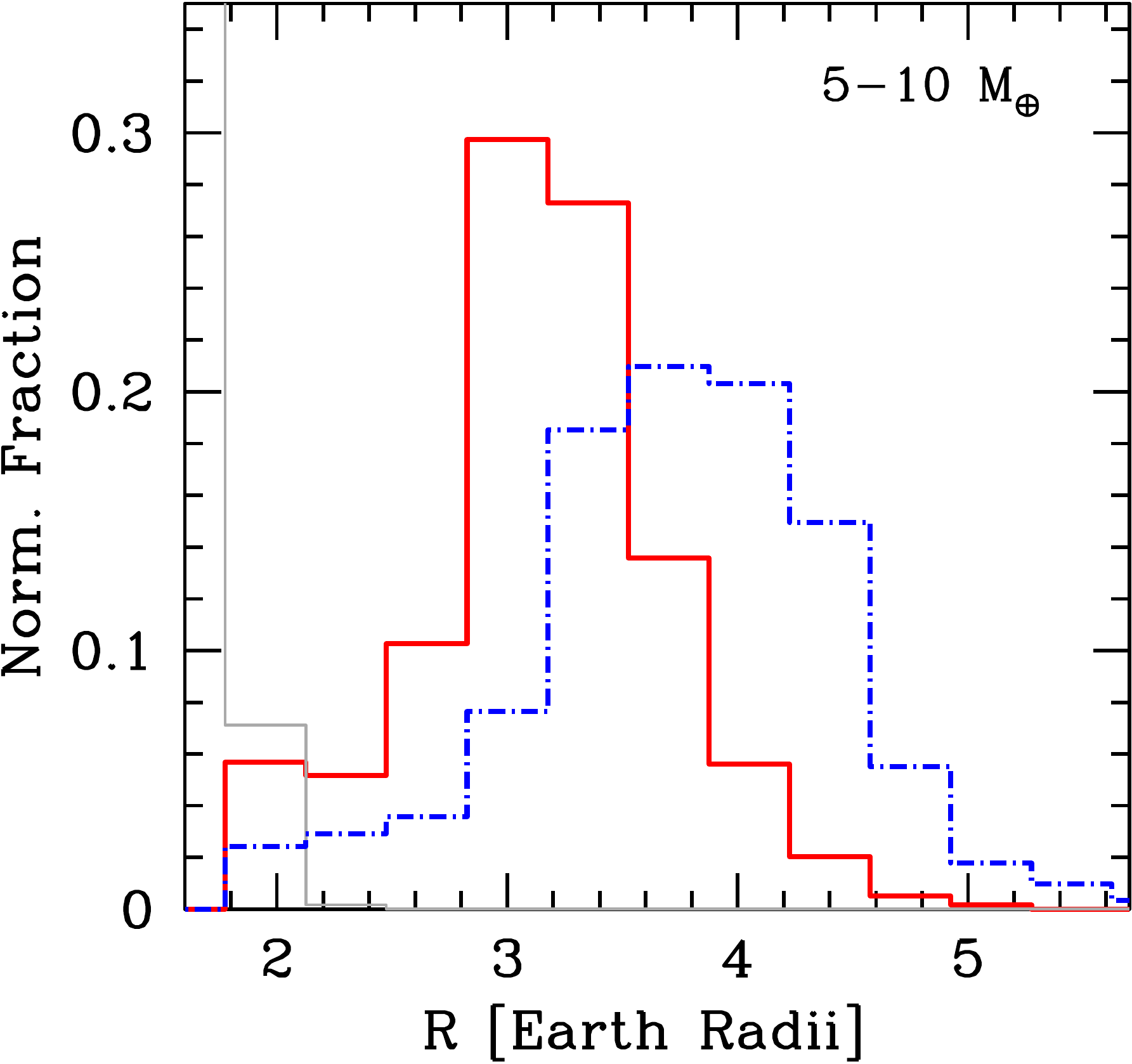}
\end{minipage}\hfill
\begin{minipage}{0.33\textwidth}
\includegraphics[width=1.0\textwidth]{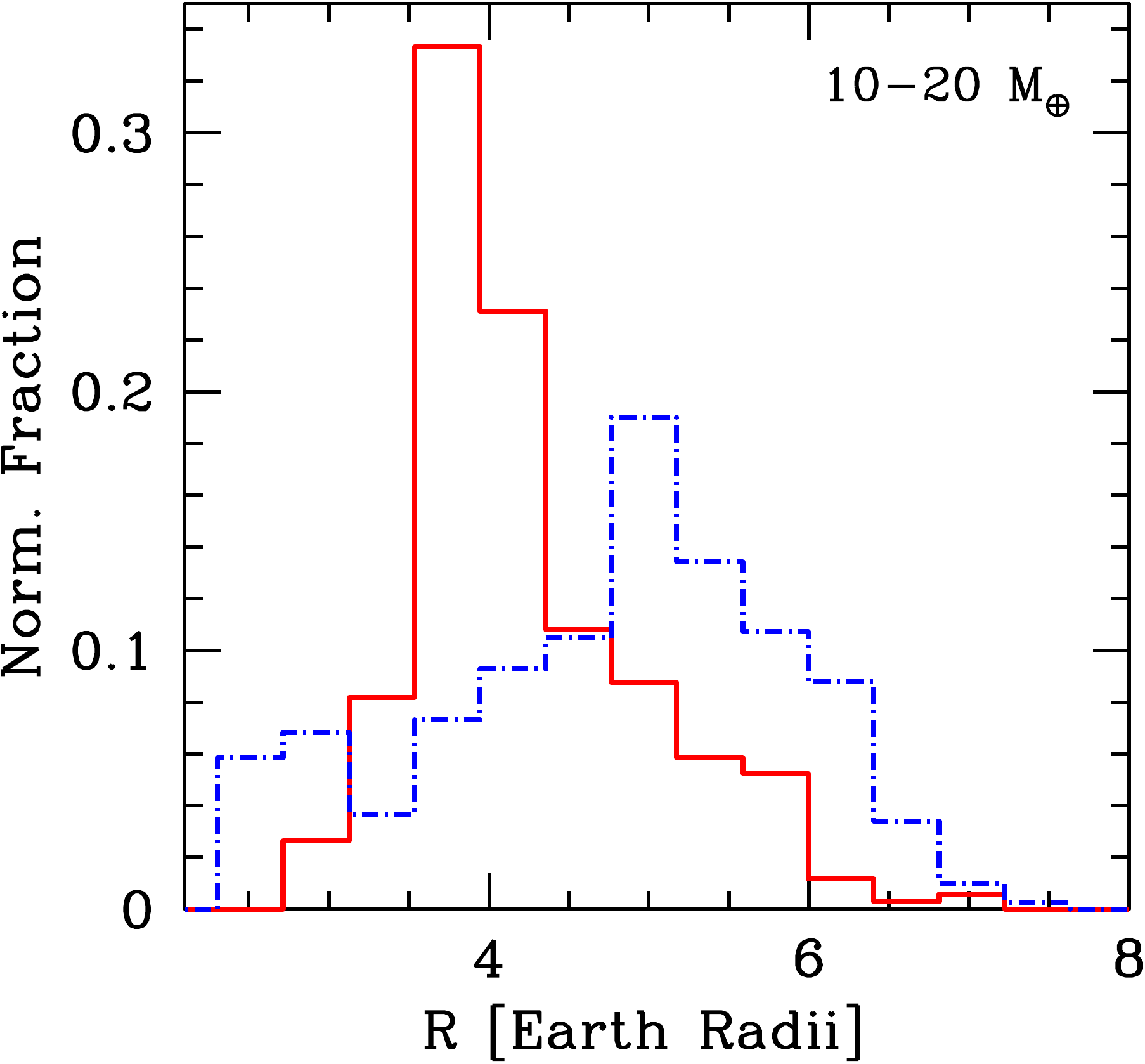}
\end{minipage}\hfill
\begin{minipage}{0.33\textwidth}
\includegraphics[width=1.0\textwidth]{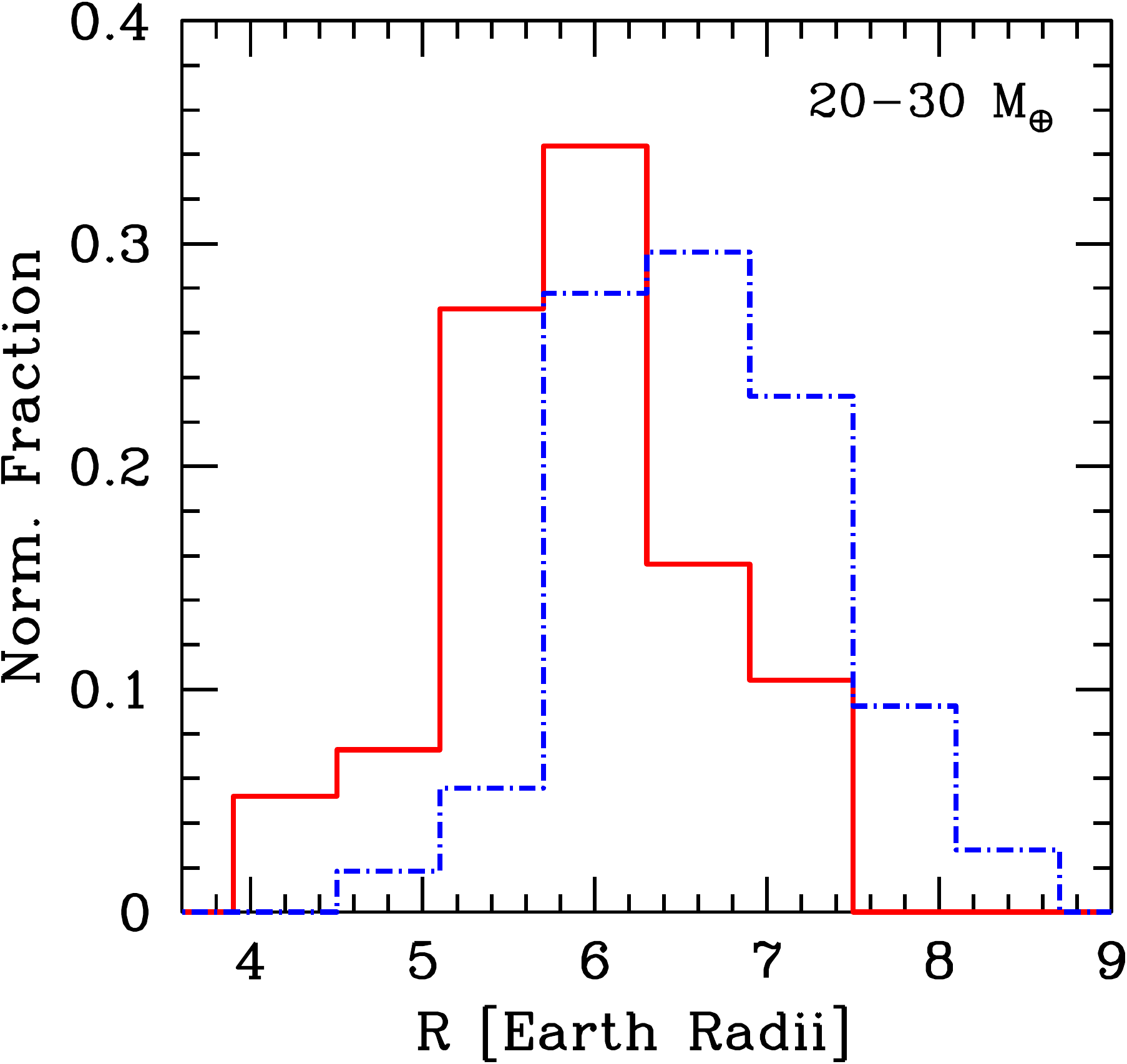}
\end{minipage}\hfill
\begin{minipage}{0.33\textwidth}
\includegraphics[width=1.0\textwidth]{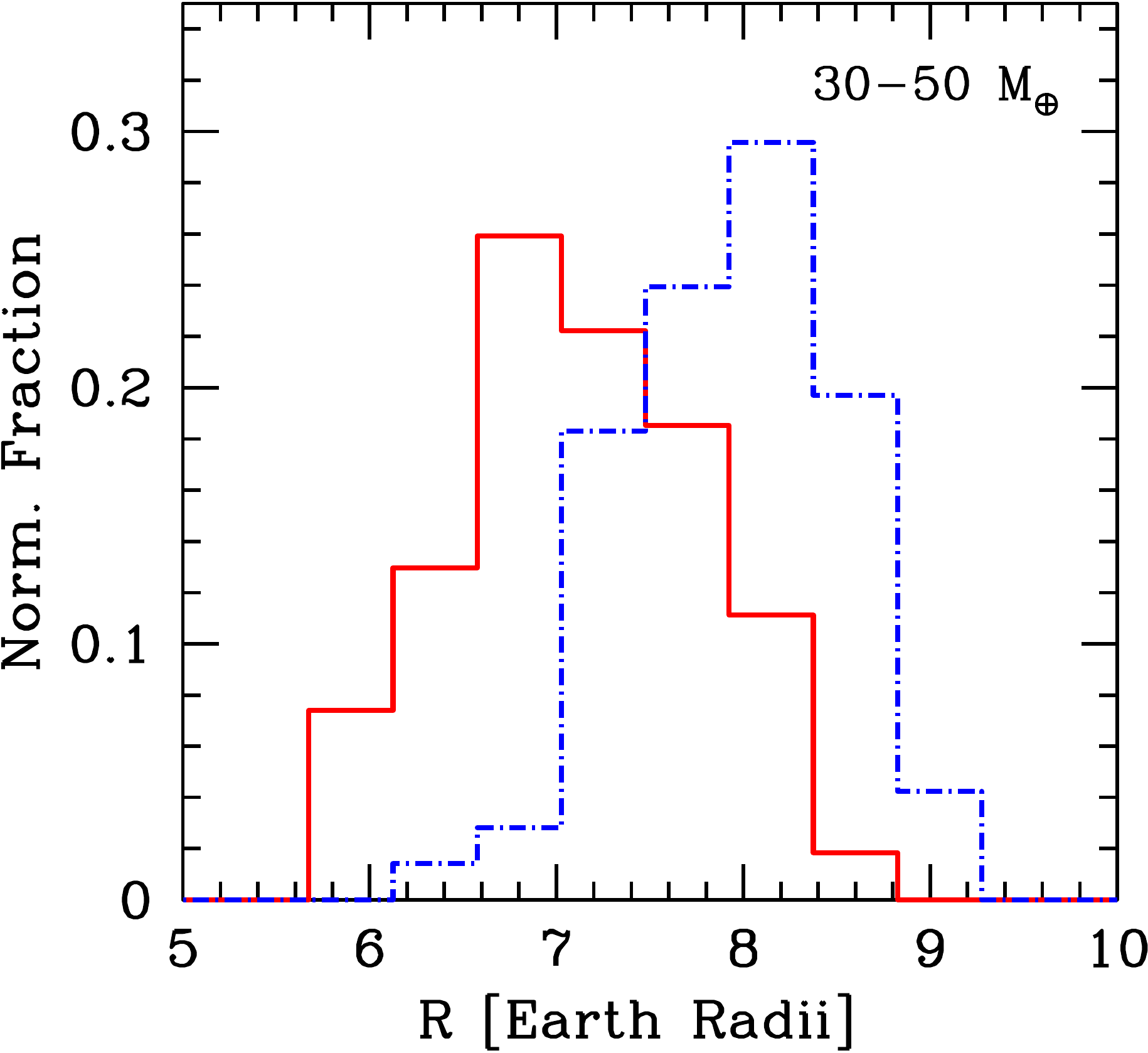}
\end{minipage}\hfill
\begin{minipage}{0.33\textwidth}
\includegraphics[width=1.0\textwidth]{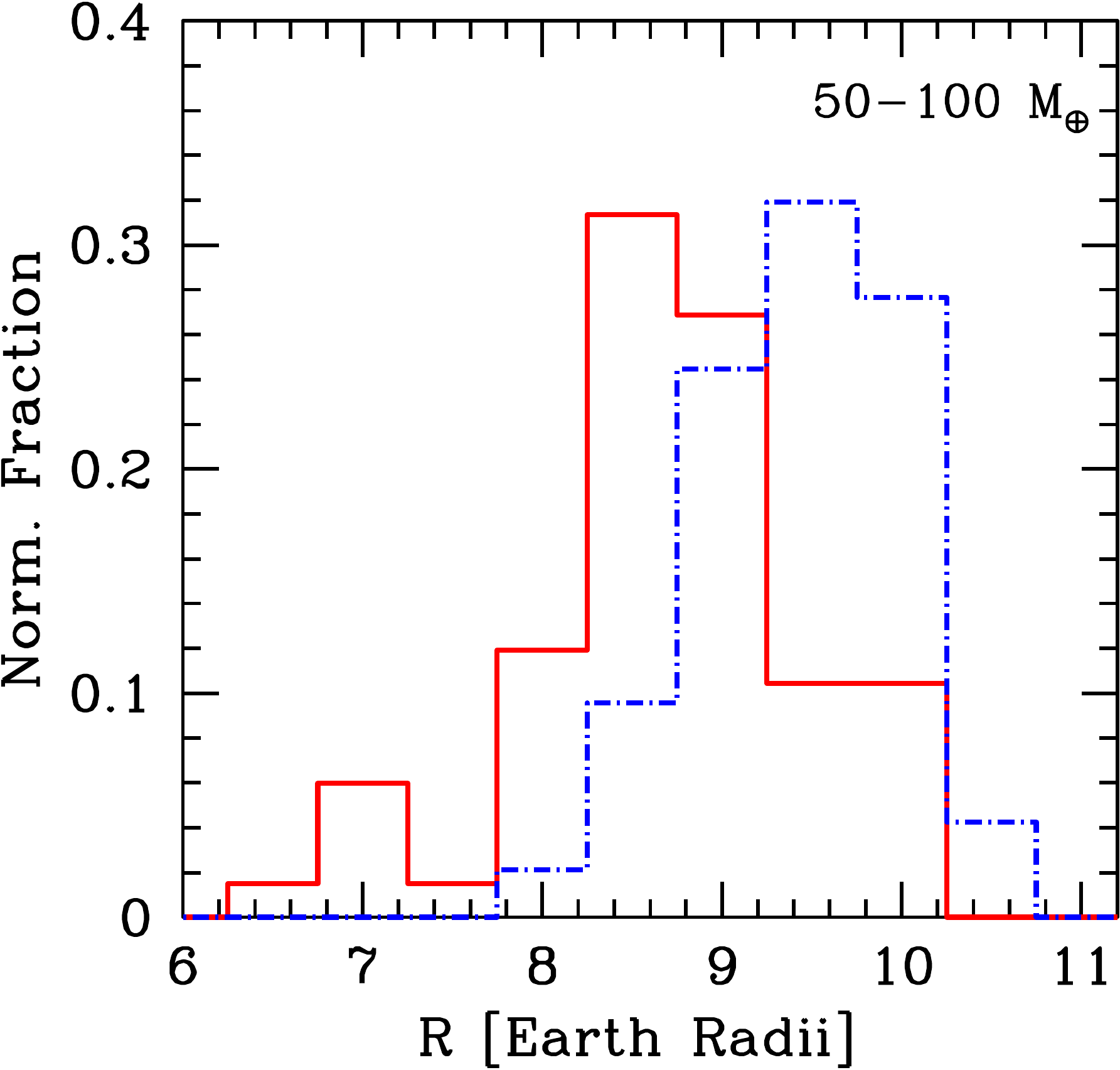}
\end{minipage}\hfill
\begin{minipage}{0.33\textwidth}
\includegraphics[width=1.0\textwidth]{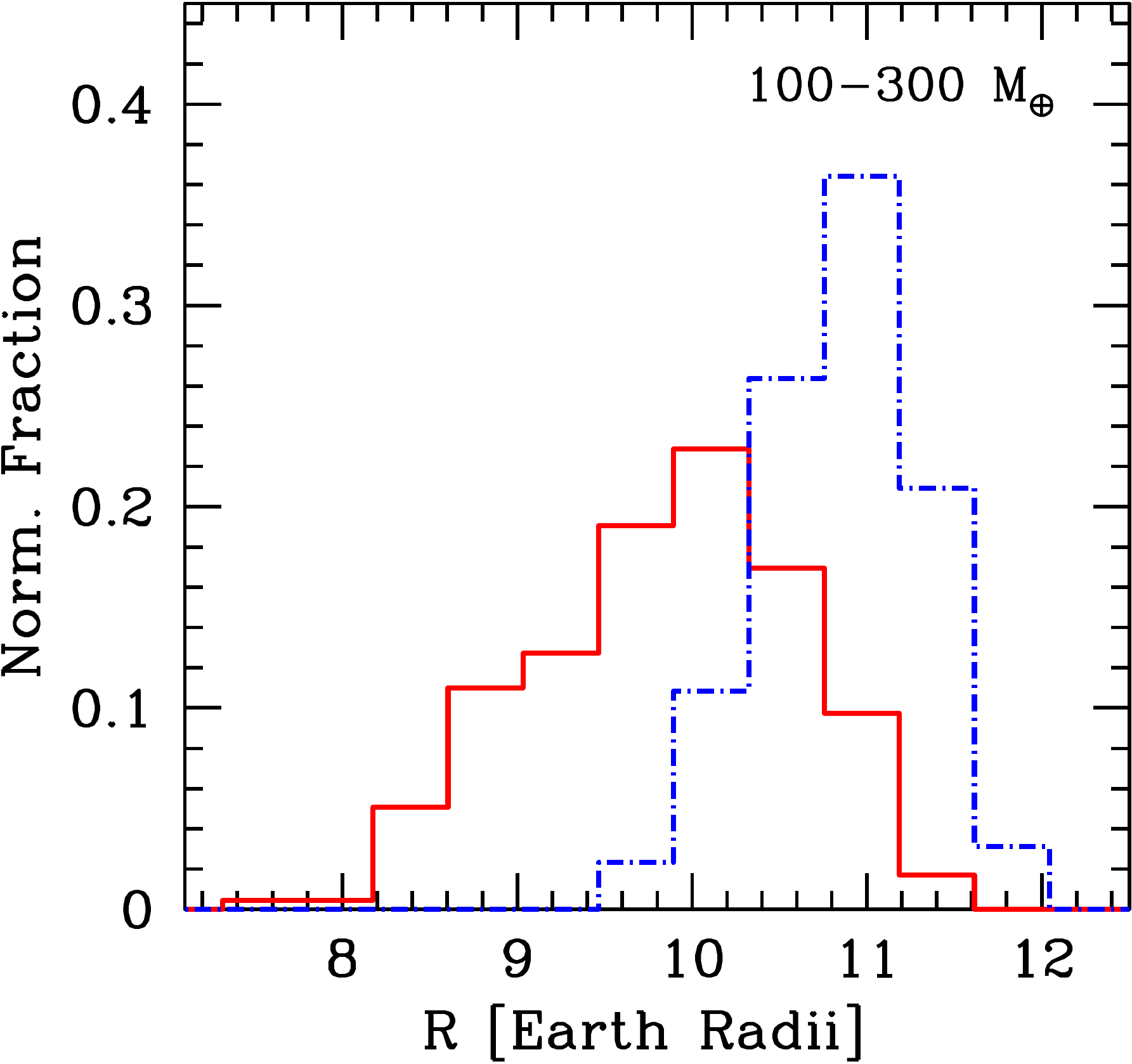}
\end{minipage}\hfill
\begin{minipage}{0.33\textwidth}
\includegraphics[width=1.0\textwidth]{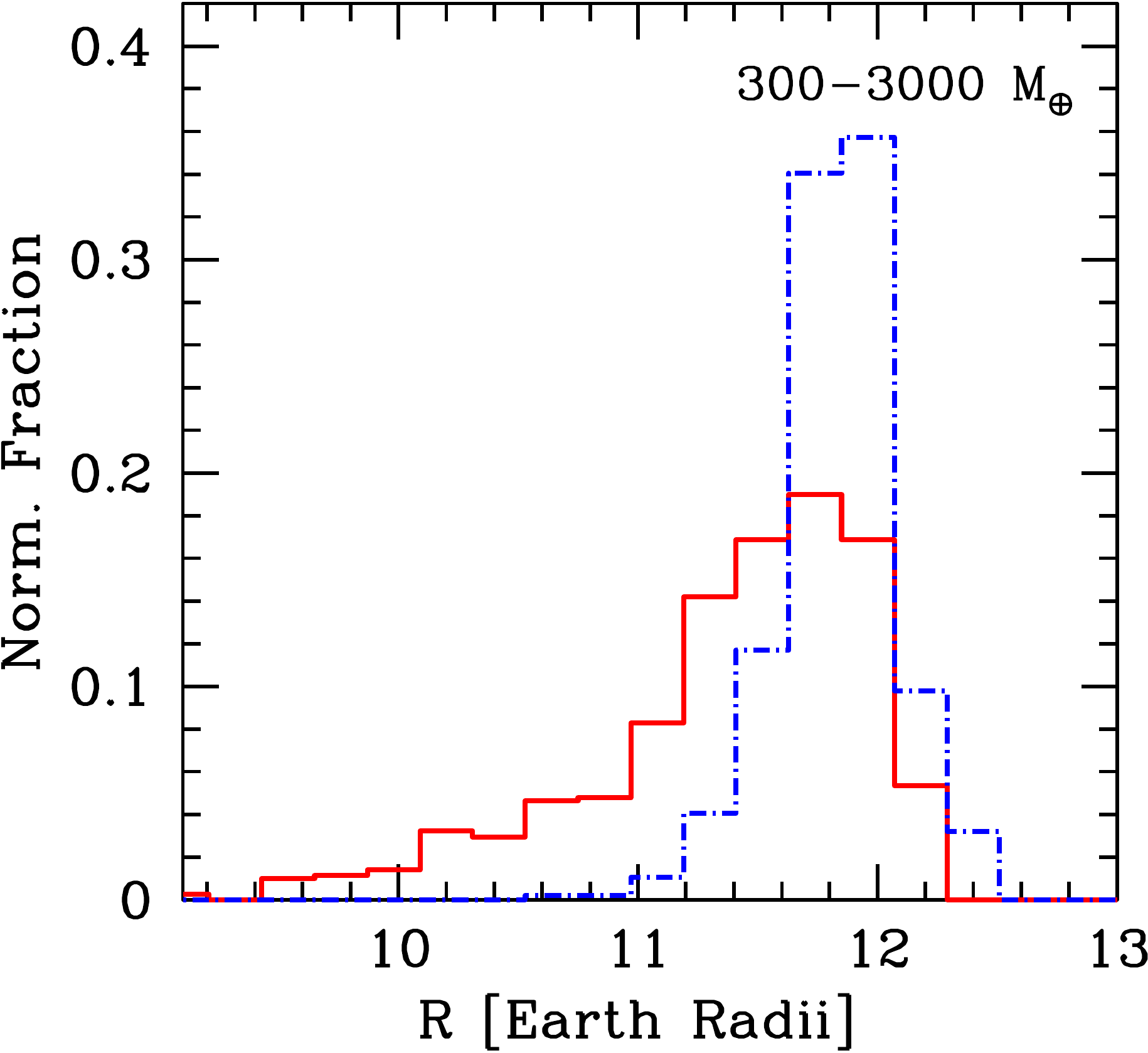}
\end{minipage}\hfill
\begin{minipage}{0.33\textwidth}
\includegraphics[width=1.0\textwidth]{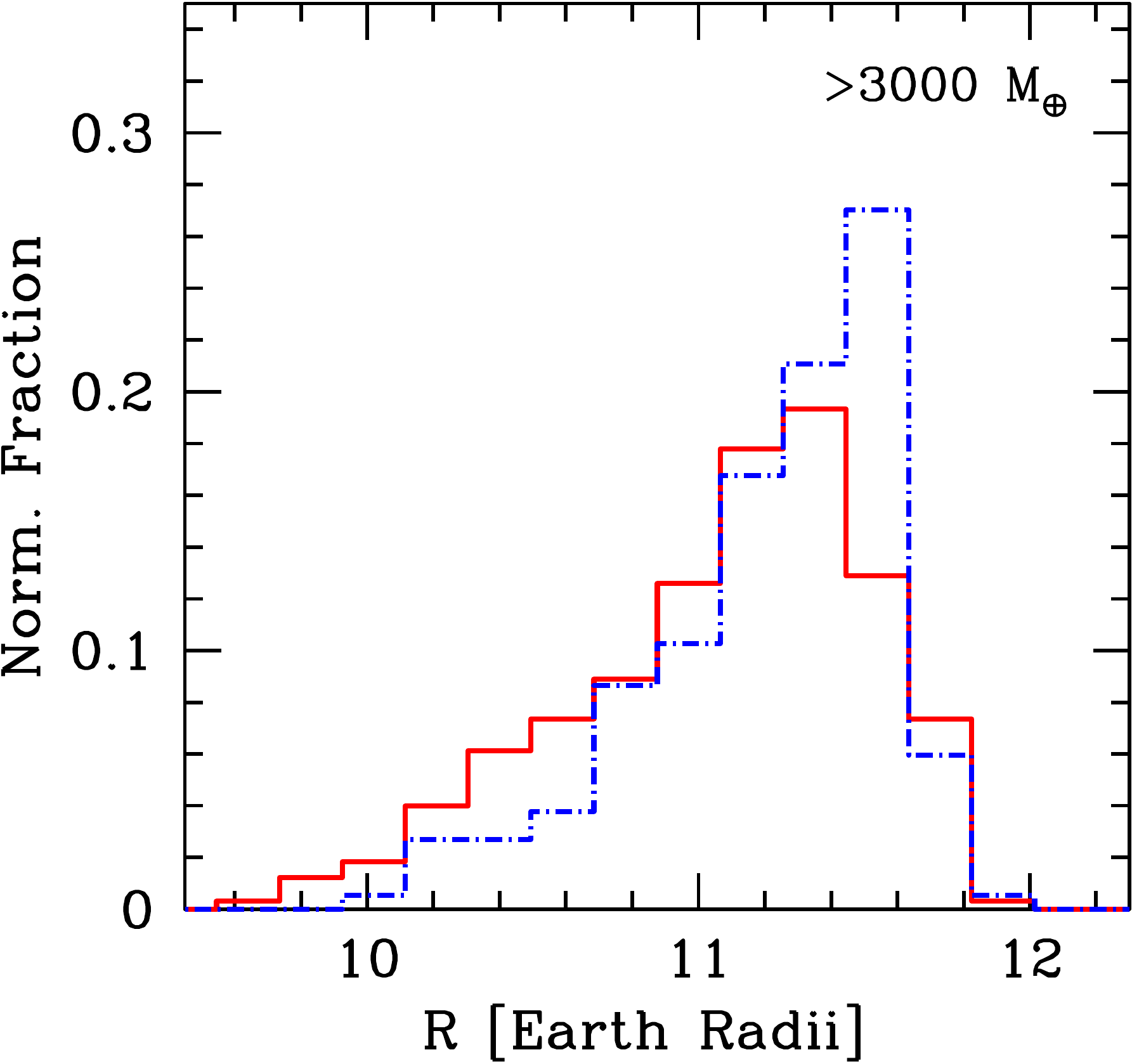}
\end{minipage}\hfill
\caption{Distribution of  radii of planets with primordial H$_{2}$/He envelopes for nine mass domains as indicated in the plot. Synthetic planets with $0.1<a\leq3$ AU are included.  The red solid line is the nominal population, while the blue dashed-dotted line is a population where  isothermal type I migration rates reduced by a factor 10 were used. In the panels showing the lowest mass domains, the fine gray lines show the distribution of the core radii. The age of the population is 5 Gyrs.}\label{fig:histosr}  
\end{center}
 \end{figure*}

In each panel, the distribution for the nominal population is shown. Additionally, we here also plot the radius distribution of another, non-nominal population. This population was calculated in an identical way as the nominal one, except that instead of the new type I migration model (Sect. \ref{typeImodel}), we  use the migration rates derived by Tanaka et al. (\cite{tanakatakeuchi2002}) for a locally isothermal disk, reduced by a global efficiency factor $\f1=0.1$. The comparison of the results for the two populations allows to quantify the impact of an aspect of the planet formation model that is still not very well understood.  

In the first two panels showing the lowest mass intervals, we also plot the core radius distribution. One sees that the total radii are typically much larger than the core radii, despite the fact that these planets are dominated by solids, as can be seen in Fig. \ref{fig:mrobs}. The effect that an H$_{2}$/He envelope of e.g.  1\%  in mass around a super-Earth  planet strongly increases the radius is well known (e.g. Rogers et al. \cite{rogersbodenheimer2011}; Fig. \ref{fig:mrrogerscomp}). We comment that the low grain opacity which is assumed during the formation phase ($\fopa=0.003$, see Mordasini et al. \cite{mordasiniklahr2011}) is responsible that already cores of e.g. 5 $\mearth$ can accrete envelopes with masses up to $\sim15\%$ of  the total mass during the lifetime of the protoplanetary disk.  

Table \ref{tab:radstat} lists the numerical values of the mean, standard deviation and skewness of the radius distribution of the different mass bins. A positive (negative) skewness indicates that  an asymmetric tail is extending to larger (smaller) radii, while a distribution which is approximately Gaussian has a vanishing skewness. 

\begin{table}
\caption{Statistical properties of the radius distribution (in Earth radii) of planets with a primordial H$_{2}$/He envelope at an age of 5 Gyrs.}\label{tab:radstat}
\begin{center}
\begin{tabular}{l c c c }
\hline \hline
Mass range $\mearth$ & mean & std. deviation & skewness \\ \hline
2-5     & 2.58 & 0.36 &  1.21           \\
5-10   & 3.15 & 0.55  & -0.12 \\    		      
10-20 & 4.23 & 0.74  &  1.04 \\  
20-30 & 5.89 & 0.75  &  -0.09  \\   		      
30-50 & 7.13 & 0.64  &  -0.01 \\
50-100 & 8.69 & 0.76  & -0.52 \\
100-300 & 9.85 & 0.74  & -0.36 \\
300-3000 & 11.39 & 0.58  & -1.15 \\
$>$3000 &  11.06 & 0.46  & -0.72 \\ \hline    	       
\end{tabular}
\end{center}
\end{table}%

The histogram and the table show that the mean is increasing with mass except for the highest mass range, as expected. Note that the mean values listed here only agree in an approximative way with the mean radius $\bar{R}$ found with eq. \ref{eq:eqMR}. This is due to two reasons: first, the two values were derived with a different ansatz (fit over the entire population versus  mean in a given mass domain), and second, the semimajor axis domains of the planets included in the analyses are different, too. The table  shows that the standard deviations can be significant, as also the total spread of the radii. 

Regarding the skewness, one sees that a clear pattern only develops for the largest masses. In the mass bin going from 300 to 1000 $\mearth$, the clearly negative skewness is due to the general shape of the planetary mass-radius relationship (see Fig. \ref{fig:mrobs}), and the fact that some planets in this mass regime contain so many solids that this has a non-negligible effect on the total radius. For the largest mass bin ($M>3000 \mearth$), the negative skewness is in contrast caused by the decrease of $R$ with $M$ at such masses.

\subsubsection{Impact of the migration model}
While the distributions of the two populations are to first order similar, we note that in almost all panels, the distribution of the non-nominal population (isothermal type I migration) is shifted towards larger radii. This means that the planets in the non-nominal population have in general a smaller heavy element fraction $Z$. This can be understood from the different migration models:  the new type I migration model allows for both in- and outward migration. In some cases, this means that the net inward migration is smaller, saving in this way planets from getting close to the star. This does however not mean that the total annulus in the disk visited by the planet is small, as the absolute migration rates are high also in the non-isothermal regimes, at least as long as a planet gets not captured into a convergence zone (cf. Lyra et al. \cite{lyrapaardekooper2010}; Mordasini et al.  \cite{mordasinidittkrist2011}). Since we follow Pollack et al. (\cite{pollackhubickyj1996}) in assuming low  planetesimal random velocities, the total amount of heavy elements that a planet can accrete is proportional to the annulus of the disk swept by it. This annulus is larger for the non-isothermal migration at full strength than for the isothermal migration reduced arbitrarily by a factor 10. This leads to the difference in the heavy element fraction.
Only for the planets more massive than 3000 $\mearth$, the difference in the heavy element content between the two populations has nearly no visible consequences any more for the total radius. These planets are too strongly dominated by gas.

When comparing the distributions found here with observations, one should keep in mind that the lower the mass range, the larger the uncertainties. Planets with a mass less than $\sim10\mearth$ will be affected in the strongest way by the limitation that we do not model any growth process once the protoplanetary disk is gone, only consider primordial atmospheres, and neglect evaporation, compositional changes due to outgassing  and non-solar opacities.

\subsection{Core and total radius}\label{coreandtotalradius}
In Sect. \ref{sect:variablecoredensity} we studied the radius of the solid cores of the planets as a function of mass, composition and external pressure. Figure \ref{fig:rcorertot} shows  besides the normal mass-radius diagram also the mass-core radius relationship. Lines indicate the theoretical $M-R$ for planets without a gaseous envelope.

\begin{figure}
\begin{center}
\includegraphics[width=\columnwidth]{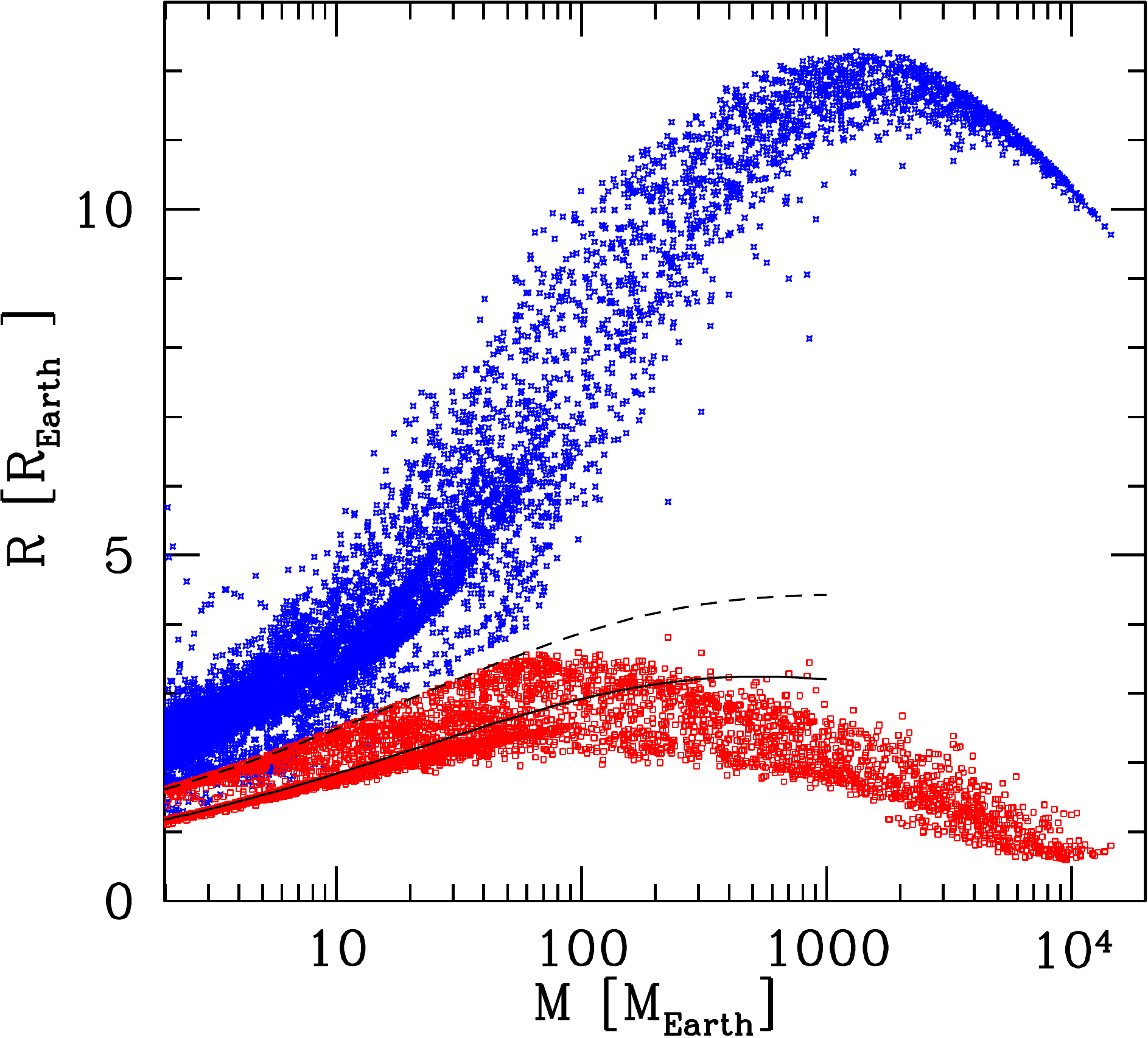}
\caption{Total and core radius as a function of mass for all planets with $M>2 \mearth$ at 5 Gyrs. The total radius is shown by blue, diagonal crosses, while the core radius is shown by red, open squares. The dashed line is the $M-R$ relation for solid planets with an ``icy'' composition (approximative solar composition beyond the iceline with 75\% ice, 17\% silicates and 8\% iron). The solid line is for solid planets with an Earth-like ``rocky'' composition   ({2:1 silicate-to-iron ratio}). These two lines show the radius if no external pressure is acting.  }\label{fig:rcorertot}
\end{center}
\end{figure}

Concerning the core radius of low-mass planets, we see that at low total masses ($M\lesssim 10 \mearth$), the core radii cluster close to two groups, which are found at (or slightly below) the dashed and solid lines which indicate a purely ``icy'' and ``rocky'' composition, respectively. This is expected, as the composition of the two classes of solids is the same in all simulations and corresponds to a solar-like composition (see Sect. \ref{sect:rcoremass}). In reality, due to the fact that stars do not have simply a scaled solar composition (e.g. Delgado Mena et al. \cite{delgadomena2010}), there will be a spread both in the composition of refractory and volatile material in the planets (Bond et al. \cite{bondobrien2010}). Giant impacts can additionally modify the iron and water fraction (e.g. Marcus et al. \cite{marcussteward2009}). At these masses ($M\lesssim 10 \mearth$), there are however also some cores which lie between the solid and the dashed line. This are planets which have migrated from outside the iceline into the inner part of the disk, so that they have a mixed composition.  The frequency of such close-in ``ocean'' planets is an important and possibly observable diagnostics of the efficiency of migration. Some core radii lie slightly below the two lines. This comes from the fact that for such cases, the pressure exerted by the gaseous envelope is already sufficient to compress the core by a small degree. 

When comparing total and core radii if low-mass planets, we see that there are some cases which have a total radius which lies between the solid and the dashed line. This are planets which are made of a ``rocky'' core and tenuous amounts of gas, resulting in a radius smaller than the one of an atmosphere-free icy planet. This is a manifestation of the well known degeneracy of the mass-radius relation for low-mass planets. The combination of mass and radius measurements with spectroscopy (e.g. with the EChO mission, Tinetti et al. \cite{tinetti2011}) is therefore very import for a better understanding of this part of the planetary mass-radius relationship.  

At masses between 10 to 100 $\mearth$, the effect of the compression of the core due to a massive envelope becomes clearly visible, so that the points indicating the actual core radii diverge from the lines indicating radii of planets without significant envelopes.  At very high masses  ($M\gtrsim 10 \mj$) cores get very strongly compressed, to sizes that are smaller than the core radii of the low-mass planets ($\rcore\lesssim 1 \rearth$). Clearly, it is  questionable whether a solid core can  survive for 5 Gyrs at the extreme conditions found in the center of such planets without being eroded as we currently assume for simplicity. Recent results on the solubility of ice in conditions as found in Jupiter's center (Wilson \& Militzer \cite{wilsonmilitzer2010}) indicate that at least for this planet, core erosion is possible. It is therefore of interest to take such effects into account. On the other hand, it is also clear that the higher the total mass, the smaller the relative fraction it contains in solids as indicated by the color coding in  Fig. \ref{fig:mrobs}. This plot shows that planets with masses higher than $\sim 10 \mj$ have a core mass fraction $Z$ of 5\% to less than 1\%, decreasing with total mass. This means that the total radius is only slightly influenced by the core mass. This is reflected in the vertical width of the $M-R$  relationship in Fig. \ref{fig:mrobs}, that becomes very narrow at such large masses.  This  finding could partially be changed if we would include different opacities (e.g.  Burrows et al. \cite{burrowsheng2011}).

\section{Summary}\label{sect:summaryconclusion}
In the first part of this paper, we have described several improvements and extensions of our combined formation and evolution model.

\subsection{Planet module}
First, we have included the  calculation of the radius of the solid core as a function of its mass, composition and (for giant planet cores) external pressure by solving the internal structure equations for a differentiated planet. We have compared our results with previously published studies (e.g. Seager et al. \cite{seagerkuchner2007}) and found very good agreement (Sect. \ref{sect:rcoremass}). We can therefore  make relatively accurate predictions for the radii of core-dominated planets, necessary for comparison with high-precision transit missions like  \textit{Kepler}. We have further found that the variable core density only has a small impact on the formation phase of a giant planet, but that it needs to be considered for correct total radii of (giant) planets at late times. These radii are essential to estimate the heavy element content of a giant planet (Sect. \ref{result:coredensity}). 

Second, we have developed a simple model for the core luminosity due to the decay of long-lived radionuclides. We assume  a chondritic composition of the mantel and take into account the temporal decay of the nuclides (Sect. \ref{sect:upgradelradio}). We have  demonstrated for of a low-mass, close-in super-Earth planet with a $\sim1\%$ primordial H$_{2}$/He atmosphere, that the radioactive decay becomes the dominant intrinsic heat source at late times. This leaves traces in the temporal evolution of the radius of the planet.  We have compared our results  with Rogers et al. (\cite{rogersbodenheimer2011}), and found good agreement (Sect. \ref{result:lradio}). This means that we can address in population synthesis calculations  the radii of  low-mass planets with tenuous primordial envelopes. This is an important subject as low-density, low-mass planets are currently detected in large numbers (Borucki et al. \cite{boruckikoch2011}), and because this question is linked to the habitability of low-mass planets.

\subsection{Disk module}
We have presented  improvements of  the protoplanetary gas disk module which is a 1+1D $\alpha$ model including the effects of stellar irradiation and photoevaporation. We now use more realistic initial conditions (Andrews et al. \cite{andrewswilner2009}), a free outer radius of the disk allowing spreading and shrinking and an updated photoevaporation model taking into account external and internal photoevaporation (Sect. \ref{standardalphamodel}). With this model we studied the stability against self-gravity and found that disks with masses up to {0.091} or {0.16} $\msun$ (without and with stellar irradiation, respectively) are stable for the characteristic radius and initial surface density slope we assume. We have also found that the radius of the lowest Toomre parameter  in these disks occurs around 25 AU, close to the location we determined analytically (Appendix \ref{result:stabilitydisk}).

\subsection{Planetary mass-radius relationship}
{In the second part of the paper, we have used the updated model  to conduct planetary population synthesis calculations. In contrast to earlier syntheses, we can now not only calculate the planetary semimajor axis-mass diagram, but also the mass-radius diagram. Also the mass-luminosity and semimajor axis-luminosity diagrams are now a self-consistent result of the model that will be presented in future work. The planetary mass-radius diagram is of similar importance as the semimajor axis-mass diagram as it allows to make a first characterization of a planet via its bulk  composition.}

{We present how the planetary mass-radius relationship for planets with primordial H$_{2}$/He envelopes (the only type of atmosphere we currently model) comes into existence during the formation phase (when planets grow) and how it then evolves over Gigayear timescales during the evolutionary phase at constant mass. During the presence of the nebula (Sect. \ref{sect:mrrformation}), two regimes exist for the radii: low-mass, subcritical cores have very large radii comparable to the Hill sphere (or Bondi) radius as their envelopes are attached to the background nebula, whereas supercritical cores have massive, collapsed envelopes with radii of a few $\rj$. Most planets however do not trigger runaway gas accretion but instead undergo a phase of rapid contraction when the nebula disappears.}

{During the evolutionary phase (Sect. \ref{sect:mrduringevo}), for $t\gtrsim0.5$ Gyrs, the basic shape of the mass-radius relationship is similar to an elongated ``S'' which can be understood from the fundamental principles of the core accretion model, and the EOS. Due to their long Kelvin-Helmholtz timescale, low-mass planets cannot bind massive gas envelopes, whereas super-critical cores necessarily must trigger runway gas accretion (at least if they form during the lifetime of the protoplanetary disk). This naturally leads to two ``forbidden'' regions in the $M-R$ plane: one at low masses with large radii, and one at high masses with small radii (Sect. \ref{sect:generalshapemrr}). }

{The radius of a planet for a given composition and age can be calculated also in purely evolutionary calculations. The spread in possible associated radii for a given mass interval (i.e. the vertical width of the mass-radius relationship), and the distribution of the radii in a specific mass range can however only be obtained from combined formation and evolution calculations.  We find that, depending on the mass domain, there is a large diversity in the associated radii. At a mass of $\sim 40 \mearth$ for example, radii  vary by a factor $\sim3$. This is mainly due to different bulk compositions, reflecting different formation histories. We present the synthetic radius distribution for various mass intervals in Sect. \ref{sect:rdistmassranges}. }

{When comparing the synthetic $M-R$ relationship with the observed one (Sect. \ref{sect:compobsmrr}), one finds good agreement with the planets of the Solar System and for actual exoplanets with $a>0.1$ AU, which corresponds to the minimal orbital distance of the synthetic planets. All exoplanets with  $a\geq0.1$ AU and the planets of the Solar System fall in the $M-R$ envelope populated with synthetic planets (Fig. \ref{fig:mrobs}) except for one case (KOI-423b).  We derive approximate heavy elements contents for the individual observed planets (Sect. \ref{sect:individobj}). The general trend is that the fraction of iron, silicates and ices decreases with increasing mass, but there is significant spread for individual objects. }

{We then show the synthetic planetary radius distribution at 5 Gyrs which is a fundamental result of population synthesis calculations (Sect. \ref{sect:planetaryrdist}). The radius distribution is bimodal and characterized by a  strong increase towards small radii, and a second, lower local maximum at a radius of about $1\rj$. The increase towards small radii is a consequence of the increase of the planetary mass function towards low $M$, and the fact that low-mass planets cannot accrete much nebular gas. This means that low-mass planets also have small radii. The second local maximum at about  $1\rj$is due to a fundamental property of matter (degeneracy) causing the radius to be nearly independent of mass in the giant planet domain. Due to this, planets from a wide mass range all fall in the same radius bin, leading to the local maximum at about 1 $\rj$.}

{A comparison of the synthetic radius distribution with the observations of the \textit{Kepler} satellite (Howard et al. \cite{howardmarcy2011})  shows a good agreement for $R\gtrsim2\rearth$, but a divergence for smaller radii (Fig. \ref{fig:Kepler}). While our quantitative results for low radii are to be taken with great caution, this could indicate that for $R\gtrsim2\rearth$, the radius distribution can be well relatively described with planets with primordial  H$_{2}$/He atmospheres as in the model, while at smaller radii, planets of a different nature dominate. This result seems to be in good agreement with analyses combining data from \textit{Kepler} and RV surveys (Wolfgang \& Laughlin \cite{wolfganglaughlin2012}; Gaidos et al. \cite{gaidosfischer2012}). We predict that in the next few years, \textit{Kepler} should find the second, local maximum at about 1 $\rj$. We also predict that the typical radii of planets with $10\lesssim M/\mearth\lesssim1000$ will decrease with distance. This stems from a positive correlation of the semimajor axis and the typical fraction of heavy elements in a planet. The reason for this is that the amount of solids in an annulus increases for our  disk surface density profile (Sect. \ref{sect:impacta}).}

{To facilitate comparison with observations, we use  the analytical expression of Traub (\cite{traub2011}) to derive the mean radius as function of mass  in the synthetic population (Sect. \ref{sect:meanrasfctofmass}), and study the mass-radius power law index $\beta$ ($M\propto R^{\beta}$). The value of $\beta$ is between 0.3 to 0.4 for planets with $M\lesssim100\mearth$. At higher masses, $\beta$ decreases to reach zero at about 4 $\mj$, and negative values at even higher masses. As we only consider primordial H$_{2}$/He atmospheres, it is expected that at low masses, the observed mass-radius relation will diverge from the synthetic result.}

\section{Conclusion}\label{sect:conclusionsfin}
With the extensions and improvements presented in Paper I and this study, we are now  able to characterize extrasolar planets  in their most important physical quantities like semimajor axis, mass, composition, radius and luminosity from one single model. These quantities are calculated in a self-consistent way for the coupled formation and evolution of a planet during its entire life, from a tiny seed embryo in the protoplanetary disk to a gigayear old planet. Our goal is  to have in the end synthetic populations that can be compared directly with all important observational techniques used to find and characterize exoplanets. We believe that a better theoretical understanding of the very complex patterns  in the various data sets (like transiting planets of very different compositions, systems of compact close-in low-mass planets, cold Neptunes beyond the iceline or massive, directly imaged giant planets at large distances)  can best be developed by comparing theoretical results to a combination of all these data sets.  {In the current case, the combined results of transit and radial velocity measurements allow to conclude that first, for $a>0.1$ AU, the actual and the synthetic mass-radius relationship are in good agreement, and second that for radii larger than roughly 2 $\rearth$, the synthetic radius distribution is similar as the one detected by \textit{Kepler}. Finally, we find that the planetary radius distribution is bimodal, with the global maximum at low radii, and a second smaller maximum at about 1 $\rj$. We predict that \textit{Kepler} should detect this maximum in the next few years.}

\acknowledgements{Christoph Mordasini acknowledges the support as an Alexander von Humboldt fellow. We are thankful for the continuous support by the Swiss National Science Foundation. Yann Alibert thanks the European Research Council for the grant 239605.  We thank Willy Benz for fruitful discussions and an anonymous referee for helpful comments.}

\appendix

\section{Stability of irradiated $\alpha$ disks against self-gravity}\label{result:stabilitydisk}
In population synthesis calculations (e.g. Ida \& Lin \cite{idalin2004}; Mordasini et al. \cite{mordasinialibert2009a}) one uses distributions of (initial) protoplanetary disk masses in order to reflect the varying initial conditions for planet formation.

As discussed in Sect. \ref{sect:stabilityselfgrav},  massive, cold disk get self-gravitationally unstable, which would invalidate the usage of a classical $\alpha$ model with one constant  $\alpha$ across the disk. Ida \& Lin (\cite{idalin2004}) and Mordasini et al. (\cite{mordasinialibert2009a}) have therefore cut off the uppermost part of the observed disk mass distribution  with the argument that they cannot be stable.  The cut off was usually done at the (often assumed) disk mass  stability limit of about a tenth of the stellar mass, without a direct calculation of this stability limit.  Observationally, high disk masses can be mimicked by residual dust in the  remains of the envelope from which the star formed, contributing to the observed flux (Andrews \& Williams \cite{andrewswilliams2005}).

In this appendix, we use our upgraded disk model to determine the maximum stable disk mass for both irradiated and non-irradiated disk, and make some more general  remarks about the stability of irradiated $\alpha$ disks against the development of spiral waves, and of clumping.

For simplicity, we use for the initial disk profile (Eq. \ref{eq:newprofilegeneral}) a power law exponent $\gamma=0.9$ and  a characteristic radius $R_{\rm c}=30$ AU. Both values correspond to the maxima of the distributions observed by Andrews et al. (\cite{andrewswilner2010}). With these values, the initial gas surface density is
\beq\label{eq:newprofile}
\Sigma(r,t=0)=\Sigma_0\left(\frac{r}{5.2\, \mathrm{AU}}\right)^{-0.9}\exp\left[-\left(\frac{r}{30\,  \mathrm{AU}}\right)^{1.1}\right].
\eeq
It is clear that this procedure faces a difficulty: it implicitly assumes that we can use the observed dust grains to trace the mass and evolution of the gas. Especially grain growth could make this assumption partially invalid (e.g. Andrews \& Williams \cite{andrewswilliams2007}). 
\begin{figure*}
\begin{center}
\includegraphics[width=\columnwidth]{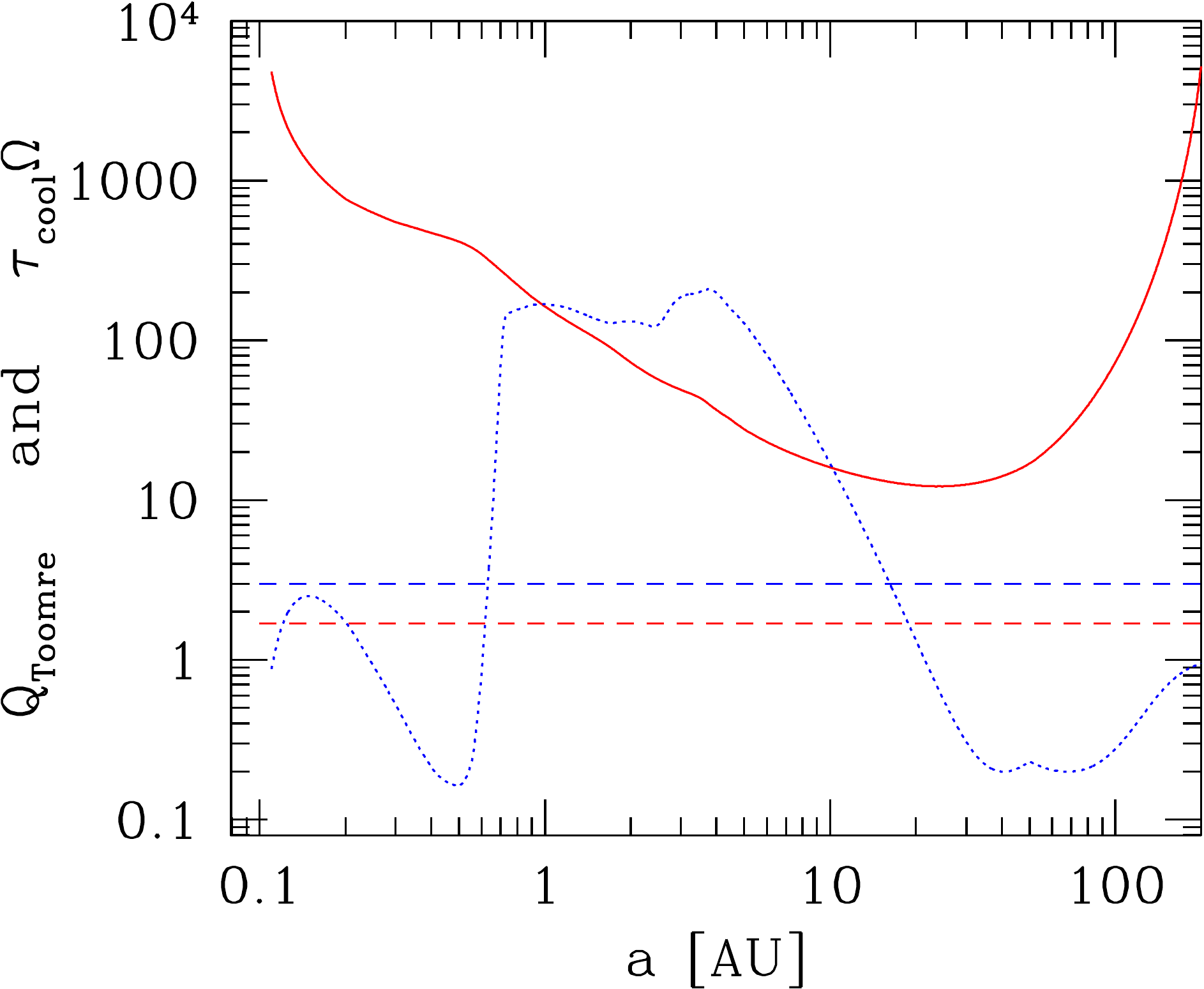}
\includegraphics[width=\columnwidth]{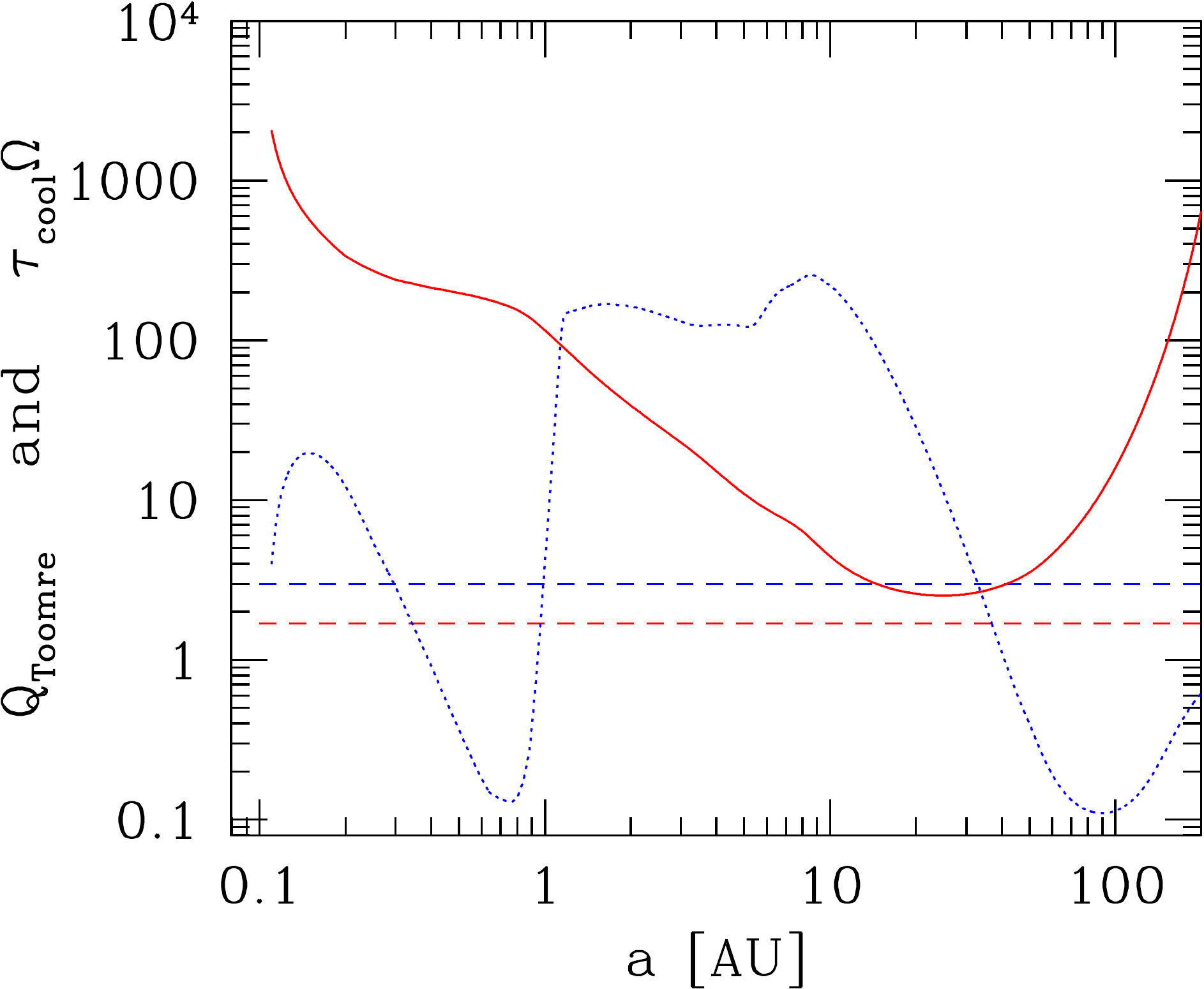}
\caption{{\footnotesize   Toomre parameter $Q_{\rm Toomre}$ (red solid line) and $\tau_{\rm cool} \Omega$ (blue dotted line) as a function of semimajor axis $a$, at a moment shortly after the beginning of disk evolution. The horizontal lines  correspond to the critical values of  $Q_{\rm Toomre,crit}${=1.7} and $\tau_{\rm cool} \Omega=3$, respectively. The left panel is for a disk with an initial mass $M_{\rm d}(t=0)=0.024 \msun$, while the right one corresponds to a higher mass, $M_{\rm d}(t=0)=0.11 \msun$.   }}\label{fig:qtcool}
\end{center}
\end{figure*}

Integration Eq. \ref{eq:newprofilegeneral} from $r=0$ to infinity using the parameters mentioned before, one finds a total initial disk mass (cf. Miguel et al. \cite{migueletal2011}) of
\beq\label{eq:totaldiskmassparams}
M_d(t=0)=0.012 \left(\frac{\Sigma_0}{100\, {\rm g / cm^{2}}}\right) M_\odot. 
\eeq
Thus, the MMSN of Weidenschilling (\cite{weidenschilling1977}, \cite{weidenschilling2005}), corresponds to $\Sigma_{0}\approx 200$ g/cm$^{2}$, while for Hayashi's (\cite{hayashi1981}) MMSN of 0.013 $\msun$, $\Sigma_{0}\approx108$ g/cm$^{2}$. It is likely that the true initial mass of the solar nebula was a few times larger than these minimal masses (Weidenschilling \cite{weidenschilling2005}). 

Note that we do not allow in the model accretion rates larger than $3\times 10^{-7}$ $\msun$/yr in order to avoid convection in the vertical direction, but reduce if necessary locally the initial $\Sigma$  in order  not to exceed this limit. Therefore, the initial gas mass for disks with $\Sigma_{0}$ larger than $\sim$1000 g/cm$^{2}$ are somewhat smaller than predicted by Eq. \ref{eq:totaldiskmassparams}. For $\Sigma_{0}=2000$ g/cm$^{2}$ the initial mass is for example about 20\% smaller than given by Eq. \ref{eq:totaldiskmassparams}. Note further that in contrast to earlier models, we can now include the total disk mass inside the computational domain, and not only the fraction contained in the innermost 30 AU (Mordasini et al. \cite{mordasinialibert2009a}).

\subsection{$Q_{\rm Toomre}$ and $\tau_{\rm cool} \Omega$ as  function of distance}
Figure \ref{fig:qtcool} shows $Q_{\rm Toomre}$ (solid red lines) and $\tau_{\rm cool} \Omega$ (blue dotted lines) as a function of distance from the star for two disks (see also Bell et al. \cite{bellcassen1997}) at $t\approx0$ (the moment we start the disk evolution). The left panel shows a 1-2 x MMSN disk ($M_{\rm d}$(t=0)=0.024 $\msun$, $\Sigma_{0}=200$ g/cm$^{2}$), while the right one shows a more massive disk with $M_{\rm d}$(t=0)=0.11 $\msun$ ($\Sigma_{0}=1000$ g/cm$^{2}$). Both examples are calculated including the effect of stellar irradiation on the disk temperature structure, $\mstar=1 \msun$, $\alpha=7\times 10^{-3}$  and the initial surface density profile discussed above. 

We see that for both disk,  $Q_{\rm Toomre}$ reaches a minimum at about 25 AU. For the MMSN disk, the minimum value is about 12, while for the more massive disk, it is 2.5. The critical value of  $Q_{\rm Toomre,crit}\approx${1.7} is not reached. Regarding the cooling, the particular structure of the curve is due to  opacity transitions. We see that in the inner parts of the disk, and outside 15-25 AU, there are parts of the disk which could in principle cool quickly enough so that $\tau_{\rm cool} \Omega<3$. The cooling timescale is  the time on which the disk would cool if we would suddenly switch off the heating mechanisms.  This is however meaningless for gravitational fragmentation due to the too high values of $Q$ everywhere in the disks. Therefore, fragmentation does not occur either. 

We have studied the overall minimal $Q_{\rm Toomre, min}$ occurring in a disks as a function of time, radius, and $\Sigma_{0}$ in order to determine the maximum $\Sigma_{0}$ which can be used. In these calculations we again assume $\alpha=7\times 10^{-3}$ and $\mstar=1 \msun$. It is found that the minimal $Q_{\rm Toomre}$ are always occurring very shortly after the start of the disk evolution (within some $10^{4}$ years), and that afterwards, $Q_{\rm Toomre}$ is always increasing. Thus, the minimal $Q_{\rm Toomre}$ are a direct consequence of the initial conditions (as expected), and  the disks evolve towards stability.  

\subsection{Radius of maximum instability}
The orbital distance where the overall minimal $Q_{\rm Toomre,min}$ are occurring correspond to about 20 AU for disks calculated without stellar irradiation, and, as also seen in the two examples in Fig. \ref{fig:qtcool}, to about 25 AU for disks with irradiation, independent of the disk mass as long as $M_{\rm d}(t=0)\lesssim 0.1 - 0.2 \msun$. For more massive disks, the distance of the minimal $Q$ increases. This result can be understood at least for the irradiated disks if we assume that in the outer parts, the temperature structure  {is given by the stellar irradiation only. We can consider two cases: First, an optically thin disk with} $T_{\rm mid}\propto r^{-1/2}$. In this case, the orbital distance $R_{\rm Toomre,min,t}$ where the minimal $Q_{\rm Toomre, min}$ occurs  is given for $\gamma<7/4$ as (using eqs.  \ref{eq:newprofilegeneral}, \ref{eq:qtoomre} and $T_{\rm mid}\propto r^{-1/2}$)
\beq\label{eq:rtoomreminp}
R_{\rm Toomre,min,t}=\left(\frac{4 \gamma-8}{4 \gamma-7}\right)^{\frac{1}{\gamma-2}} R_{\rm c}.
\eeq
This corresponds to ratios $R_{\rm Toomre,min,t}/R_{\rm c}$=1/4, 3/4, 0.791, 0.886 for  {a disk surface density exponent (Eq. \ref{eq:newprofilegeneral})} $\gamma$=3/2, 1, 0.9, and 1/2. 

{Second, we can consider a passively irradiated disk without viscous dissipation, where for orbital distances between about 0.4 and 84 AU, $T_{\rm mid}\propto r^{-3/7}$ (Chiang \& Goldreich \cite{chianggoldreich1997}). In this case, one finds an orbital distance $R_{\rm Toomre,min,p}$ where the minimal $Q_{\rm Toomre, min}$ occurs at}
\beq\label{eq:rtoomremint}
R_{\rm Toomre,min,p}=\left(\frac{7 \gamma-14}{7 \gamma-12}\right)^{\frac{1}{\gamma-2}} R_{\rm c}.
\eeq
{This corresponds to ratios $R_{\rm Toomre,min,p}/R_{\rm c}$=0.18, 0.71, 0.76, 0.87 for $\gamma$=3/2, 1, 0.9, and 1/2. We see that the results for an optically thin and a passively irradiated disk are similar. This is due to the fact that the temperature depends in both cases in a similar way on the distance.}

For $R_{\rm c}=30$ AU and $\gamma=0.9$, this leads {for a passively irradiated disk (which is the situation assumed here, see Fouchet et al. \cite{fouchetalibert2011}) to a $R_{\rm Toomre,min,p}=22.8$ AU,} close to the result ($\approx 25$ AU) seen in the simulations. The somewhat larger value in the simulations is likely due to  residual viscous heating.

In Figure \ref{fig:qminsigma} we plot the overall minimum $Q_{\rm Toomre,min}$ (i.e. the lowest value reached at any distance and time) as a function of  $\Sigma_{0}$ (or equivalently the initial disk mass), for disks with (red solid) and without (blue dotted) irradiation. In both cases, $Q$ follows a simple power law which scales as $\Sigma_{0}^{-1}$, as expected from Eq. \ref{eq:qtoomre}. 

One  sees that $Q$ is higher in disks with irradiation, which is expected as these disks are hotter and thus more stable, in particular in the outer regions where $Q$ becomes small, and where viscous heating is not important.
  
\begin{figure}
\begin{center}
\includegraphics[width=\columnwidth]{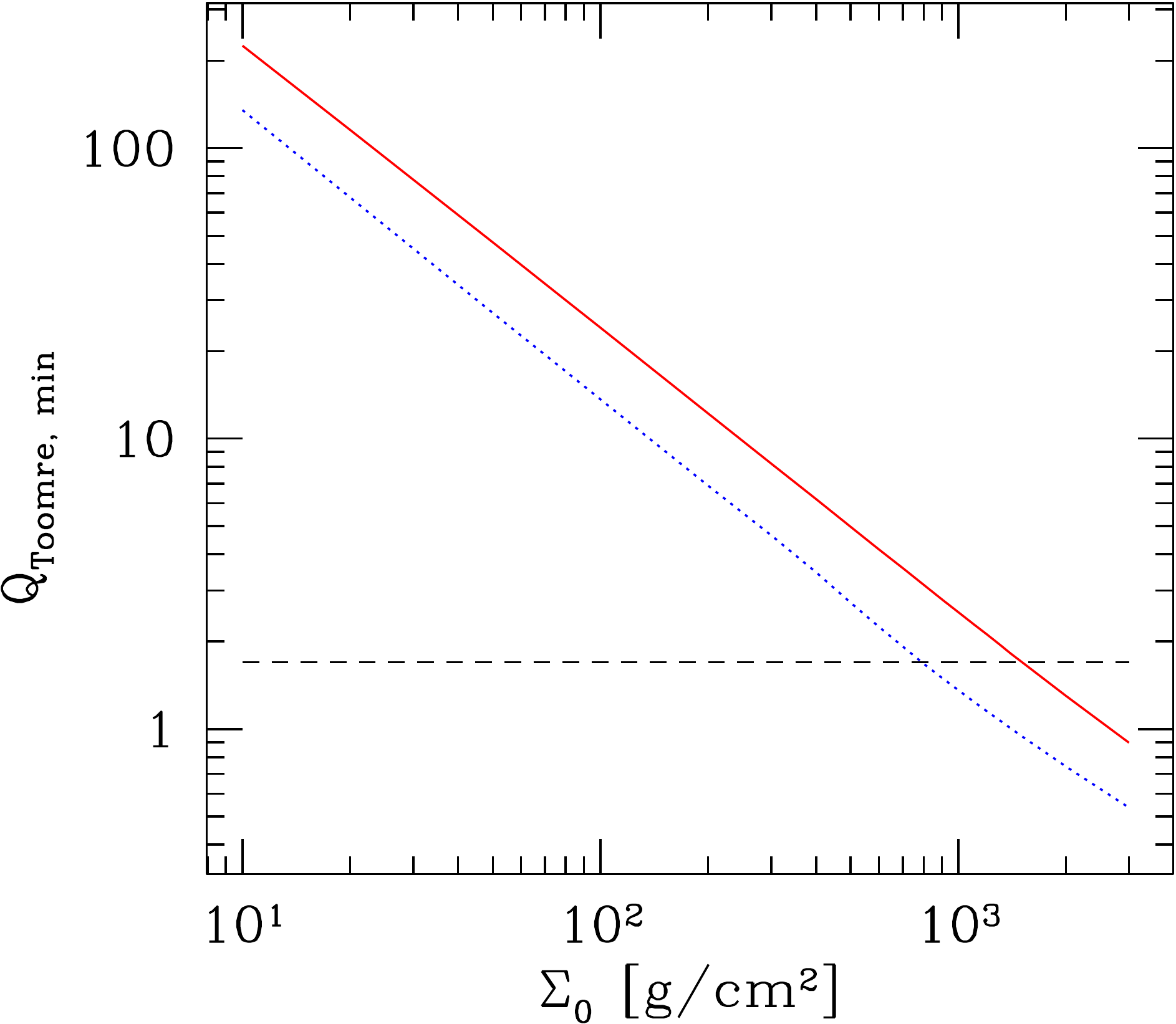}
\caption{{\footnotesize Overall minimal Toomre parameter $Q_{\rm Toomre,min}$ as a function of initial gas surface density at 5.2 AU, $\Sigma_{0}$ which is a direct proxy of the initial disk mass (Eq. \ref{eq:totaldiskmassparams}). The red solid line is for disks with irradiation, the dotted blue one is for disks where viscosity is the only heating source. The dashed line gives the critical value of {1.7}.  }}\label{fig:qminsigma}
\end{center}
\end{figure}

From Fig. \ref{fig:qminsigma}  we find a maximum initial gas surface density  $\Sigma_{\rm 0, max}$ (measured at 5.2 AU) leading to stable disks ($Q_{\rm Toomre,crit}$=1.7) for disks without irradiation of  {790} g/cm$^{2}$, corresponding to an initial disk mass of  {0.091} $\msun$, while for disks with irradiation, the maximum allowed value is  {1510} g/cm$^{2}$, corresponding to an initial mass of  {0.16} $\msun$. 

\subsection{Global instability}\label{sect:globalinst}
{At disk masses higher than $\sim$10-25\% the mass of the central object, the possibility of  global gravitational instability must be considered, too (e.g. Harsono et al. \cite{harsonoalexander2011}). } 

{The particular case of global instability to m=1 modes (one-armed spirals) has been studied numerically and analytically in Adams et al. (\cite{adamsruden1989}) and Shu et al. (\cite{shutremaine1990}). If this mode is amplified by the SLING mechanism (which is due to the fact that for a one-armed spiral, the star is displaced through the conservation of the center of mass), there is a finite threshold for the instability to set in, which is a function of the Toomre criterion.  For the case that $Q_{\rm Toomre}=1$ at the outer disk edge, it is  possible to derive analytically a critical disk mass, below which the disk is gravitationally stable to all modes. In a linear stability analysis, Shu et al. (\cite{shutremaine1990}) find for this situation a critical "maximum-mass solar nebula" at $\mdisk/(\mdisk+\mstar)=3/(4 \pi)$, corresponding to $\mdisk/\mstar=0.31$. }

{The  analytical analysis of Shu et al. (\cite{shutremaine1990}) suffers from a number of limitations (cf. Noh et al. \cite{nohvishniac1992}  for a discussion), and might  not  be applicable in the strict sense in the context here, as first, our disks have a smooth outer edge, so that they might not reflect the density waves, and second, for our disks, the minimal $Q$ does no occur at the outer disk edge, but further in (Eq. \ref{eq:rtoomremint}). If we compare the critical disk mass of Shu et al. (\cite{shutremaine1990}), $\mdisk/\mstar=0.31$, despite these limitation with the most massive disk stable to local  disturbances for the model here ($\mdisk/\mstar=0.16$), we find that the disks should also be globally stable.}  

{Since the works of Adams et al. (\cite{adamsruden1989}) and Shu et al. (\cite{shutremaine1990}), many studies have revisited the question of (global) disk stability. While a direct connection to the SLING mechanism can in general not be made (Nelson et al. \cite{nelsonbenz1998}), it is nevertheless found that at constant $Q$, with increasing $\mdisk/\mstar$, the character of the instability changes. At low $\mdisk$, thin, multi-armed structures develop which are characterized by high order patterns (m$\gtrsim5$), whereas at high $\mdisk$, global, low order (m=2-5) instabilities dominate (Nelson et al. \cite{nelsonbenz1998}). The transition between the two regimes occurs at approximately $\mdisk/\mstar$ = 0.2 - 0.4. These results are confirmed at much higher numerical resolution by  Harsono et al. (\cite{harsonoalexander2011}) and Lodato \& Rice (\cite{lodatorice2004}). The latter authors  find that, as long as the disk mass is less than 0.25$\mstar$ and the disk aspect ratio $H/R\lesssim0.1$, local effects dominate. Our massive disks are characterized by $H/R$ between 0.06 to 0.09 in their outer parts, so that they fulfill this criterion.}

{These results indicate that the transition from local to global instabilities might rather be a gradual one, and not setting in at a specific disk mass, as originally advocated by Shu et al. (\cite{shutremaine1990}). The disk mass where global effects become important seem to be at about 0.25$\mstar$ (Nelson et al. \cite{nelsonbenz1998}; Lodato \& Rice \cite{lodatorice2004}; Harsono et al. \cite{harsonoalexander2011}), which is comparable to the original criterion by Shu et  al. (\cite{shutremaine1990}). These results indicate that our most massive, Toomre-stable disk with 0.16 $\msun$ should also be stable to global modes. We must however keep in mind that the numerical simulations mentioned here did not use exactly the same surface density and temperature profile. Therefore, in order to get a firm conclusion, dedicated hydrodynamic simulations are necessary.}  

\section{Characteristic disk evolution under the influence of photoevaporation}
\label{sect:diskcharactevo}

\begin{figure*}
\begin{center}
\includegraphics[width=\columnwidth]{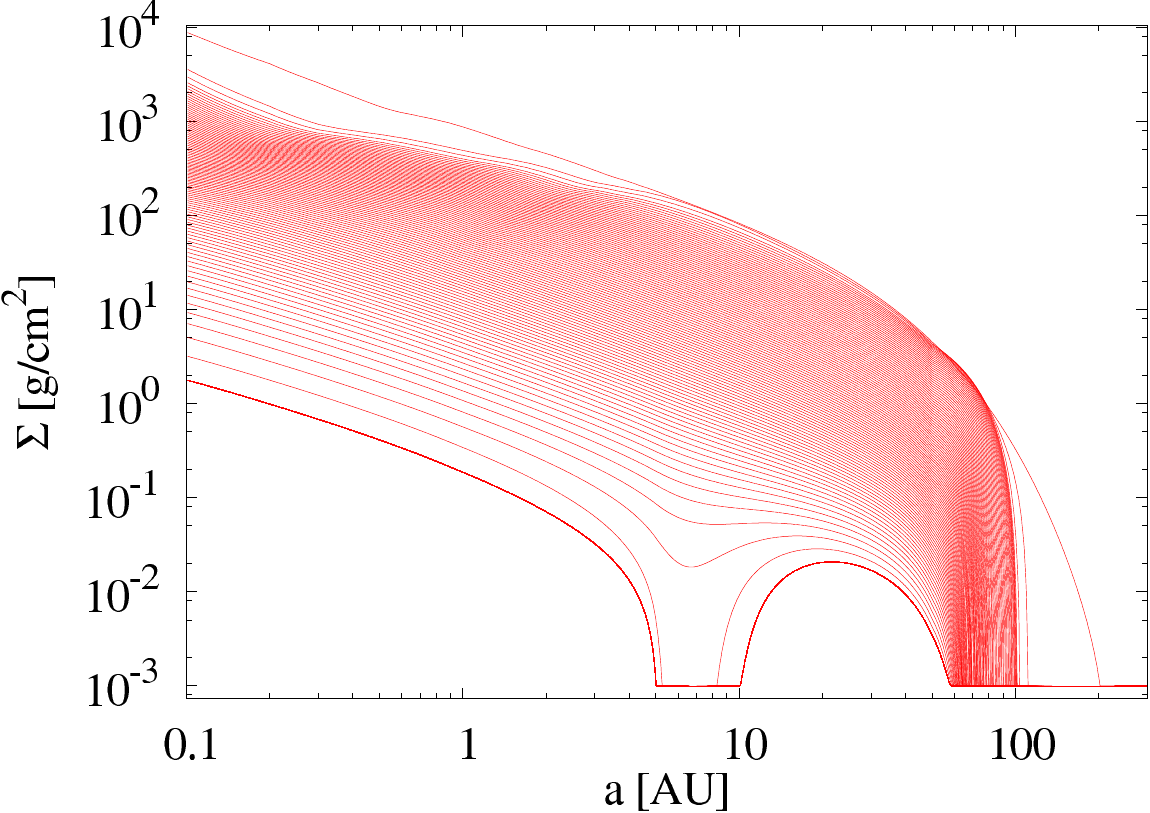}
\includegraphics[width=\columnwidth]{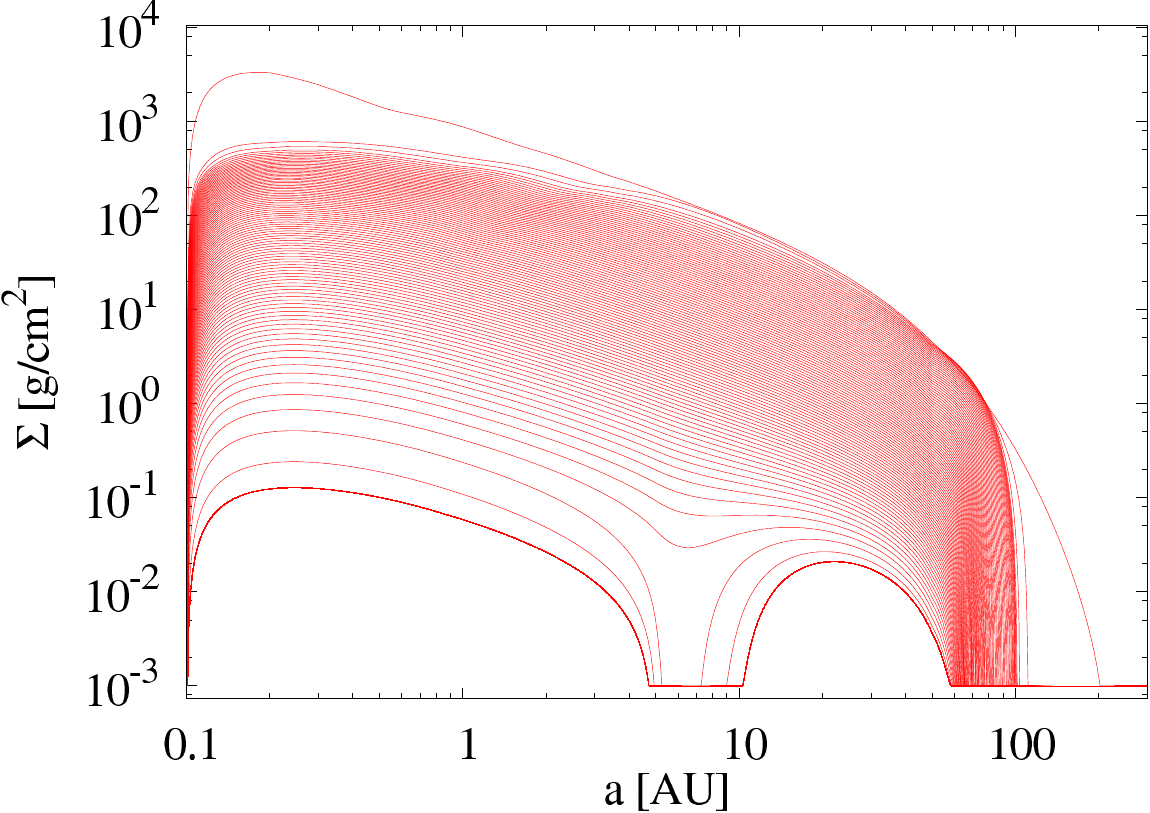}
\caption{{\footnotesize Gas surface density $\Sigma$ in a protoplanetary disk as a function of distance and time. The surface density is plotted in intervals of $2\times 10^{4}$ years. The uppermost line shows a state shortly after the beginning of disk evolution. The lowermost  line is the profile when the calculation is stopped. The two panels differ from each other by the boundary condition that is used at the inner edge of the disk at 0.1 AU, as described in the text. All other parameters are identical and given as $\mstar=1\msun$,  $\alpha=7\times 10^{-3}$ and $\Sigma_{0}=200$ g/cm$^{2}$.} }\label{fig:sigmaevo}
\end{center}
\end{figure*}

In this appendix we illustrate the characteristic evolution of the protoplanetary disks under the combined action of viscosity and photoevaporation as found in the update model, in order to check our results against previous studies. Figure \ref{fig:sigmaevo} shows the gas surface density as a function of time for a disk with $\alpha=7\times 10^{-3}$, $\Sigma_{0}=200$ g/cm$^{2}$, and a mean external photoevaporation rate over the lifetime of the disk of about $7\times10^{-9}$ $\msun$/yr. The effects of irradiation on the thermal structure of the disk is taken into account as described in Fouchet et al. (\cite{fouchetalibert2011}). These initial conditions result in a disk lifetime of 2.0 Myrs (remaining disk mass $10^{-5}\msun$). In the figure, a line is plotted all $2\times 10^{4}$ years, and the minimal allowed surface density is set to $10^{-3}$ g/cm$^{2}$.

The two panels of the figure differ only by the inner boundary condition at $R_{\min}=0.1$ AU. In the left panel, we use the same boundary conditions as in our earlier models which means that the flux through the innermost cell instantaneously adopts to the flux coming from further out. This is equivalent to the statement that the disk structure would continue to some smaller radius inside 0.1 AU, which is not modeled. In the right panel, $\Sigma$ is forced to fall to zero at 0.1 AU. Physically, we can associate the two situations with different sizes of the magnetospheric cavity, i.e. with a weak and a strong magnetic field of the host star, respectively (e.g. Bouvier et al. \cite{bouvieralencar2006}). A detailed description of the structure of the disk close to the host star taking into account magnetic fields will be presented in oncoming work (Cabral et al. in prep.). 

While the inner boundary condition has a very strong impact on the migration behavior of low-mass planets close to the star (Benitez-Llambay et al. \cite{benitezllambay2011}),  it has otherwise no effect on the characteristic evolution of the disk.

\subsection{Phases during disk evolution}
Figure \ref{fig:sigmaevo} shows that the evolution of the disk can approximatively be separated in four phases:
\begin{enumerate}
\item The initial exterior radius specified by the initial conditions gets quickly reduced by external photoevaporation to a radius where mass removal due to photoevaporation, and the viscous spreading of the disk are in a quasi-equilibrium as described by Adams et al. (\cite{adamshollenbach2004}). In the specific case, the radius decreases from initially about 200 AU to $\sim100$ AU. In the inner part, the disk very quickly evolves from the initial profile towards equilibrium.
\item In the second, dominant phase a quasi self-similar evolution of the disk occurs. The inner part of the disk ($r\lesssim10$ AU) is in near equilibrium (i.e. the mass accretion rate is nearly constant as a function of radius), and the slope of the gas surface density $\gamma$ is approximately -0.9 (but varying between -0.4 and -1.5 due to opacity transitions). The outer radius is slowly moving inwards, from about 100 AU to 60 AU. For more massive disk, this equilibrium radius is further out.
\item Once the surface density has fallen to about 0.01 - 0.1 g/cm$^{2}$ at $\sim$10 AU, a gap opens somewhat outside of $R_{g,{\rm II}}$. The evolution of the disk now speeds up which correspond to the so called ``two-timescale'' behavior (Clarke et al. \cite{clarkegendrin2001}).
\item Quickly afterwards, the total disk mass has fallen to 10$^{-5} \msun$, where we stop the calculation. Note that the evolution at very small $\Sigma$ is not important for our purpose of planet formation modeling. Therefore, we currently do not include the effect of the direct radiation field which would clear the disk quickly from inside out once the gap has opened (Alexander \& Armitage \cite{alexanderarmitage2009}). 
\end{enumerate}
Such an evolution is very similar to the findings of previous studies (Matsuyama et al. \cite{matsuyamajohnstone2003}; Clarke et al. \cite{clarkegendrin2001}). Note that the evolution can become more complex if there is additionally a planet accreting significant amounts of gas. This is the case if the planetary core becomes massive enough to trigger gas runaway accretion, so that a giant planet forms. Such a planet then effectively acts as a sink cell in the disk (see Paper I), so that another gap would form at its position towards the end of disk evolution, even if we neglect the tidal gap formation. 

With the inclusion of a detailed model for the photoevaporation, but also of the calculation of the luminosity of the planets in the gas runaway accretion phase (where forming giant planets can be bright, cf. Paper I), we are able to address  new observational constraints. As shown by Fouchet et al. (\cite{fouchetalibert2011}) we can use the disk structure to calculate the SED, in which we can now include the contribution from the  star, the disk but also of the growing planet. From a giant planet undergoing rapid gas accretion, we expect itself two contributions, similar as for an accreting star. A contribution in the IR coming from the internal luminosity, and a hard component from the accretion shock.  This will be addressed in a dedicated study (Mordasini et al. in prep). With upcoming observational facilities like ALMA, observing planet formation as it happens (Wolf et al. \cite{wolfgueth2002}; Klahr \& Kley \cite{klahrkley2006}) will become possible, and put a whole new class of constraints on formation models. 


\end{document}